\newcommand{\PRE}[1]{{#1}} 
\documentclass[12pt,aps,prd,preprint,tightenlines,superscriptaddress, amsmath,amssymb, nofootinbib]{revtex4}
\pdfoutput=1

\RequirePackage[colorlinks=true
,urlcolor=blue
,anchorcolor=blue
,citecolor=blue
,filecolor=blue
,linkcolor=blue
,menucolor=blue
,pagecolor=blue
,linktocpage=true
,pdfproducer=medialab
,pdfa=true
]{hyperref}

\usepackage{amsmath,amssymb,amsthm,amsfonts}
\allowdisplaybreaks
\usepackage{slashed} 
\usepackage{graphicx}
\usepackage{subfigure}
\usepackage{dcolumn}
\usepackage{hyperref}      
\usepackage{bm}
\usepackage{epsfig}
\usepackage{epstopdf}
\usepackage{setspace}
\usepackage[usenames, dvipsnames]{color}
\usepackage{slashed}
\usepackage{comment}

\newcommand{\be}{\begin{equation}\begin{aligned}}
\newcommand{\ee}{\end{aligned}\end{equation}}
\newcommand{\beq}{\begin{equation}}
\newcommand{\eeq}{\end{equation}}
\newcommand{\beqa}{\begin{eqnarray}}
\newcommand{\eeqa}{\end{eqnarray}}

\newcommand{\ifb}{\text{fb}^{-1}}
\newcommand{\iab}{\text{ab}^{-1}}

\newcommand{\mev}{\text{MeV}}
\newcommand{\gev}{\text{GeV}}
\newcommand{\tev}{\text{TeV}}

\newcommand{\mb}{\text{mb}}

\newcommand{\mm}{\text{mm}}
\newcommand{\cm}{\text{cm}}
\newcommand{\m}{\text{m}}
\newcommand{\km}{\text{km}}
\newcommand{\mrad}{\text{mrad}}
\newcommand{\murad}{\mu\text{rad}}

\newcommand{\kg}{\text{kg}}

\newcommand{\ns}{\text{ns}}
\newcommand{\s}{\text{s}}

\renewcommand{\eqref}[1]{Eq.~(\ref{#1})}
\newcommand{\eqsref}[2]{Eqs.~(\ref{#1}) and (\ref{#2})}
\newcommand{\secref}[1]{Sec.~\ref{sec:#1}}
\newcommand{\secsref}[2]{Secs.~\ref{sec:#1} and \ref{sec:#2}}

\newcommand{\figref}[1]{Fig.~\ref{fig:#1}}

\newcommand{\Figref}[1]{Figure~\ref{fig:#1}}

\newcommand{\appref}[1]{Appendix~\ref{sec:#1}}
\newcommand{\appsref}[2]{Appendices~\ref{sec:#1} and \ref{sec:#2}}

\newcommand{\slas}[1]{\! \not{\! \! #1}}
\newcommand{\slass}[1]{\! \not{\! #1}}

\newcommand{\name}{FASER}
\newcommand{\fullname}{{\bf F}orw{\bf A}rd {\bf S}earch {\bf E}xpe{\bf R}iment}

\newcommand{\lmin}{L_{\text{min}}}
\newcommand{\lmax}{L_{\text{max}}}

\begin{document}

\preprint{UCI-TR-2017-08}

\title{\PRE{\vspace*{1.0in}}
\name: \fullname\ at the LHC
\PRE{\vspace*{.4in}}}

\author{Jonathan L. Feng}
\email{jlf@uci.edu}
\affiliation{Department of Physics and Astronomy, University of
California, Irvine, CA 92697-4575 USA
\PRE{\vspace*{.1in}}}

\author{Iftah Galon}
\email{iftachg@uci.edu}
\affiliation{Department of Physics and Astronomy, University of
California, Irvine, CA 92697-4575 USA
\PRE{\vspace*{.1in}}}

\author{Felix Kling}
\email{fkling@uci.edu}
\affiliation{Department of Physics and Astronomy, University of
California, Irvine, CA 92697-4575 USA
\PRE{\vspace*{.1in}}}

\author{Sebastian Trojanowski\PRE{\vspace*{.2in}}}
\email{strojano@uci.edu }
\affiliation{Department of Physics and Astronomy, University of
California, Irvine, CA 92697-4575 USA
\PRE{\vspace*{.1in}}}
\affiliation{National Centre for Nuclear Research,\\Ho{\. z}a 69, 00-681 Warsaw, Poland
\PRE{\vspace*{.4in}}}


\begin{abstract}
\PRE{\vspace*{.2in}} New physics has traditionally been expected in the high-$p_T$ region at high-energy collider experiments.  If new particles are light and weakly-coupled, however, this focus may be completely misguided: light particles are typically highly concentrated within a few mrad of the beam line, allowing sensitive searches with small detectors, and even extremely weakly-coupled particles may be produced in large numbers there.  We propose a new experiment, \fullname, or \name, which would be placed downstream of the ATLAS or CMS interaction point (IP) in the very forward region and operated concurrently there.  Two representative on-axis locations are studied: a far location, $400~\m$ from the IP and just off the beam tunnel, and a near location, just $150~\m$ from the IP and right behind the TAN neutral particle absorber.  For each location, we examine leading neutrino- and beam-induced backgrounds.  As a concrete example of light, weakly-coupled particles, we consider dark photons produced through light meson decay and proton bremsstrahlung.  We find that even a relatively small and inexpensive cylindrical detector, with a radius of $\sim10~\cm$ and length of $5-10~\m$, depending on the location, can discover dark photons in a large and unprobed region of parameter space with dark photon mass $m_{A'}\sim10~\mev-1~\gev$ and kinetic mixing parameter $\epsilon \sim 10^{-7} - 10^{-3}$.  \name\ will clearly also be sensitive to many other forms of new physics.  We conclude with a discussion of topics for further study that will be essential for understanding \name's feasibility, optimizing its design, and realizing its discovery potential.
\end{abstract}


\maketitle

\section{Introduction}
\label{sec:introduction}

The search for new physics at the LHC has primarily focused on high-$p_T$ physics at the ATLAS~\cite{Aad:2008zzm} and CMS~\cite{Chatrchyan:2008aa} experiments.  This is not surprising, since new particles have traditionally been expected to be heavy.  There are also experiments exploring the very forward region, including ATLAS/ALFA/AFP/ZDC~\cite{White:2010zzd, Grinstein:2016sen, AbdelKhalek:2016tiv}, CMS/CASTOR/HFCAL~\cite{Gottlicher:2010zz}, LHCf~\cite{Adriani:2008zz}, and TOTEM~\cite{Anelli:2008zza}, but their physics programs are typically thought of as complementary to those of ATLAS and CMS, focusing on standard model (SM) topics, such as the structure of the proton and hadronic interactions, and providing precise measurements of the LHC luminosity.

If new particles are light and weakly coupled, however, the focus at the LHC on high-$p_T$ searches may be completely misguided. In the case of weakly-coupled physics, extraordinary event rates are required to discover very rare events. On the other hand, although the cross section for TeV-mass, strongly-interacting particles at the 13 TeV LHC is typically picobarns or less, the total inelastic $pp$ scattering cross section is $\sigma_{\text{inel}}(13~\tev) \approx 75~\mb$~\cite{Aaboud:2016mmw, VanHaevermaet:2016gnh} (see also results for $7~\tev$~\cite{Aad:2011eu, Antchev:2013iaa, Antchev:2013gaa, Antchev:2013haa, Chatrchyan:2012nj, Abelev:2012sea, Aad:2014dca, Aaij:2014vfa}, and $8~\tev$~\cite{Antchev:2013paa, Aaboud:2016ijx}),
with most of it in the very forward direction. This implies
\begin{equation}
N_{\text{inel}} \approx 2.3 \times 10^{16} \ (2.3 \times 10^{17})
\label{eq:ppcollisions}
\end{equation}
inelastic $pp$ scattering events for an integrated luminosity of $300~\ifb$ at the LHC ($3~\iab$ at the HL-LHC).  Even extremely weakly-coupled new particles may therefore be produced in sufficient numbers in the very forward region.  Moreover, such particles may be highly collimated, as they are typically produced within $\theta \sim \Lambda_{\text{QCD}} / E \sim\mrad$ of the beam line, where $\Lambda_{\text{QCD}} \simeq 250~\mev$ and $E \sim 100~\gev - 1~\tev$ is the energy of the particle.  This implies that even $\sim 100~\m$ downstream, such particles have only spread out $\sim 10~\cm$ in the transverse plane.  A small, inexpensive detector placed in the very forward region may therefore be capable of extremely sensitive searches, provided a suitable location can be found and the signal can be differentiated from the SM background.

Given this potential, we propose a new experiment, \fullname, or \name.\footnote{The acronym recalls another marvelous instrument that harnessed highly collimated particles and was used to explore strange new worlds.}  \name\ would be placed in the very forward region downstream of the ATLAS or CMS interaction point (IP). We study two representative locations for \name, both of which are on the beam collision axis, but are just off the beam line: a ``far'' location $400~\m$ downstream from the IP, where the beam is curved, and a ``near'' location $150~\m$ downstream from the IP, just behind the TAN neutral particle absorber, where the beam lines are split into two beam pipes. The far location requires a minimum of digging off the main LHC tunnel or may even make use of existing side tunnels. The near location is in the main tunnel. In addition, in \appref{off_axis_detector}, we discuss the possibility of placing the detector at a location 100 m from the IP in the main tunnel, but slightly off the beam collision axis. The feasibility and cost of placing a new experiment at these locations remains to be seen. Our goal here is to highlight the significant new physics opportunities of even a small, inexpensive detector at these locations, determine the virtues and drawbacks of the various locations, and motivate more detailed studies.

To examine the physics potential of \name\ and determine the detector requirements, we consider a concrete and well-studied example of light, weakly-coupled physics: dark photons $A'$.  For masses $m_{A'} \sim \mev - \gev$ and kinetic mixing parameter $\epsilon \sim 10^{-7} - 10^{-3}$, dark photons are produced in significant numbers in light meson decays and through proton bremsstrahlung, and they then decay with long lifetimes to electrons and other light SM particles.  The chain of processes
\begin{equation}
pp \to A' \, X \, , \quad A' ~\text{travels}~\sim {\cal O}(100)~\m \, , \quad A' \to e^+e^-, \, \mu^+ \mu^- \ ,
\end{equation}
then leads to the signal of two highly energetic, charged tracks created hundreds of meters downstream in the very forward region at ATLAS or CMS.

Assuming a sufficiently strong magnet to separate the opposite-charge tracks, the signal of two $\sim \tev$ charged tracks pointing back through rock or absorbers to the IP is spectacular.  We estimate leading backgrounds, including beam-induced backgrounds and those from neutrinos, and find that, given certain detector assumptions, these can be differentiated from the signal.  For dark photons, the physics potential can be realized by a small cylindrical detector with a total volume of $\sim 1~\m^3$ in the far location or $\sim 0.03~\m^3$ in the near location.  With both of the on-axis detector locations we consider, one may observe up to thousands of dark photon events and discover or exclude dark photons in a large swath of unprobed parameter space with $m_{A'} \sim 10~\mev - 500~\gev$ and $\epsilon \sim 10^{-7} - 10^{-3}$. 

The \name\ concept is complementary to other ideas to search for long-lived particles produced at ATLAS and CMS, including old proposals to look for late decays of weak-scale particles~\cite{Feng:2004yi,Hamaguchi:2004df,DeRoeck:2005cur}, as well as more recent proposals, such as MoEDAL-MAPP~\cite{Pinfold:2017dot}, MATHUSLA~\cite{Chou:2016lxi,Curtin:2017izq,Evans:2017lvd}, MilliQan~\cite{Ball:2016zrp}, and CODEX-b~\cite{Gligorov:2017nwh}.  All of these target long-lived particles produced in heavy particle decays. The \name\ detector specifically targets long-lived particles produced, for example, in light meson decays, which are collimated along the beam line, and so can be detected with a relatively small and inexpensive detector.  In terms of its physics objectives, \name\ is more similar to low-energy collider and beam dump experiments that have been proposed to search for dark photons and related light new particles~\cite{Battaglieri:2017aum}. Of course, \name\ differs from these, in that it makes use of the LHC and the ATLAS and CMS interaction regions and may run relatively inexpensively and concurrently with those existing programs, while benefiting from the large center-of-mass energy.

In \secref{lhcinfrastructure} we discuss the ATLAS and CMS very forward infrastructure and our representative detector locations.  In \secref{decays} we then review the properties of dark photons, and in \secref{production} we present the production rates and distributions of dark photons in the very forward regions at the LHC.  We discuss the \name\ detector requirements and SM backgrounds in \secsref{signalanddetector}{background}, respectively.  Given a possible realization of \name, we then determine the discovery potential for dark photons and present our results in \secref{results}. Our conclusions and outlook are summarized in \secref{conclusions}.  In \appsref{off_axis_detector}{brem}, we present results for the representative off-axis detector location and details of our proton bremsstrahlung rate calculation, respectively.

\section{LHC Very Forward Infrastructure}
\label{sec:lhcinfrastructure}

\begin{figure}[tb]
\centering
\includegraphics[width=0.99\textwidth]{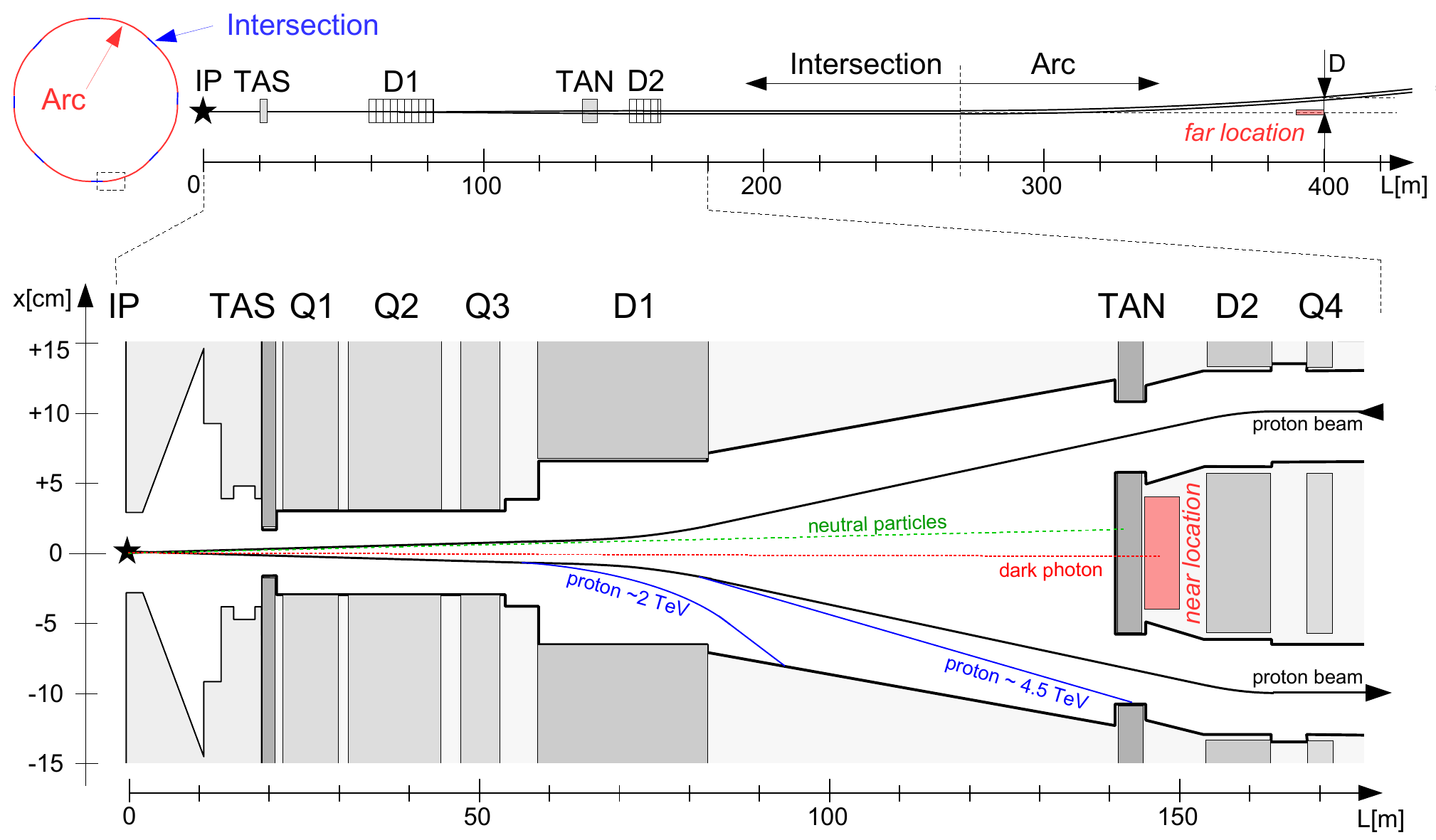}
\caption{Schematic drawings of the LHC ring and the current very forward infrastructure downstream from the ATLAS and CMS interaction points, along with the representative far and near on-axis detector locations for \name. Note the extreme difference in the transverse and longitudinal scales in the lower figure. Details of the geometry and sample tracks have been taken from Refs.~\cite{Mokhov:2003ha,Mokhov:2006nd,Fornieri:2015}. See the text for details.}
\label{fig:infrastructure}
\end{figure}

A schematic drawing of the LHC ring is shown in \figref{infrastructure}, along with a current expanded view of the region downstream from the ATLAS and CMS IPs and the infrastructure common to both. The LHC ring consists of eight straight segments, called intersections, and eight curved segments, called arcs. The IPs of CMS and ATLAS are at the center of $545~\m$ long intersections on opposite sides of the LHC ring. The stable particles emitted from the IPs in the very forward direction, including charged pions, muons, protons, neutrons, and photons, and also possibly dark photons and other new particles, propagate inside the beam pipe. About $20~\m$ downstream they encounter the TAS front quadrupole absorber~\cite{Hoyer1,Hoyer2}, a $1.8~\m$ long copper block with an inner radius of $17~\mm$, which shields the superconducting quadrupole magnets (Q1, Q2, and Q3) behind it from the forward radiation. The two proton beams are then separated by the inner beam separation dipole magnet D1 at a distance of $59-83~\m$ downstream~\cite{Aad:2013zwa}. The D1 magnet also deflects other charged particles produced at the IP. Note that multi-TeV charged particles are only slightly deflected from the proton beam and therefore can travel $\sim 10-100~\m$ before colliding with the beam pipe, as indicated by the blue sample tracks in \figref{infrastructure}. At the distance of $\sim140~\m$ downstream, the neutral particles produced at the IP are absorbed by the $3.5~\m$ thick TAN neutral particle absorber~\cite{Adriani:2015iwv}. In passing through the TAN, the two proton beams transition from a single beam pipe for both beams to individual beam pipes for each beam. At this point the horizontal separation between the inner edges of the beam pipes is $96~\mm$~\cite{Adriani:2006jd}. Finally, $153-162~\m$ downstream, the proton beams encounter the outer beam separation dipole magnet D2, which aligns the proton beams to be parallel. After the D2 magnet the proton beams have a horizontal separation of $194~\mm$~\cite{Aad:2013zwa}. 

New light particles may be predominantly produced in the very forward direction with very little transverse momentum relative to the beam collision axis. A forward detector, placed on the beam collision axis downstream from the IP, can then be sensitive to such new physics, provided it does not interfere with the beam lines. A careful examination of \figref{infrastructure} reveals two promising possibilities for on-axis detector locations. The first possibility is to place the detector far enough downstream that the beams have entered an arc. The straight intersection segment of the LHC ring ends at $L^*= 272~\m$ downstream from the IP and the arc's radius of curvature is $R = 3.13~\km$~\cite{Aad:2008zzm}.  The distance between the center of our detector and the beam pipe is, therefore, 
\be
\label{eq:tunnel_detector_dist}
D \approx \frac{(L-L^*)^2}{2R} \approx 1.6~\m \  \left(\frac{L-L^*}{100~\m}\right)^2 \ ,
\ee
where $L$ is the distance of the detector from the IP.  We therefore consider a far location for an on-axis detector, where the far end of the detector is $L=400~\m$ downstream and approximately $D \approx 2.6~\m$ from the beam pipe. The beam is roughly 1 m from the edge of the tunnel at this location, and so particles traveling on-axis must pass through many meters of rock to get to this location, providing essential shielding from background, as discussed in \secref{background}.  If shielding can be added in the main tunnel, lower values of $L$ and $D$ may be possible.  As we see in \secref{signalanddetector} for the case of dark photons, moving the detector even slightly closer, say to $L = 350~\m$ and $D \approx 1.0~\m$, may significantly improve signal rates.  We note that the tunnel TI18 (See Fig.~2.13 in Ref.~\cite{Buning:2004wk}), which was previously used by LEP as a connection between SPS and the main tunnel, might already be at (or close to) an optimal location for the far detector.

The second possibility we consider is to place the detector closer to the IP in an intersection region of the LHC ring. An appealing near location is $L=150~\m$, between the TAN, the D2 magnet, and the two proton beam pipes. This closer location is sensitive to new particles with shorter lifetimes and lower momenta, which, as we see, can greatly improve the signal yield and compensate for the fact that the detector's size is limited by the other detector components.  In this location, the role of the TAN as the D2 magnet radiation shield is leveraged to also provide shielding for \name.

In addition to these far and near locations, there is also accessible space on-axis between the beam pipes at $L \approx 180~\m$, near the TOTEM detector downstream from CMS, and at $L \approx 220~\m$, near the ALFA detector downstream from ATLAS.  These are also possible locations for \name, and may have lower backgrounds than our representative near location.  However, the near location at $L = 150~\m$ is expected to have larger signal rates, and so we limit our consideration to it here.

Finally, we note that the two proton beams cross at a small angle of $285~\murad$ relative to one another in the vertical (horizontal) plane at the ATLAS (CMS) IP~\cite{Aaboud:2016rmg,Bayatian:2006nff}.  At the far location $400~\m$ downstream (near location $150~\m$ downstream), this shifts the location of the center of an on-axis detector by $5.7~\cm$ ($2.1~\cm$).  The beam crossing angle is expected to increase to $590~\murad$ during the HL-LHC era~\cite{Apollinari:2015bam}, resulting in corresponding shifts of $12~\cm$ ($4.4~\cm$).  Throughout our analysis below, we assume that our detector is placed exactly on-axis with the correct offset included in either the ATLAS or CMS location.  The distinction between the vertical and horizontal offsets for ATLAS and CMS may play a role in optimizing the location of \name, however, especially in the HL-LHC era.  We note that there are many other possible changes for the HL-LHC era.  Below, we comment on particularly relevant changes that are currently under discussion, but for our calculations, for concreteness, we assume the current LHC beam and infrastructure configurations.

\section{Dark Photon Decays}
\label{sec:decays}

Dark photons~\cite{Okun:1982xi,Galison:1983pa,Holdom:1985ag,Boehm:2003hm,Pospelov:2008zw} provide a concrete and well-studied example of light, weakly-coupled new particles. They arise when the SM is supplemented by a hidden sector, which may be motivated, for example, by the need for dark matter. If the hidden sector contains a (broken) $U(1)$ symmetry, the hidden gauge boson generically mixes with the SM photon through the renormalizable coupling $\tilde{F}^{\mu \nu} \tilde{F}'_{\mu \nu}$, where $\tilde{F}_{\mu\nu}$ and $\tilde{F}'_{\mu\nu}$ are the field strengths of the SM and hidden gauge bosons, respectively.  After a field re-definition to remove this kinetic coupling, the resulting Lagrangian is
\begin{align}
	\mathcal L &=
	-\frac 14 F_{\mu\nu}F^{\mu\nu}
	-\frac 14 F'_{\mu\nu}F'^{\mu\nu}
	+ \frac{1}{2} m_{A'}^2 A'^2
	+ \sum_f \bar{f} (i \slass{\partial} - e q_f \slas{A} - \epsilon e q_f \slas{A'} - m_f)  f \ ,
	\label{eq:Lagrangian}
\end{align}
where $F_{\mu\nu}$ and $F'_{\mu\nu}$ are the field strengths of the photon $A$ and dark photon $A'$, respectively, the dark photon has mass $m_{A'}$ and kinetic mixing parameter $\epsilon$, and $f$ represents SM fermions with electric charges $q_f$ and masses $m_f$.  

The dark photon may decay to $e^+e^-$ pairs throughout the parameter space we study.  The partial decay width is
\begin{equation}
\Gamma_e \equiv \Gamma (A' \to e^+ e^-) 
=  \frac{\epsilon^2 e^2 m_{A'}}{12 \pi} 
\Biggl[ 1 - \left( \frac{2 m_e} {m_{A'}} \right)^{\! \! 2} \ \Biggr]^{1/2} 
\Biggl[ 1 + \frac{2 m_e^2}{m_{A'}^2} \Biggr] \ .
\end{equation}
For $m_{A'} > 2 m_{\mu}$, decays to muons and a number of hadronic states are also possible.  We assume that there are no non-SM decays.  In this case, the full dark photon decay width is 
\begin{align}
\Gamma_{A'} = \frac{\Gamma_e }{B_e(m_{A'}) } \ ,
\end{align}
where $B_e(m_{A'})$ is the branching ratio to $e^+e^-$ pairs of a dark photon with mass $m_{A'}$. The function $B_e(m_{A'})$ may be extracted from measurements of $e^+ e^-$ scattering at center-of-mass energy equal to $m_{A'}$. It varies from $40\%$ to $100\%$ for dark photon masses between $1$ and $500~\mev$~\cite{Buschmann:2015awa}. 

In the limit $E_{A'} \gg m_{A'} \gg m_e$, the dark photon decay length is
\begin{equation}
\label{eq:ap_decay_length}
\bar{d} = c \, \frac{1}{\Gamma_{A'}} \, \gamma_{A'} \beta_{A'} \approx (80~\m ) \ B_e \left[ \frac{10^{-5}}{\epsilon} \right]^2 
\left[ \frac{E_{A'}}{\tev} \right] \left[ \frac{100~\mev}{m_{A'}} \right]^2 \ ,
\end{equation}
where we have normalized $\epsilon$ and $E_{A'}$ to typical values that yield observable event rates.  We find that for $m_{A'} \sim 10 - 100~\mev$ and $\epsilon \sim 10^{-5}$, dark photons with $E_{A'}\sim \tev$ have a decay length of ${\cal O} (100)~\m$, the length scale of the LHC accelerator infrastructure in the intersection.

\section{Dark Photon Production in the Forward Region}
\label{sec:production}

Dark photon couplings to fermions, shown in \eqref{eq:Lagrangian}, are inherited from photon couplings with the modification $e\to\epsilon e$. As a result, the dark photon production mechanisms follow those of the photon, up to mass-related effects. For $m_{A'}$ in the sub-GeV range, $pp$ collisions at center-of-mass energy $\sqrt{s}=13~\tev$ give rise to three dominant sources of forward dark photons: rare decays of mesons to dark photons, proton bremsstrahlung of dark photons in coherent proton scattering, and direct dark photon production in QCD processes. Throughout the rest of the paper, we use two representative parameter-space points to illustrate the dark photon kinematics: 
\be
\label{eq:ps_points}
 \textbf{Low-mass point: }  m_{A'}&=20~\mev \, , \ \epsilon=10^{-4}\ , \\
 \textbf{High-mass point: }  m_{A'}&=100~\mev \, , \ \epsilon=10^{-5} \ .
\ee
These points have not been excluded by current dark photon searches, but, as we see, are within the region that may be probed by \name.

\subsection{Meson Decays}
\label{sec:meson_decays}

Light hadrons $h$, which are abundantly produced in $pp$ collisions, act as dark photon sources via the decay $h\to A'X$, provided there are SM decay modes $h\to \gamma X$ and $m_h - m_X > m_{A'}$. Of particular interest are the light neutral mesons $\pi^0$ and $\eta$, which are produced in large multiplicities and decay to two photons with large branching fractions. These decay modes are induced by the chiral anomaly of the light quark flavor group and have branching fractions~\cite{Batell:2009di}
\begin{eqnarray}
\label{eq:BR_pi0_to_ap}
B(\pi^0 \to A' \gamma) &=& 2\epsilon^2
\left(1 - \frac{ m^2_{A'} }{m_{\pi^0}^2} \right)^3
B(\pi^0 \to \gamma \gamma) \ , \\
 \label{eq:BR_eta_to_ap}
B(\eta \to A' \gamma) &=& 2\epsilon^2
\left(1 - \frac{ m^2_{A'} }{m_\eta^2} \right)^3 B(\eta\to \gamma \gamma) \ ,
\end{eqnarray}
where $B(\pi^0 \to \gamma \gamma) \simeq 0.99$, and $B(\eta \to \gamma \gamma) \simeq 0.39$~\cite{Olive:2016xmw}. The former is dominant at $m_{A'} < m_{\pi^0}$, while the latter is relevant for $m_{\pi^0} < m_{A'} < m_{\eta}$. The decays of heavier hadrons also contribute to dark photon production, but they typically suffer from small branching ratios to photons and suppressed production multiplicities in $pp$ collisions. Examples of interesting decay modes of heavier mesons are $B(\rho^0\to\pi^+\pi^-\gamma)\simeq 10^{-2}$, $B(\rho\to\pi\gamma)\simeq 4.5\times 10^{-4}$, $B(\omega\to\pi^0\gamma)\simeq 0.084$, $B(\eta'\to\rho^0 \gamma)\simeq 0.289$, $B(J/\psi \to \gamma gg)=0.088$, and $B(\Upsilon \to \gamma gg)= 0.022$. However, in this work, we do not expect such contributions to dramatically improve our results, and therefore do not include them.  

Determination of the forward dark photon event yield requires a reliable estimate of the forward $\pi^0$ and $\eta$ spectra and multiplicities in high-energy $pp$ collisions. Such estimates, which have traditionally relied on data from ultra-high-energy cosmic-ray experiments, have been greatly improved in recent years with the availability of forward high-energy scattering data from the LHC experiments~\cite{N.Cartiglia:2015gve}: ATLAS/ALFA/AFP/ZDC, CMS/CASTOR/HFCAL, LHCf, and TOTEM. Three Monte-Carlo simulation tools that have been tuned to match this data, EPOS-LHC~\cite{Pierog:2013ria}, QGSJET-II-04~\cite{Ostapchenko:2010vb}, and SIBYLL 2.3~\cite{Ahn:2009wx,Riehn:2015oba}, are available via the CRMC simulation package~\cite{CRMC}. 

We have compared the predictions of the three codes for $\pi^0$ and $\eta$ production in proton-proton collisions at $\sqrt s = 13~\tev$. \Figref{multiplicity} (left) shows the normalized per-event multiplicity distributions.  For completeness, we also show the EPOS-LHC multiplicity predictions for various additional hadrons in \figref{multiplicity} (right). \Figref{pionptheta} shows the distribution of produced $\pi^0$ and $\eta$ mesons in the $(\theta, p)$ plane, where $\theta$ and $p$ are the meson's angle with respect to the beam axis and momentum, respectively.   Given that the simulations have been tuned to the LHC data, the consistency of the results comes as no surprise, with the mild differences stemming from the physics assumptions employed in each model.  For example, QGSJET-II-04 does not include strange mesons. In the remainder of this work, we use EPOS-LHC to derive our results. 

The clustering of events in \figref{pionptheta} around the (log-log) line $p \, \theta \approx p_T = \Lambda_{\text{QCD}} \simeq 0.25~\gev$ is indicative of the characteristic momentum transfer scale and is an important consistency check. The added value of the simulations is the estimation of the spread around this line. Particularly interesting is the large multiplicity of high-momentum mesons with $p > 100~\gev$ at small angles $\theta < 10^{-3}$, which are efficient sources of forward, high-momentum dark photons.  

\begin{figure}[tbp]
\includegraphics[width=0.47\textwidth]{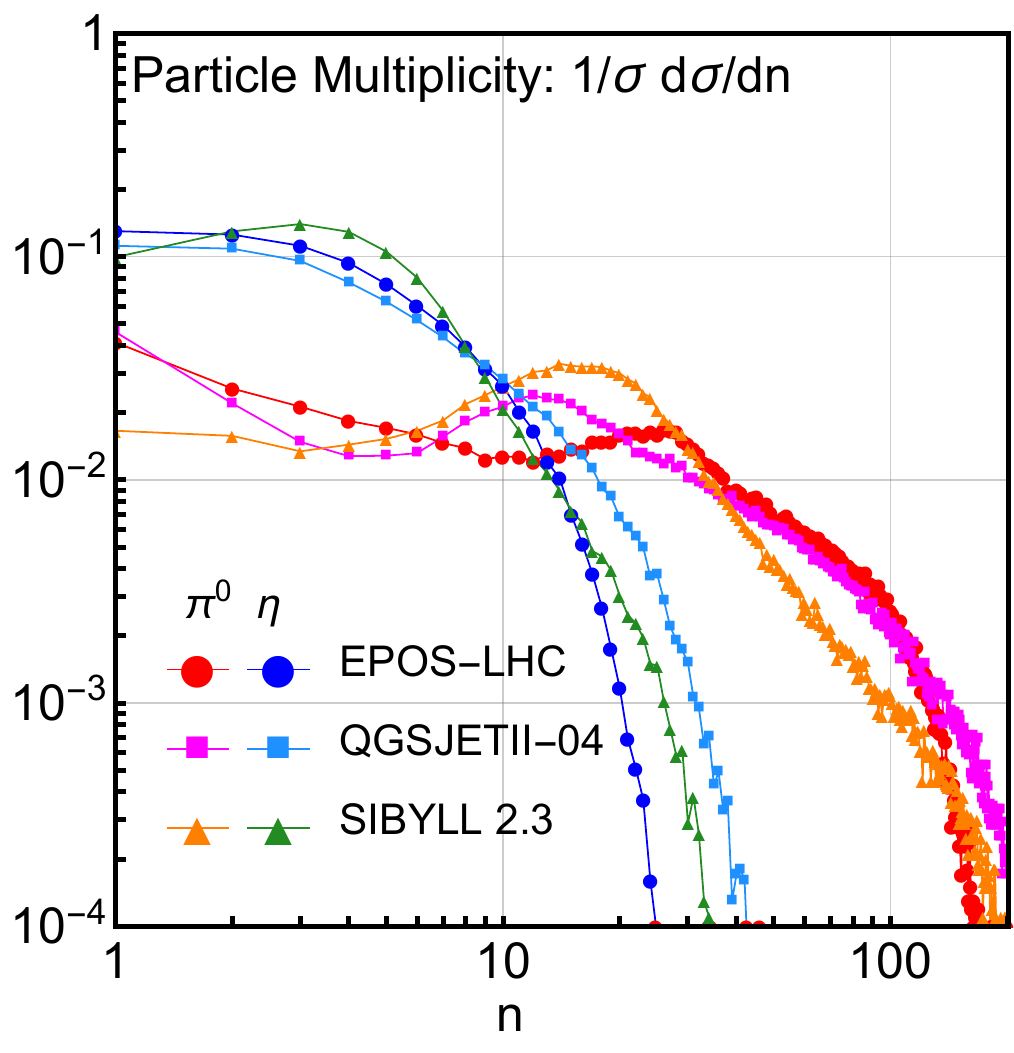} \quad
\includegraphics[width=0.47\textwidth]{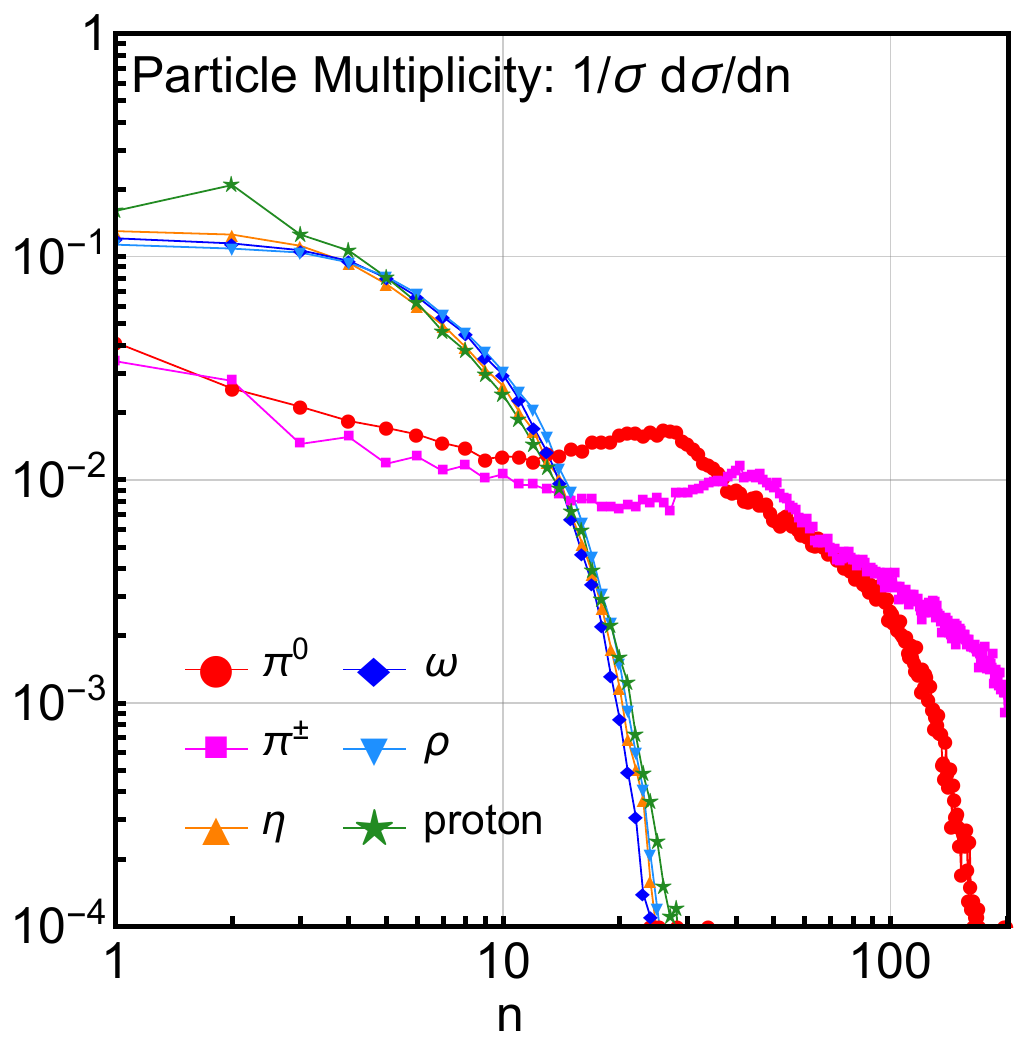}
\caption{Particle multiplicities in $13~\tev$ $pp$ collisions at the LHC.  Left: $\pi^0$ and $\eta$ multiplicities from EPOS-LHC~\cite{Pierog:2013ria} (circles), QGSJET-II-04~\cite{Ostapchenko:2010vb} (squares), and SIBYLL 2.3~\cite{Ahn:2009wx,Riehn:2015oba} (triangles). Right: $\pi^0$, $\pi^{\pm}$, $\eta$, $\omega$, $\rho$, and $p$ multiplicities from EPOS-LHC~\cite{Pierog:2013ria}.}
\label{fig:multiplicity}
\end{figure}

\begin{figure}[tbp]
\includegraphics[width=0.32\textwidth]{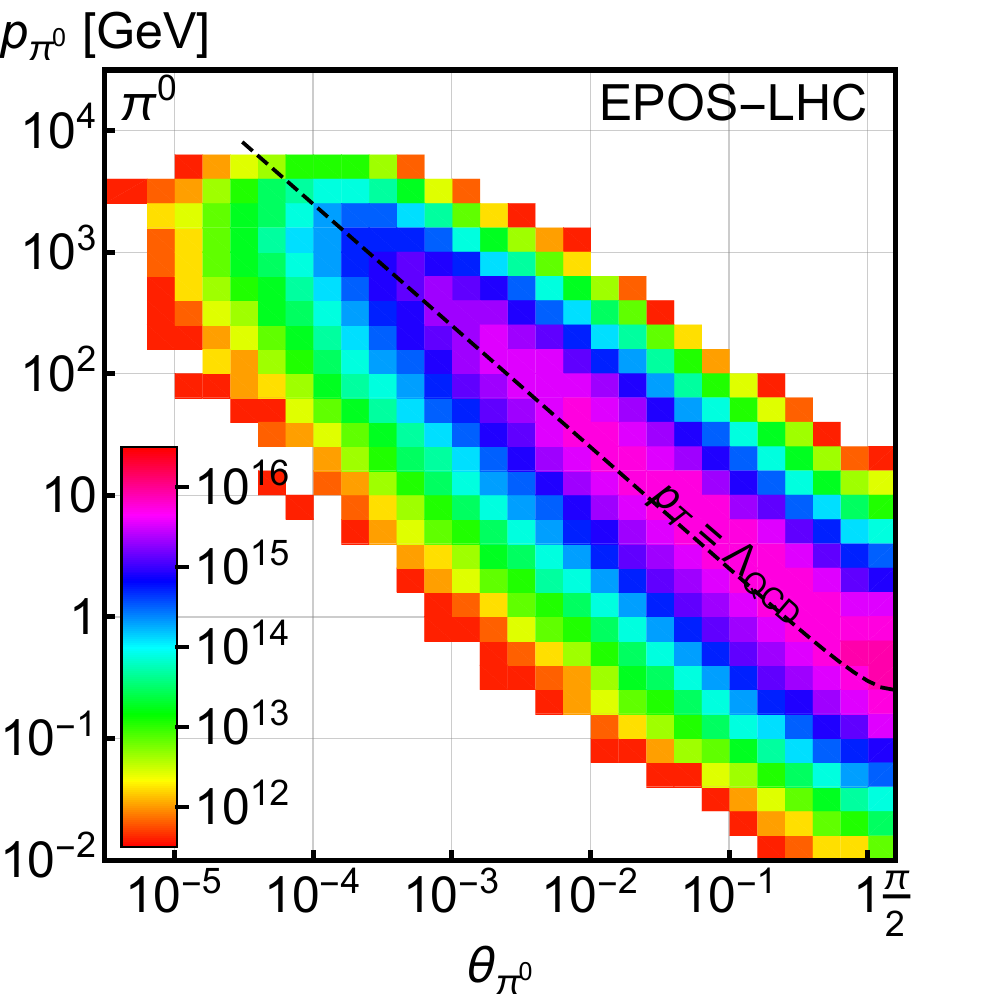}
\includegraphics[width=0.32\textwidth]{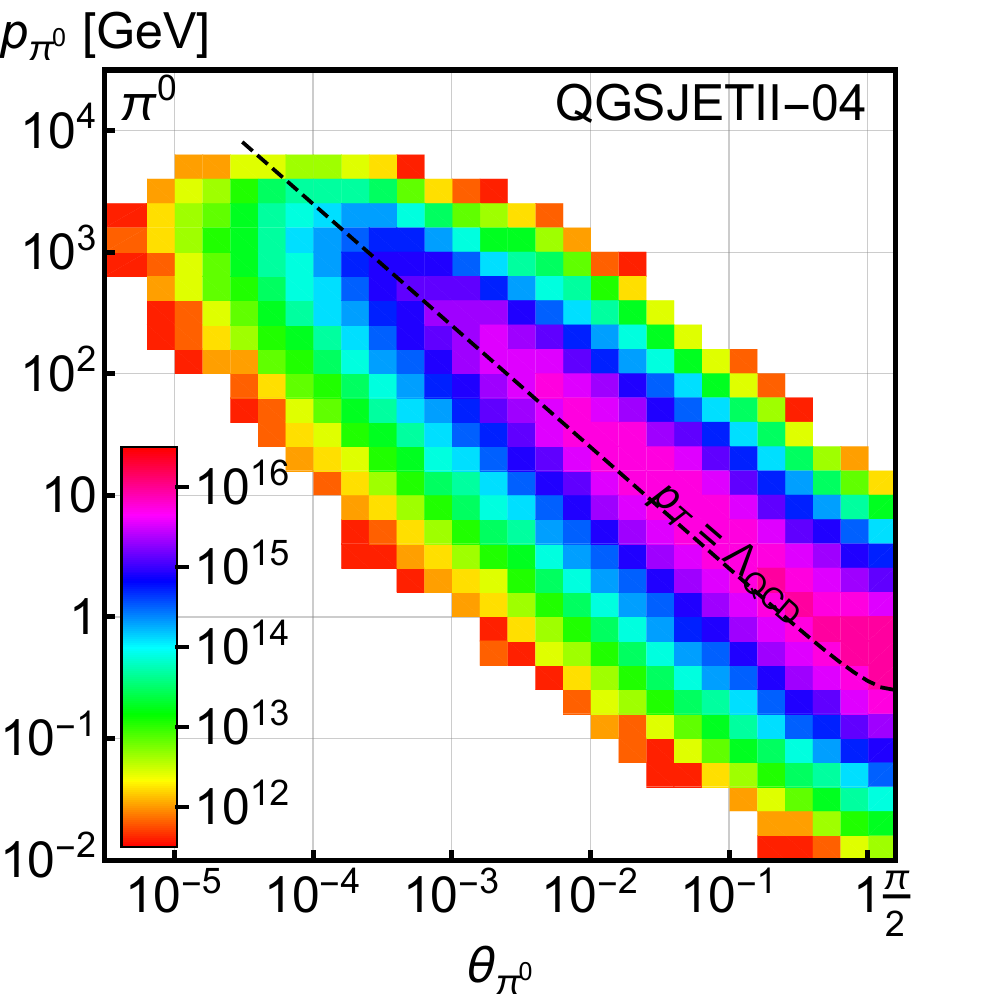}
\includegraphics[width=0.32\textwidth]{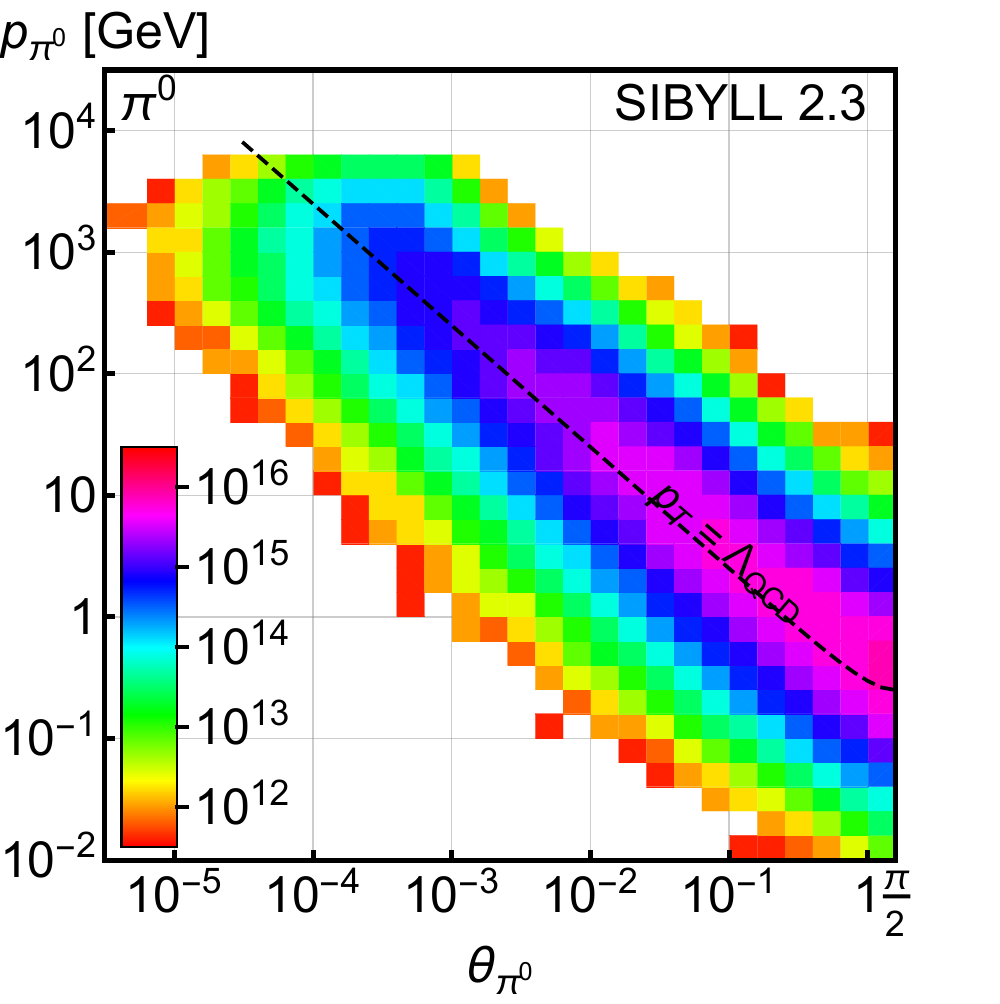}\\
\includegraphics[width=0.32\textwidth]{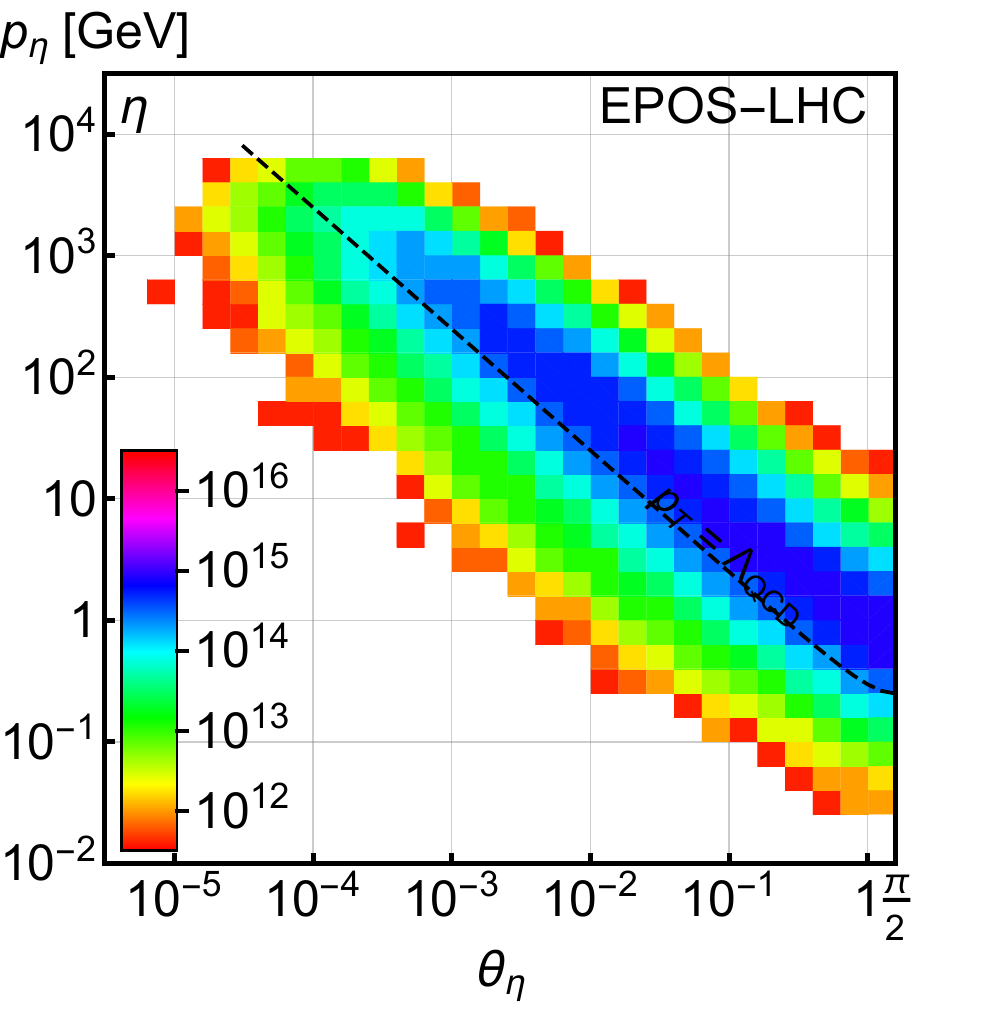}
\includegraphics[width=0.32\textwidth]{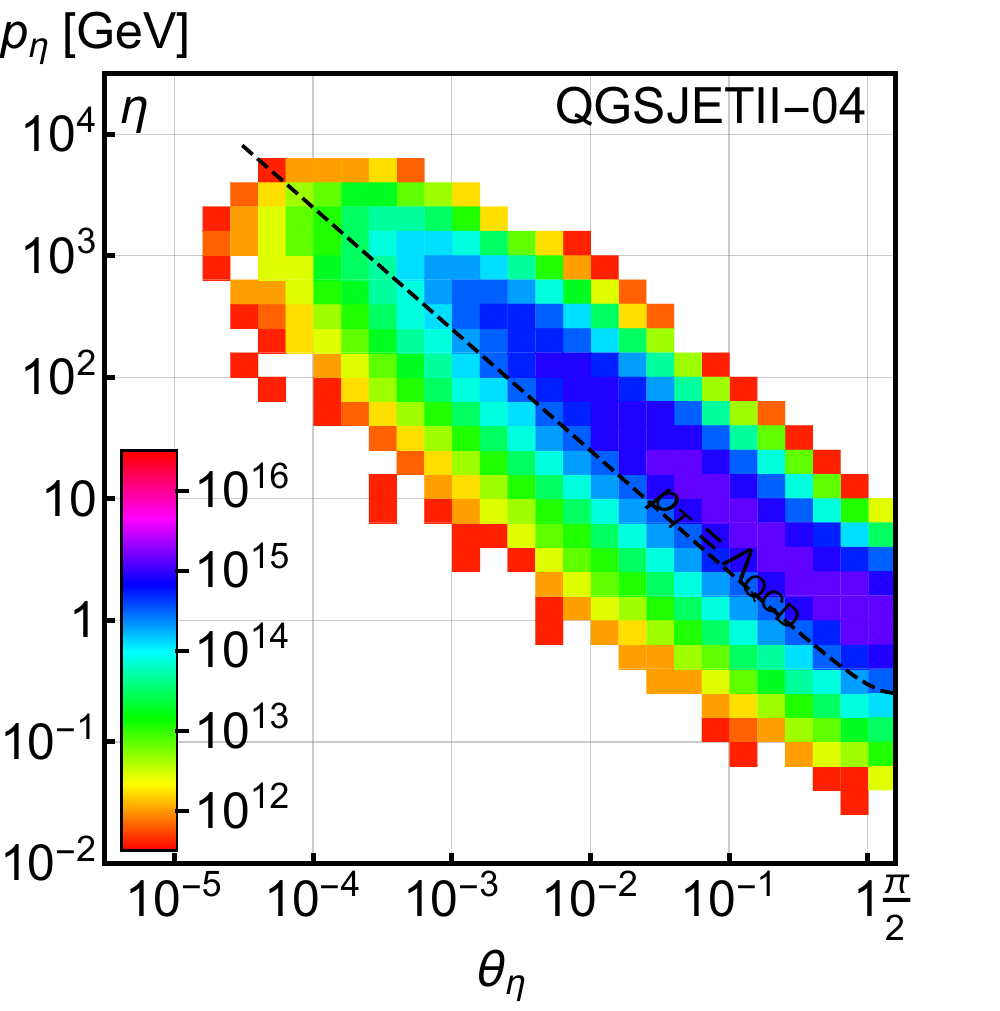}
\includegraphics[width=0.32\textwidth]{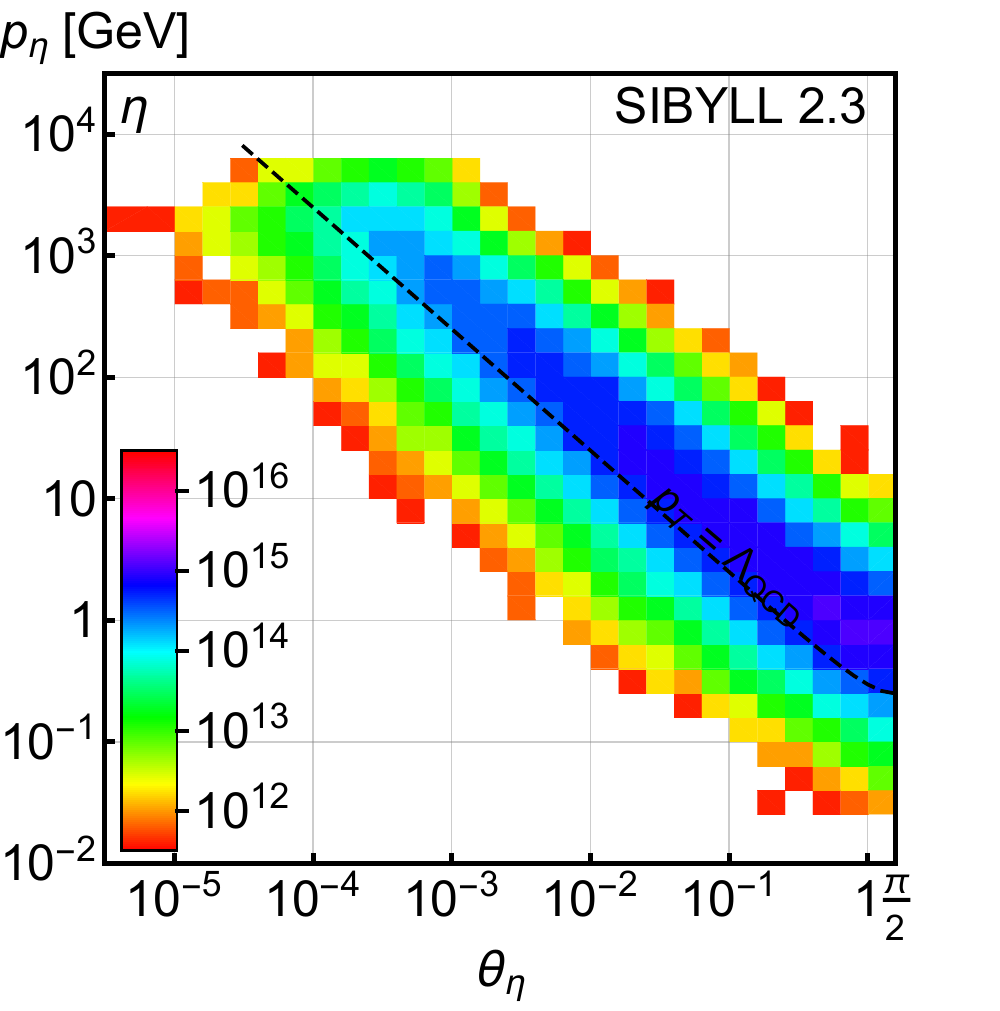}
\caption{Distribution of $\pi^0$ (top) and $\eta$ (bottom) mesons in the $(\theta, p)$ plane, where $\theta$ and $p$ are the meson's angle with respect to the beam axis and momentum, respectively.  The different panels show results from the simulation codes EPOS-LHC~\cite{Pierog:2013ria} (left), QGSJET-II-04~\cite{Ostapchenko:2010vb} (center) and SIBYLL 2.3~\cite{Ahn:2009wx,Riehn:2015oba} (right).  The total number of mesons is the number produced in one hemisphere ($0 < \cos \theta \le 1$) in 13 TeV $pp$ collisions at the LHC with an integrated luminosity of $300~\ifb$.  The bin thickness is $1/5$ of a decade along each axis.  The dashed line corresponds to $p_T = p \sin \theta = \Lambda_{\text{QCD}} \simeq 250~\mev$. }
\label{fig:pionptheta}
\end{figure}

To derive the dark photon distributions from the meson distributions, we decay the $\pi^0$ and $\eta$ mesons in the Monte-Carlo sample, scaling the yield according to \eqsref{eq:BR_pi0_to_ap}{eq:BR_eta_to_ap} and normalizing to $300~\ifb$. The $\pi^0$ and $\eta$ are pseudoscalars, and so dark photons are produced isotropically in the mesons' rest frames.  To avoid mis-sampling in the Monte-Carlo, for each $\pi^0$ and $\eta$, we perform a fine-grid scan over the $A'$ angles in the meson's rest frame and normalize accordingly. The dependence of these results on the $(m_{A'}, \epsilon)$ parameter-space point is shown in the $(\theta, p)$ distributions of \figref{aprimeptheta} for the two representative points of \eqref{eq:ps_points}. The left and middle columns represent the contributions from $\pi^0 \to A' \gamma$ and  $\eta \to A' \gamma$, respectively; the right column displays the proton bremsstrahlung contribution to be discussed below. Most important for this study, \figref{aprimeptheta} shows that significant numbers of forward dark photons with momenta $\sim 1~\tev$ and decay lengths $\bar{d} \sim 100~\m$ are expected. 

\begin{figure}[tbp]
\includegraphics[width=0.32\textwidth]{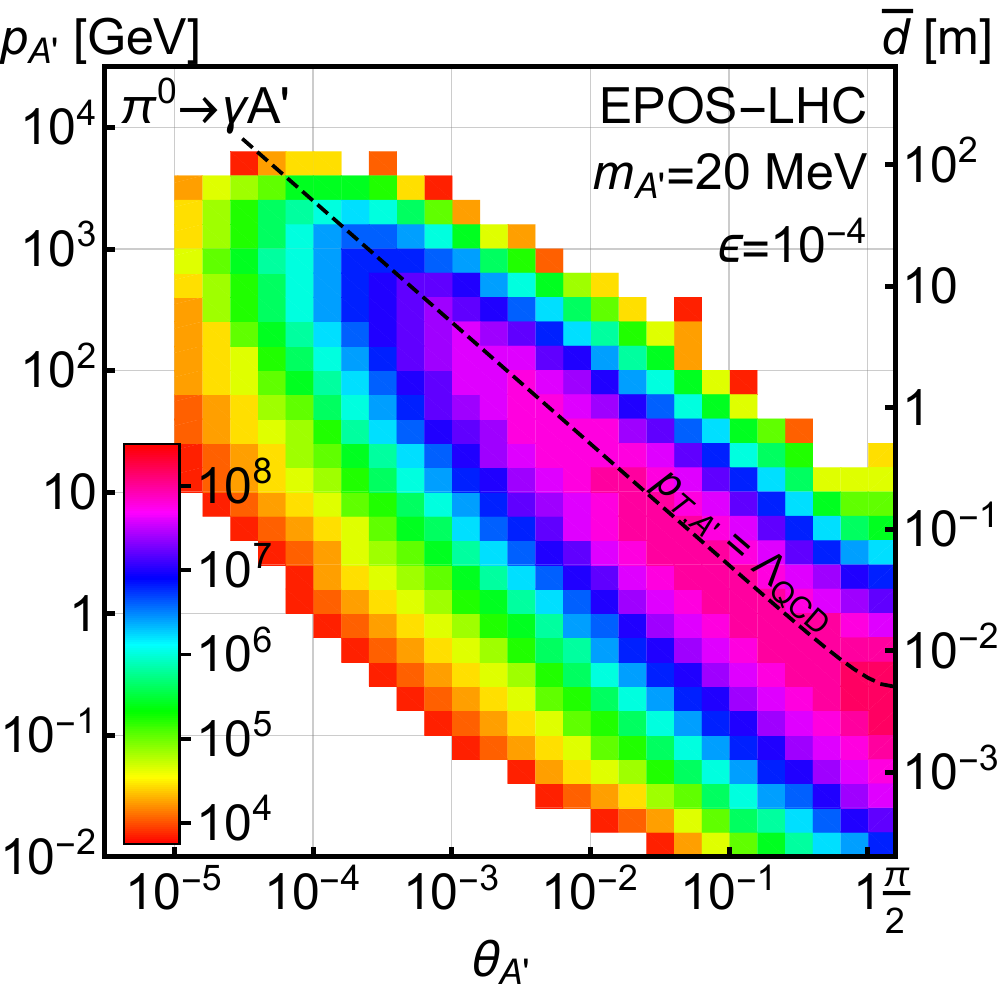}
\includegraphics[width=0.32\textwidth]{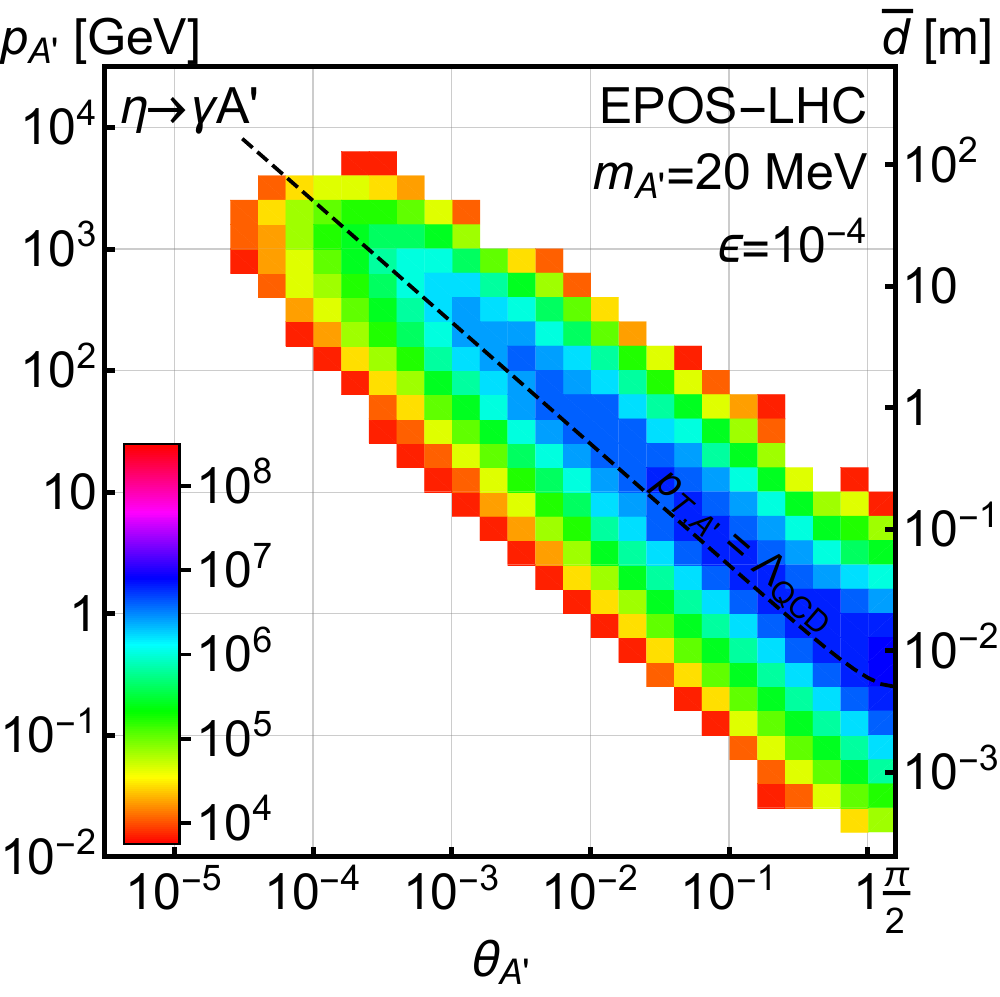}
\includegraphics[width=0.32\textwidth]{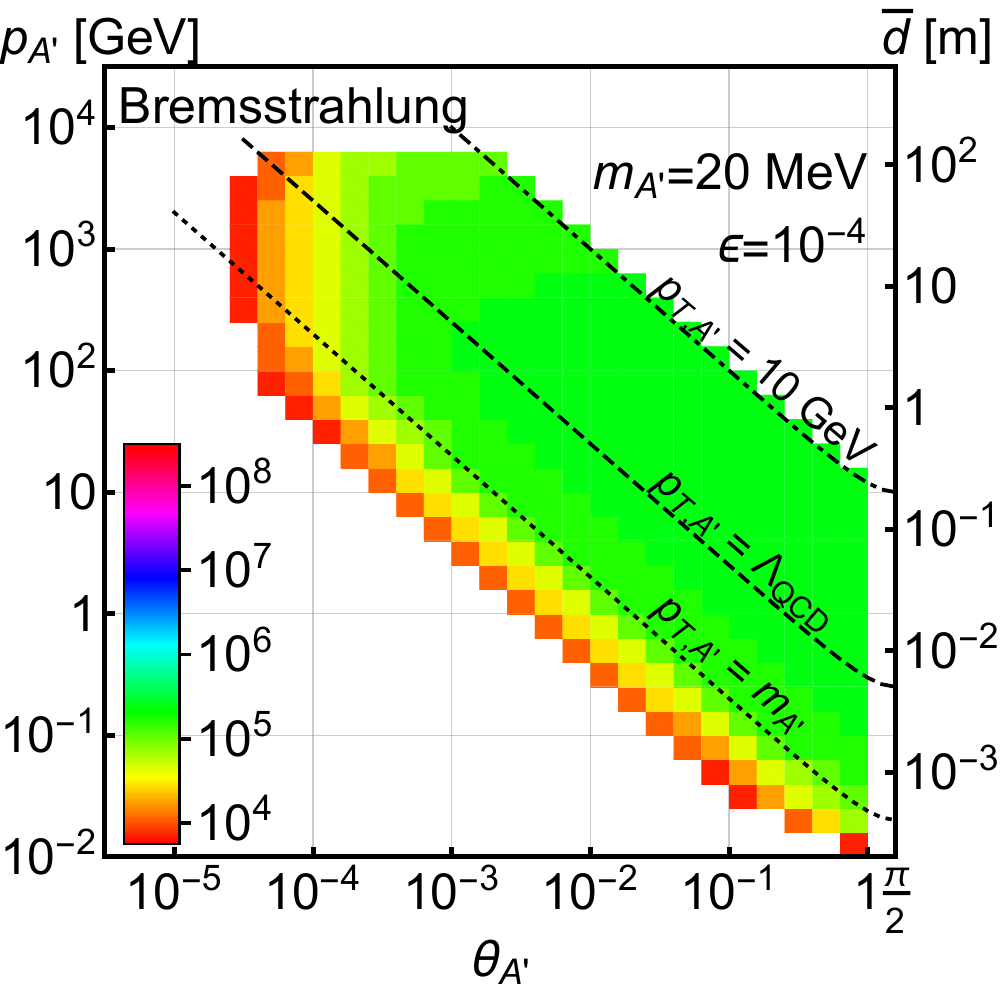}\\
\includegraphics[width=0.32\textwidth]{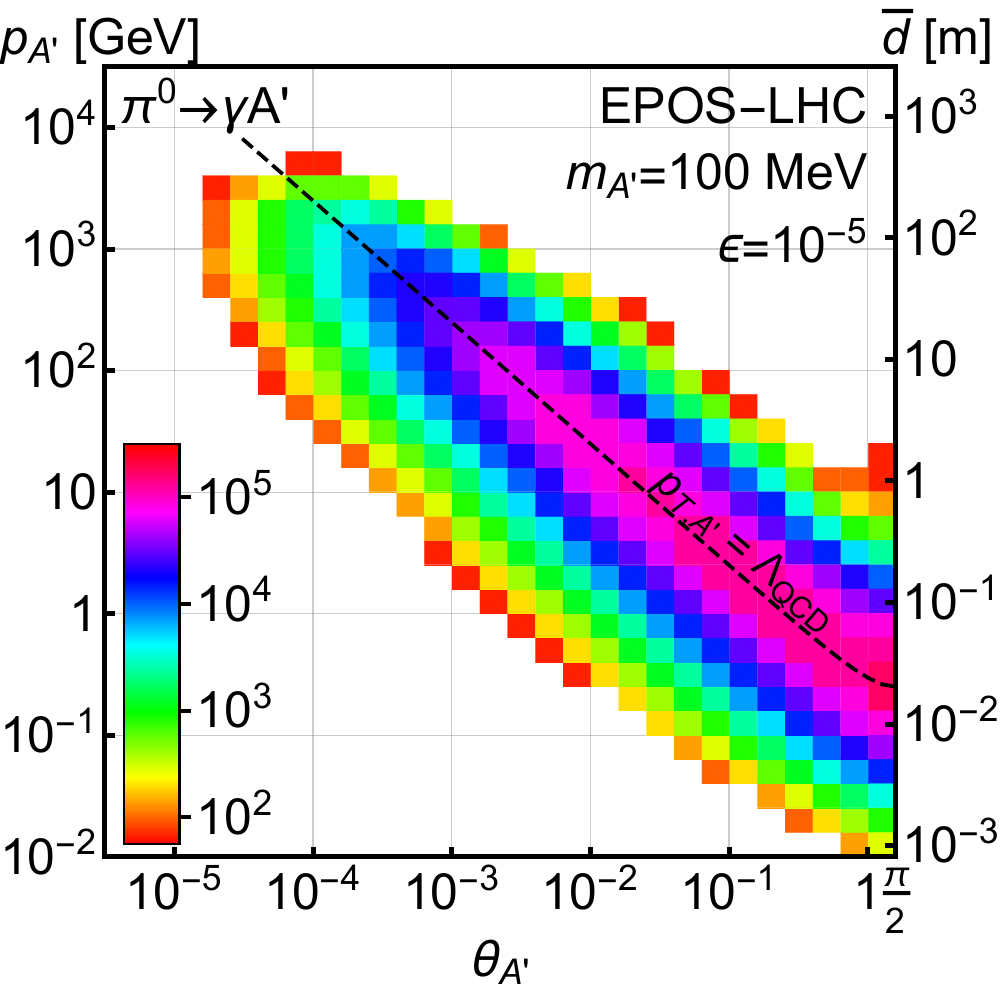}
\includegraphics[width=0.32\textwidth]{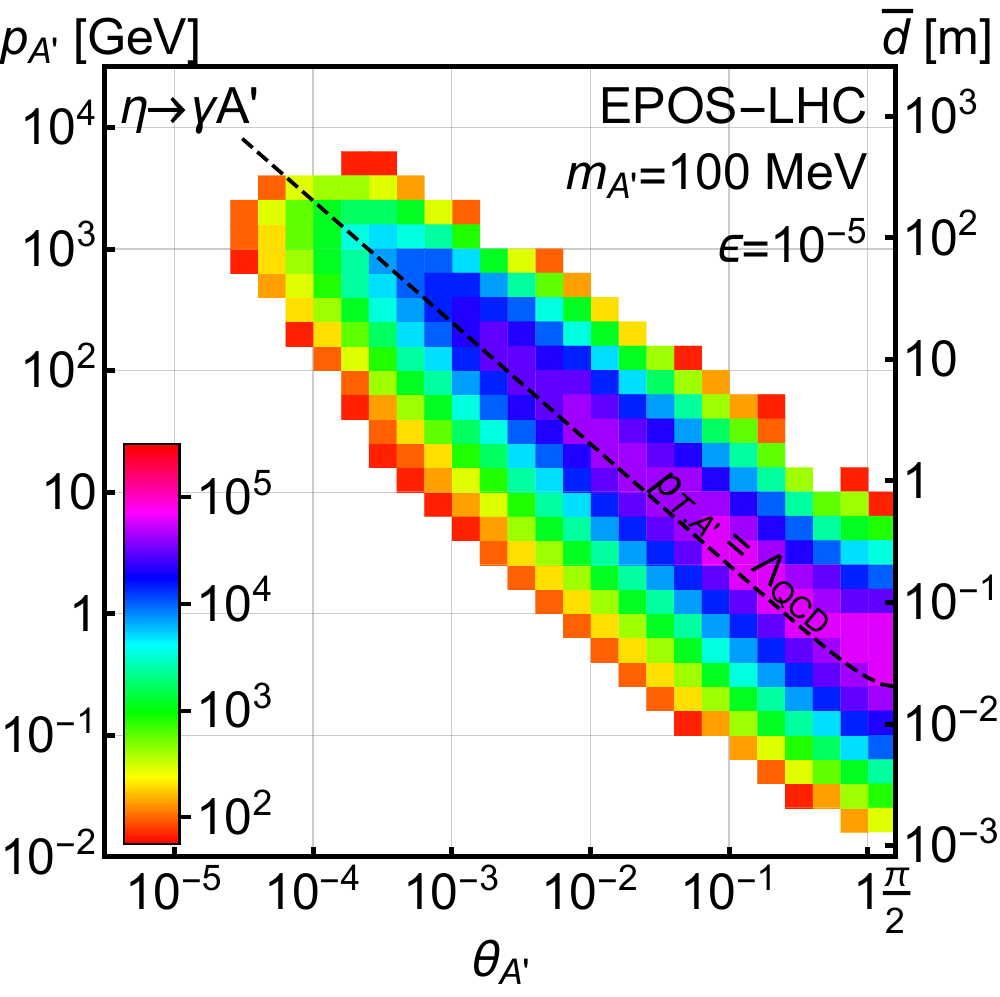}
\includegraphics[width=0.32\textwidth]{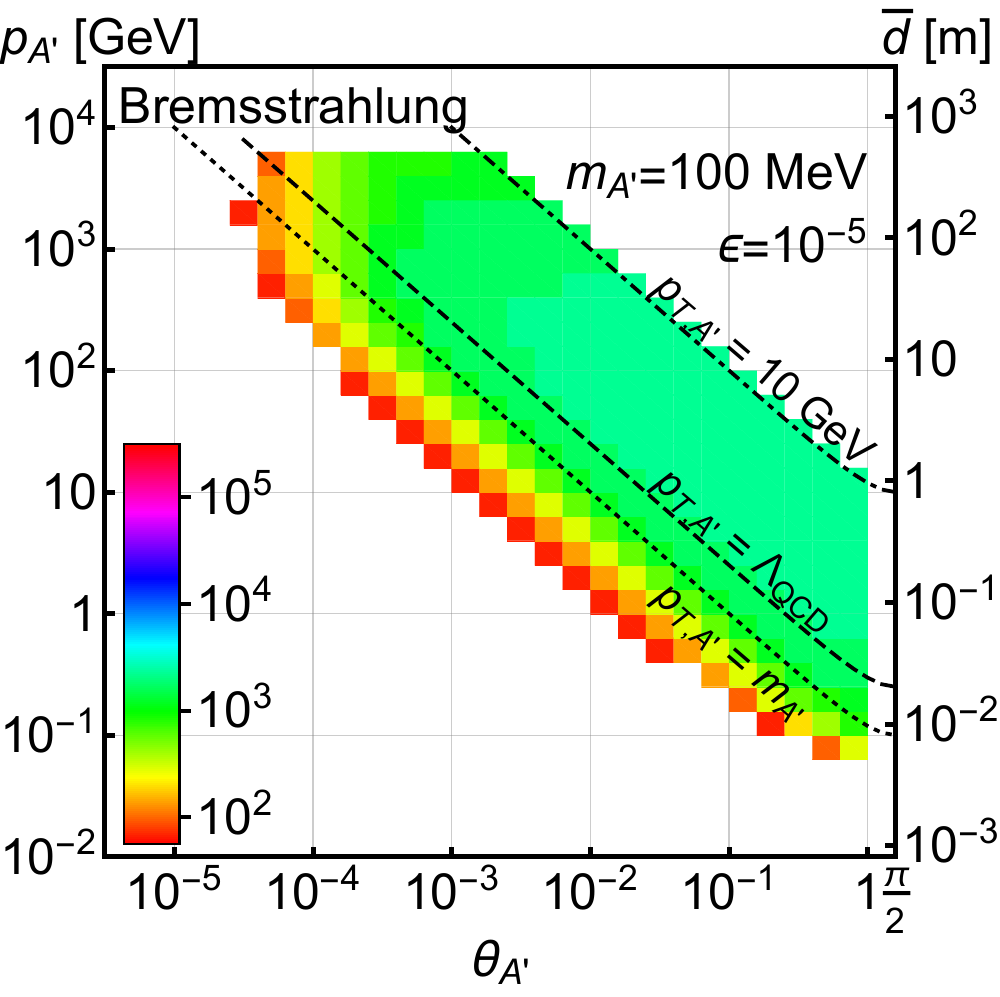}\\
\caption{Distribution in the $(\theta, p)$ plane, where $\theta$ and $p$ are the angle with respect to the beam axis and momentum, respectively, for dark photons produced by $\pi^0$ decays (left), $\eta$ decays (center), and proton bremsstrahlung (right), for $A'$ parameters $(m_{A'}, \epsilon) = (20~\mev, 10^{-4})$ (top) and $(100~\mev, 10^{-5})$ (bottom). The right-hand axis indicates the dark photon's decay length; see \eqref{eq:ap_decay_length}.  The total number of dark photons is the number produced in one hemisphere ($0 < \cos \theta \le 1$) in $13~\tev$ $pp$ collisions at the LHC with an integrated luminosity of $300~\ifb$.  The bin thickness is $1/5$ of a decade along each axis.  The black dashed, dotted, and dash-dotted lines correspond to $p_{T,A'} = \Lambda_{\text{QCD}} \simeq 250~\mev$, $m_{A'}^2$, and $10~\gev$, respectively.}
\label{fig:aprimeptheta}
\end{figure}

The similarity between the $A'$ distributions and those of their parent mesons comes as no surprise. Neglecting ${\cal O}\left(m_{\pi^0, \eta}\right)$ effects, the $A'$ lab-frame momentum is $|\vec p_{A'}^{\text{ lab}}|\approx\frac12 p_{\pi^0, \eta} [ 1+\cos\theta_{A'} +(m_{A'}^2 / m_{\pi^0, \eta}^2) (1-\cos\theta_{A'}) ]$, and follows in the meson direction. Here, $\theta_{A'}$ is the $A'$ polar angle in the meson rest frame, and $\cos\theta_{A'}$ is uniformly distributed, since pseudo-scalar mesons decay isotropically.  The broadening of the distributions in \figref{aprimeptheta} along the diagonal direction, relative to the meson distributions in \figref{pionptheta}, is, then, a result of the linear smearing of the meson $p_T$ with $\cos\theta_{A'}$. 

\subsection{Proton Bremsstrahlung}
\label{sec:protonbrem}

Proton bremsstrahlung of dark photons in high energy $pp$ collisions, $pp\to p\,A'\,X$, is another important source of forward-going $A'$s.  This type of signal contribution has been extensively discussed in the context of fixed target and beam dump experiments, which inject an energetic proton beam onto a heavy nucleus target~\cite{Blumlein:2013cua}. The common lore in the dark photon sensitivity reach estimate for experiments such as U70~\cite{Blumlein:2013cua}, SHiP~\cite{Gorbunov:2014wqa,Graverini:2016}, and SeaQuest~\cite{Gardner:2015wea} is to apply the generalized Fermi-Weizsacker-Williams (FWW) approximation~\cite{Fermi:1924tc,Williams:1934ad,vonWeizsacker:1934nji}. In our case, dark photons arise in collisions of identical particles in the center-of-mass frame (lab) frame.  We give a detailed discussion of proton bremsstrahlung in \appref{brem} and only outline the general features that make this potential signal contribution very attractive for future dark photon searches at \name.

To derive the dark photon spectrum from proton bremsstrahlung by applying the FWW approximation, we treat the protons as coherent objects, and therefore only allow proton momentum transfers up to $\Lambda_{\text{QCD}}$, and dark photon $p_T$ up to $10~\gev$. The resulting $(\theta, p)$ distribution is given in the right-hand panels of \figref{aprimeptheta} for the two representative points of \eqref{eq:ps_points}. In the high-momentum, forward (low-$p_T$) regions (where the FWW approximation is valid), the expected event yield is comparable and can even exceed that from meson decays, even though the proton bremsstrahlung cross section is far below that of meson production.  This is due to the different characteristics of the two processes: although the $A'$ spectrum from meson decays is centered around $p_T \sim \Lambda_{\text{QCD}}$ and decreases (roughly exponentially) at high-$p_T$, the dark photon bremsstrahlung spectrum follows the characteristics of photon bremsstrahlung and peaks around the collinear cutoff, with $p_T\approx m_{A'}$ and the high-$p_T$ tail of the distribution suppressed by $\sim 1/p_T^2$ (see \eqref{eq:wsplitting}).  For the two representative parameter-space points, $m_{A'} < \Lambda_{\text{QCD}}$, so for a given dark photon momentum $p_{A'}$, the events cluster around $\theta_{A'}\sim \Lambda_{\text{QCD}}/p_{A'}$ for meson decays, but peak around $\theta_{A'}\sim m_{A'}/p_{A'}$ in the case of bremsstrahlung. In the latter case, though, also events from regions with larger $p_T$ can contribute non-negligibly up to an experimental upper limit on $\theta_{A'}$.

\subsection{Direct Dark Photon Production}
\label{sec:directproduction}

Dark photons can also be produced directly through $q\bar{q}\to A'$ or the related QCD scattering processes $q\bar{q}\to gA'$, $qg \to qA'$, and $\bar qg \to \bar qA'$. These processes can have large cross sections and could be the dominant dark photon production mode for large dark photon masses $m_{A'}\agt 1~\gev$~\cite{deNiverville:2012ij,Alekhin:2015byh}. However, the estimation of the corresponding production rates suffers from large theoretical uncertainties, mainly coming from the evaluation of parton distribution functions (PDFs) $f(x,Q^2)$ at low $Q^2$ and low $x$.

In direct production, the partonic center-of-mass energy, $\hat{s} = x_1 x_2 s$, is bounded from below by the dark photon mass $\hat{s} > m_{A'}^2$. Given that $\sqrt{s} = 13~\tev$ at the LHC, the relevant momentum fractions for the present case are as low as $x=6 \times 10^{-9} \ (m_{A'} / 1~\gev )^2$. At the relevant scale $Q^2 \sim m_{A'}^2$, the available PDFs are highly uncertain.  For example, some of them are not well-defined, and others diverge or become negative, e.g., NNPDF~\cite{Ball:2012cx}.\footnote{In comparison, estimates for fixed target experiments are on much firmer footing. For example, at the SHiP experiment, the center-of-mass energy is only $\sqrt{s}=20~\gev$, and so the relevant momentum fractions are $x>2.5 \times10^{-3} \ (m_{A'} / 1~\gev )^2$, which are under better theoretical control at $Q^2 \sim 1~\gev^2$.}

Another difficulty arises if one is interested in properly simulating this production in the forward region. At angles below $\theta=1~\mrad$, the characteristic parton transverse momentum, $p_T \sim \Lambda_{\text{QCD}}$, should be taken into account, an option that is not common in MC generators simulating hard processes. Given these large theoretical uncertainties, we have decided not to include the direct dark photon production channel in this study, although this contribution may potentially significantly improve the reach for $\sim \gev$ dark photon masses.

\section{Signal and Detector Considerations}
\label{sec:signalanddetector}

\subsection{Signal Rates and Detector Geometry}
\label{sec:general_considerations}

\Figref{aprimeptheta} of the previous section shows the large yield of very forward, high-momentum dark photons that propagate ${\cal O}(100~\m)$ before decaying.  We now determine the signal rates for detectors placed at the on-axis locations discussed in \secref{lhcinfrastructure}. We consider cylindrically-shaped detectors with radius $R$ and depth $\Delta = \lmax - \lmin$, where $\lmax$ ($\lmin$) is the distance from the IP to the far (near) edge of the detector along the beam axis. The probability of a dark photon to decay inside the detector volume is then given by 
\be
\label{eq:P_decay_in_volume}
\mathcal{P}_{A'}^{det} (p_{A'},\theta_{A'})
= ( e^{-\lmin/\bar{d}} - e^{-\lmax/\bar{d}} ) \
\Theta(R-\tan\theta_{A'}\lmax) \ ,
\ee
where the first term is the probability that the dark photon decays within the $(\lmin, \lmax )$ interval, and the second term enforces the angular acceptance of the detector. 

We first consider an on-axis detector placed at the far location after the intersection, as described in \secref{lhcinfrastructure}. Following the discussion below \eqref{eq:tunnel_detector_dist}, to avoid an overlap of the detector with the LHC infrastructure in the tunnel, the detector should be located at a minimal distance $L \approx 350~\m$ from the IP. For the two dark photon parameter-space points considered, \figref{aprimeptheta} shows that dark photons with decay lengths $\bar{d}\sim100~\m$ make a $\theta_{A'} \sim 0.1~\mrad$ angle with the beam axis. As a benchmark design for this detector, we therefore consider the detector geometry
\be
\label{eq:far_benchmark}
 \textbf{detector at far location: } \lmax=400~\m,\, \Delta=10~\m,\, 
 R=20~\cm \ .
\ee

\begin{figure}[t]
\includegraphics[width=0.32\textwidth]{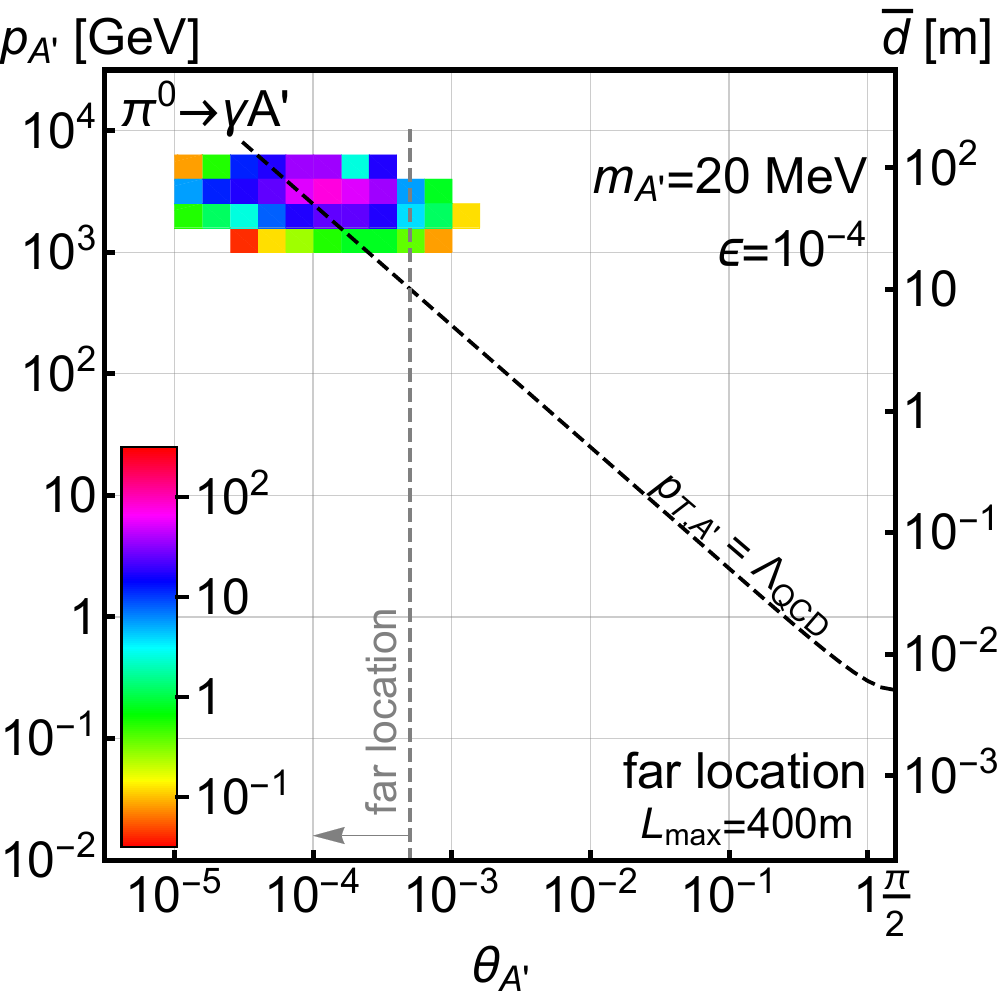}
\includegraphics[width=0.32\textwidth]{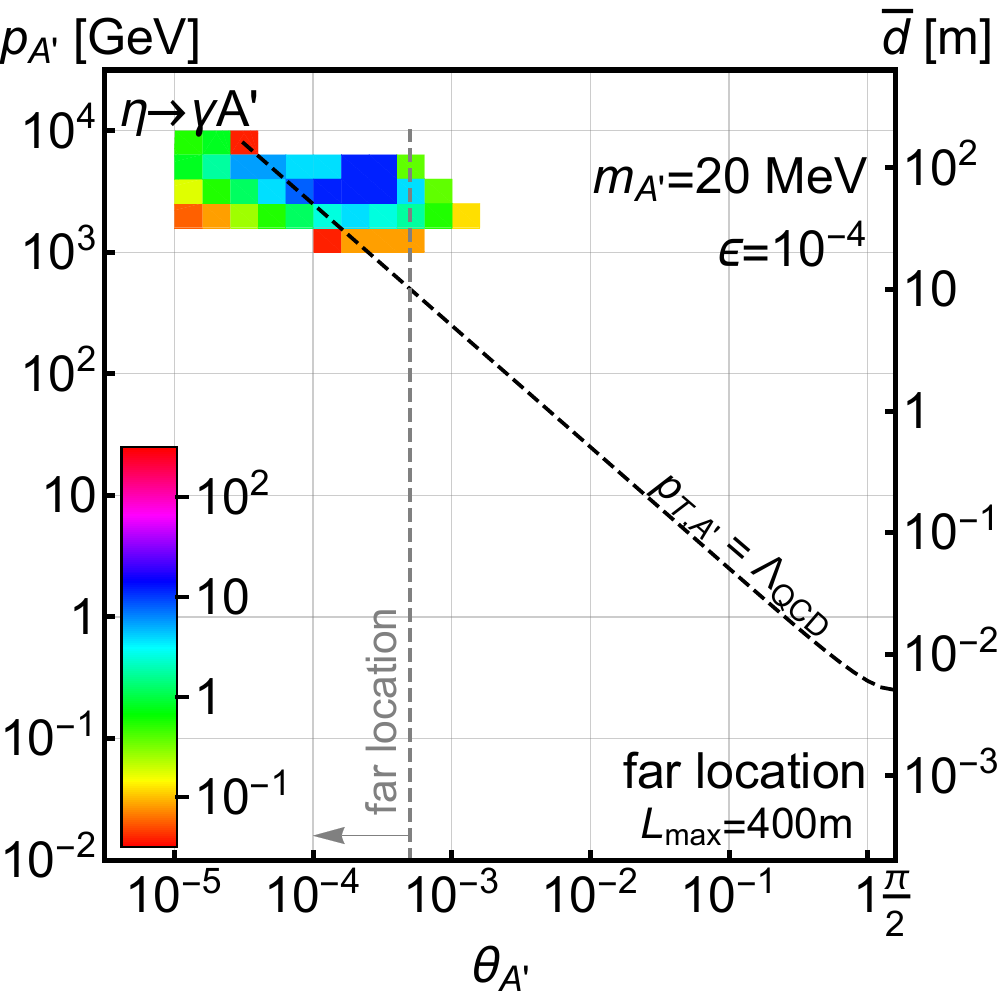} 
\includegraphics[width=0.32\textwidth]{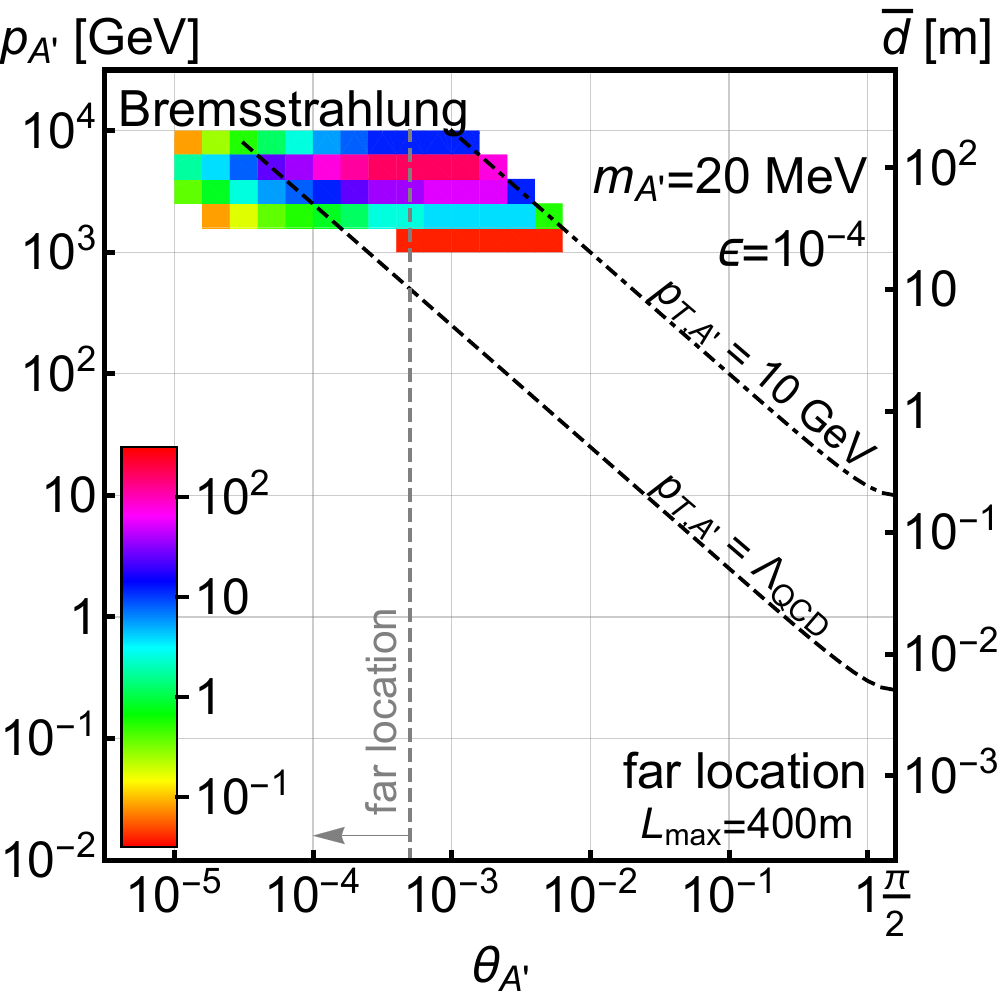}
\includegraphics[width=0.32\textwidth]{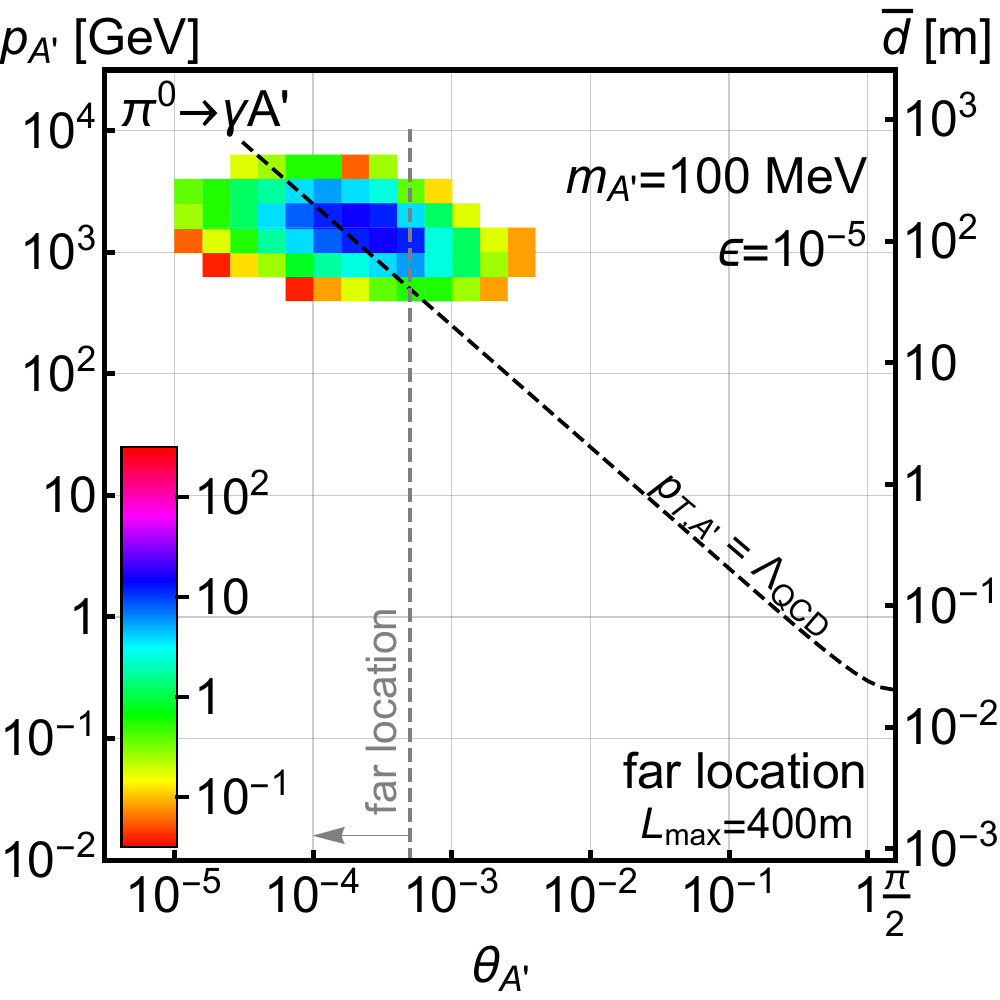}
\includegraphics[width=0.32\textwidth]{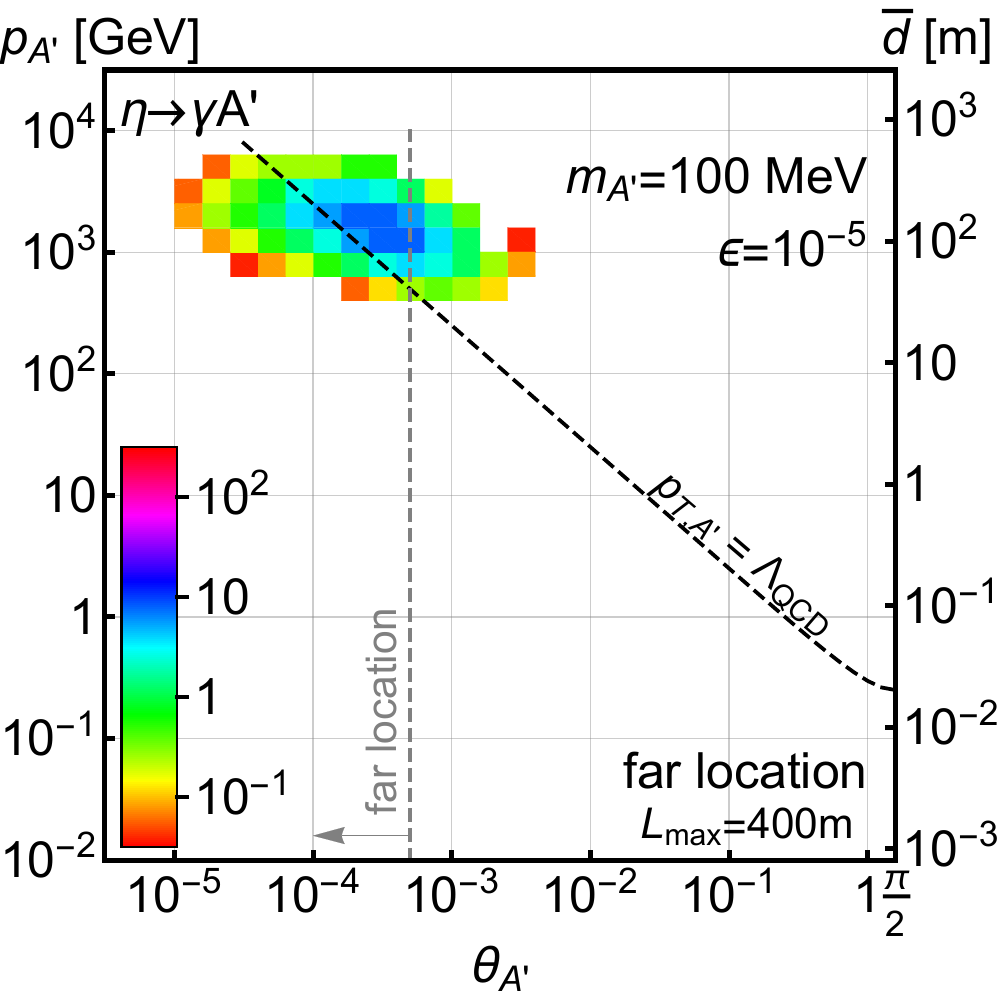}
\includegraphics[width=0.32\textwidth]{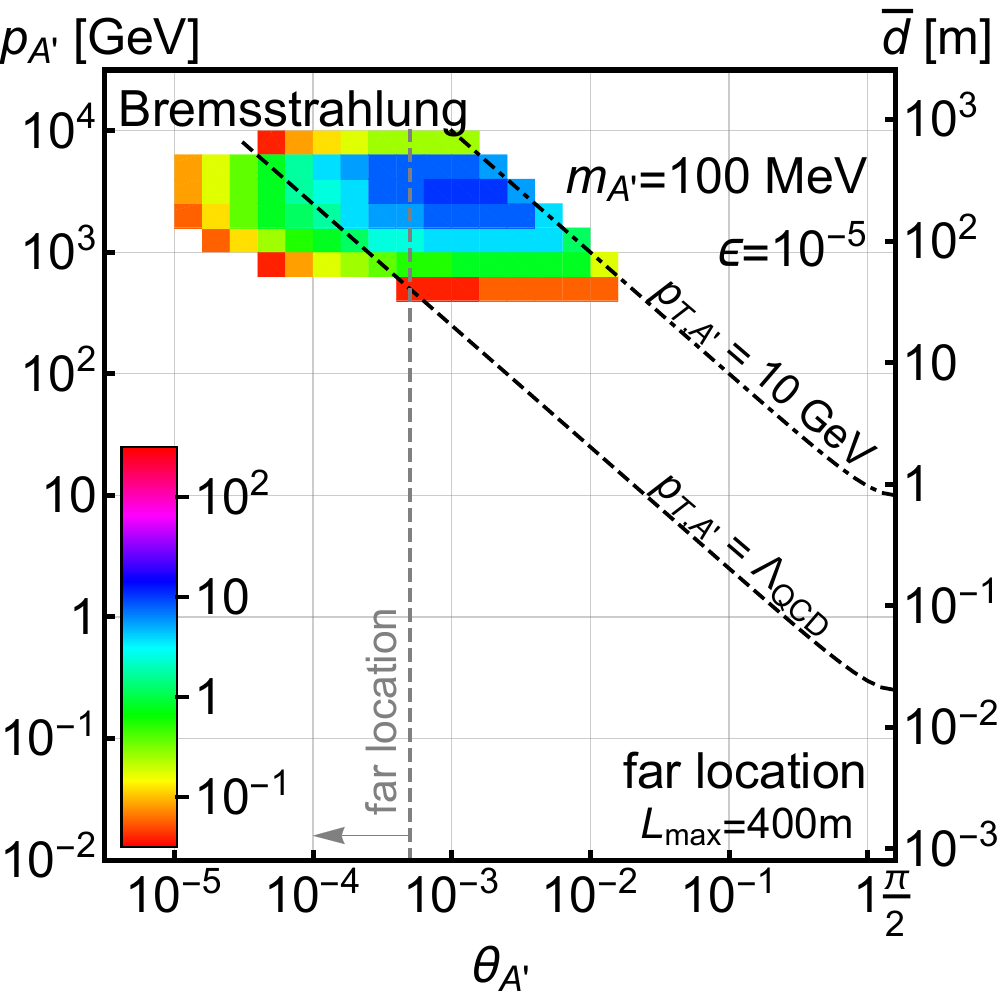}
\caption{Distribution in the $(\theta, p)$ plane, where $\theta$ and $p$ are the angle with respect to the beam axis and momentum, respectively, for dark photons that decay in the interval $(\lmin, \lmax) = (390~\m, 400~\m)$ (the far detector location) and are produced by $\pi^0$ decays (left), $\eta$ decays (center), and proton bremsstrahlung (right) for $A'$ parameters $(m_{A'}, \epsilon) = (20~\mev, 10^{-4})$ (top) and $(100~\mev, 10^{-5})$ (bottom). The total number of $A'$s is the number produced in one hemisphere ($0 < \cos \theta \le 1$) in 13 TeV $pp$ collisions at the LHC with an integrated luminosity of $300~\ifb$.  The bin thickness is $1/5$ of a decade along each axis.  The dashed and dashed-dotted lines correspond to $p_{T,A'} = \Lambda_{\text{QCD}} \simeq 250~\mev$ and $10~\gev$, respectively.  In each plot the right $y$-axis indicates the dark photon's characteristic decay length $\bar{d}$ (see \eqref{eq:ap_decay_length}). The angular coverage of the detector is indicated via vertical gray dashed lines. 
}
\label{fig:aprimeptheta_weighted-onaxis}
\end{figure}

\Figref{aprimeptheta_weighted-onaxis} shows the $(\theta, p)$ distributions for dark photons that decay within the $(\lmin,~\lmax)$ range of \eqref{eq:far_benchmark} for the two representative $(m_{A'}, \epsilon )$ points given in \eqref{eq:ps_points}. Here, we have applied the first term (written in the parentheses) of \eqref{eq:P_decay_in_volume} to the distributions of \figref{aprimeptheta}, but ignored the angular cut. We see that the dark photon signal from meson decays is characterized by an energy $E_{A'} \agt 1~\tev$ and an angle $\theta_{A'}<1~\mrad$. In contrast, the dark photon bremsstrahlung signal also occurs at larger angles $\theta_{A'}>1~\mrad$. However, both of them will eventually be limited by the experimental cut on $\theta_{A'}$ that comes from the detector design, $\theta_{A'}<\theta_{A'}^{max}=20~\cm/400~\m=0.5~\mrad$.

In \figref{nevents_L_d} we explore the far detector signal rate's dependence on the various detector parameters, properly taking into account both the decay length and angular acceptance conditions.  In the left panel, we examine the signal yield as a function of $\lmax$, keeping the remaining detector characteristics in \eqref{eq:far_benchmark} fixed. Even for high-momentum dark photons with $p_{A'} \ge 100~\gev$, the signal decreases exponentially with $\lmax$, so for these dark photon models, it is preferable to place the detector as close as possible to the IP.  In the right panel, we fix $\lmax = 400~\m$, but vary the detector radius $R$.  As can be seen, the benchmark radius $R=20~\cm$ captures most of the dark photon decays.\footnote{Note that the flattening of the bremsstrahlung contribution at large $R$ is due to the transverse momentum cut imposed on the dark photon $p_T<10~\gev$, which ensures the validity of the FWW approximation. This cut has no impact on our sensitivity reach plots, as discussed in \appref{brem}.}  Increasing $R$ above 20 cm would not improve the yield much, but decreasing it below 10 cm would result in a rather drastic drop in sensitivity.  Both effects can be understood by referring to \figref{aprimeptheta_weighted-onaxis}: varying $R$ changes the angular coverage of the detector and moves the ``far location'' line in the figure to include more or fewer events for a given detector location.

\begin{figure}[t]
\centering
\includegraphics[width=0.47\textwidth]{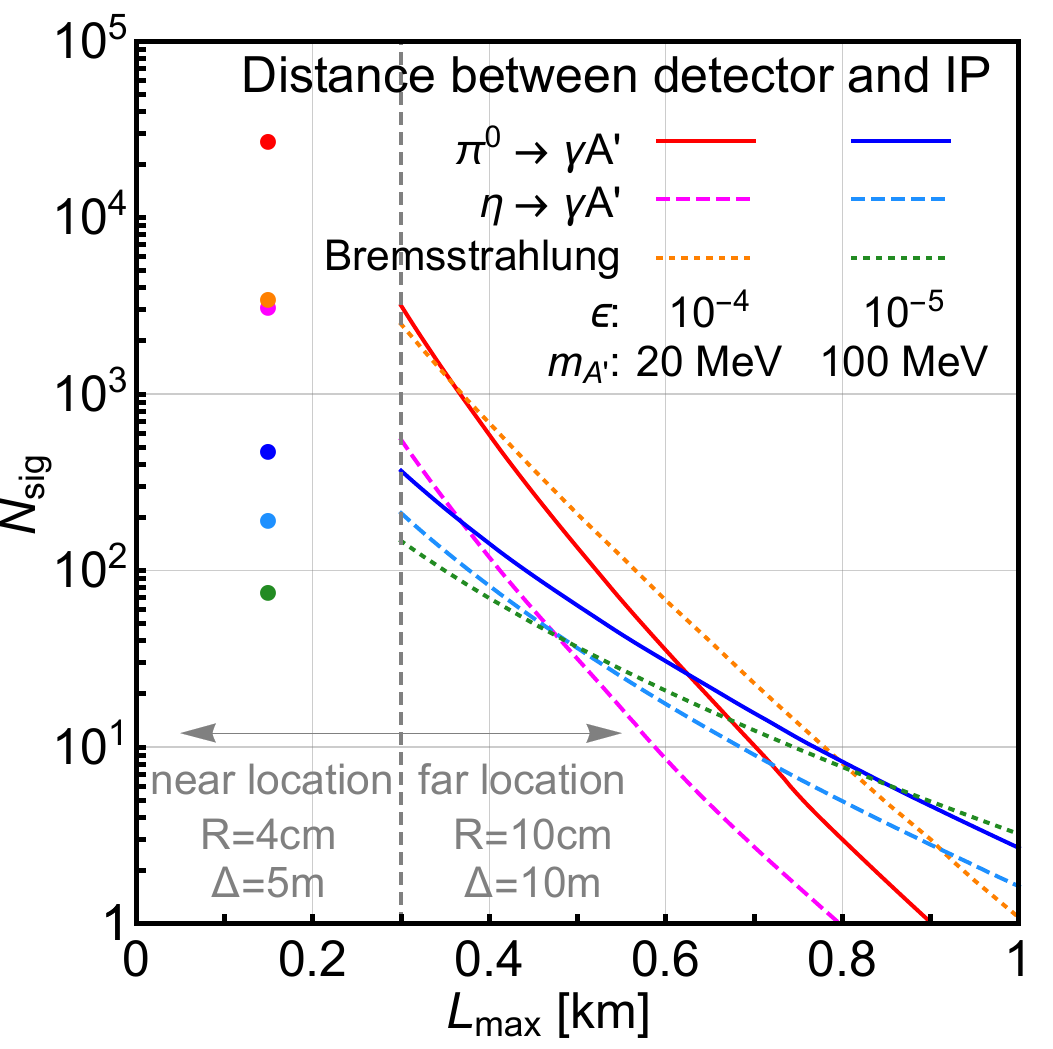} \quad
\includegraphics[width=0.47\textwidth]{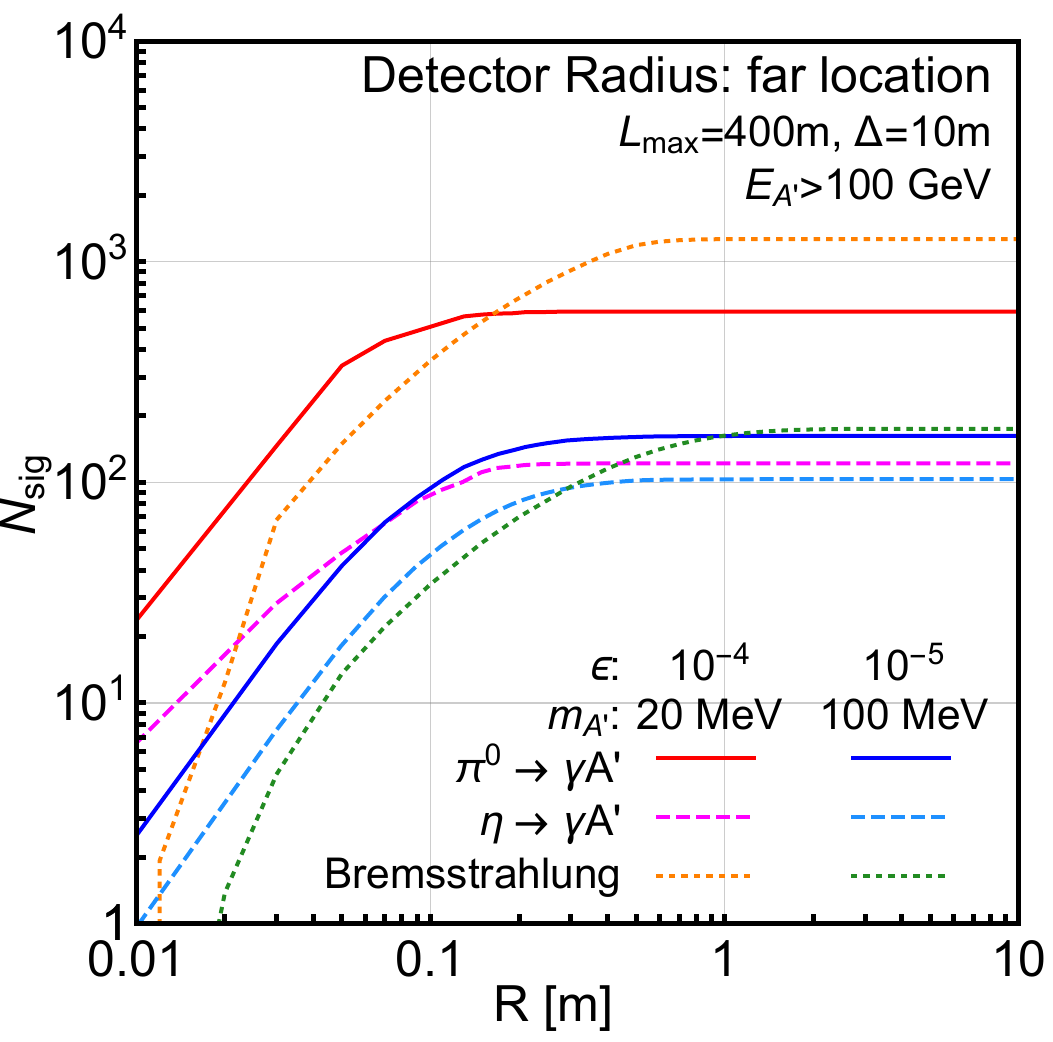}
\caption{Left: $N_{\text{sig}}$, the expected number of signal events, for two representative $(m_{A'}, \epsilon )$ points as a function of the distance between the IP and detector, $\lmax$, for the near and far detector benchmark design (see text). Right: $N_{\text{sig}}$ for the far detector location as a function of the detector radius $R$.}
\label{fig:nevents_L_d}
\end{figure}

We now consider the after-TAN, near location described in \secref{lhcinfrastructure}.  This location is closer to the IP, and therefore increases the signal acceptance of the detector. In this unique location, the TAN shields the detector from the IP direction. On the other hand, requiring \name\ to be positioned between the TAN, the D2 magnet, and the two beam pipes limits the size of such a detector. For this near detector location, we assume the detector geometry
\be
\label{eq:near_benchmark}
\textbf{detector at near location: } \lmax=150~\m,\, \Delta=5~\m,\, R=4~\cm \ .
\ee
The depth $\Delta$ is limited by the distance between the TAN and the D2 magnet.  Note, however, that for HL-LHC running, it is expected that the TAN absorbers will be replaced by TAXN absorbers, which will be moved towards the IP by roughly 10 m, while the D2 magnet remains fixed~\cite{Apollinari:2015bam}.  The depth $\Delta$ of the near detector could then be much larger, with correspondingly larger signal event rates.  

In \figref{aprimeptheta_weighted_TAN} we repeat the analysis of \figref{aprimeptheta_weighted-onaxis} for the TAN-shielded near detector case with the values of $(\lmin , \lmax)$ given in \eqref{eq:near_benchmark}. Compared to the far detector location, the angular coverage of the near detector design is reduced by roughly half to $\theta_{A'}^{\text{max}}\approx 4~\cm/150~\m= 0.27~\mrad$. On the other hand, the detector at the near location benefits from capturing less-energetic dark photons with $E_{A'}\sim\text{few}~100~\gev$. As a result, for $\lmax = 150~\m$, the predicted signal yield of the near detector is significantly better than for the far detector, as can be seen in the left panel of \figref{nevents_L_d}.

\begin{figure}[t]
\includegraphics[width=0.32\textwidth]{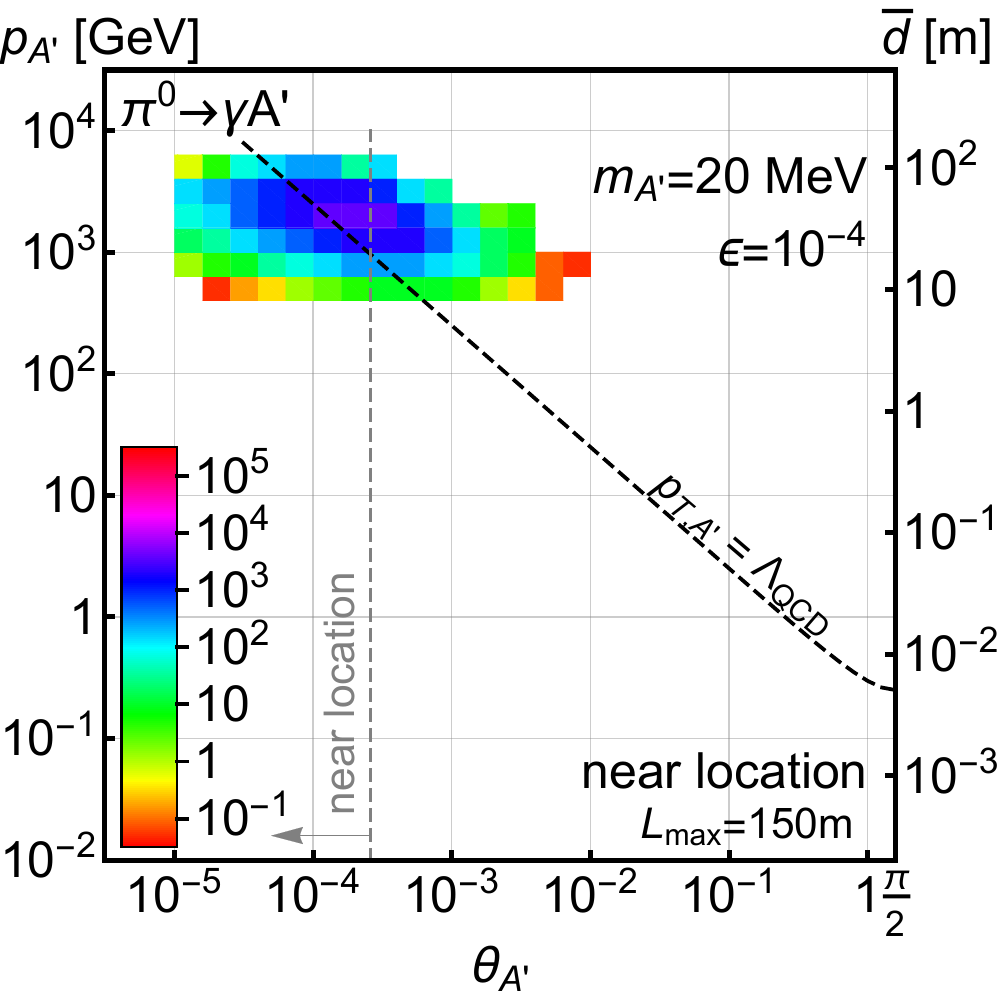}
\includegraphics[width=0.32\textwidth]{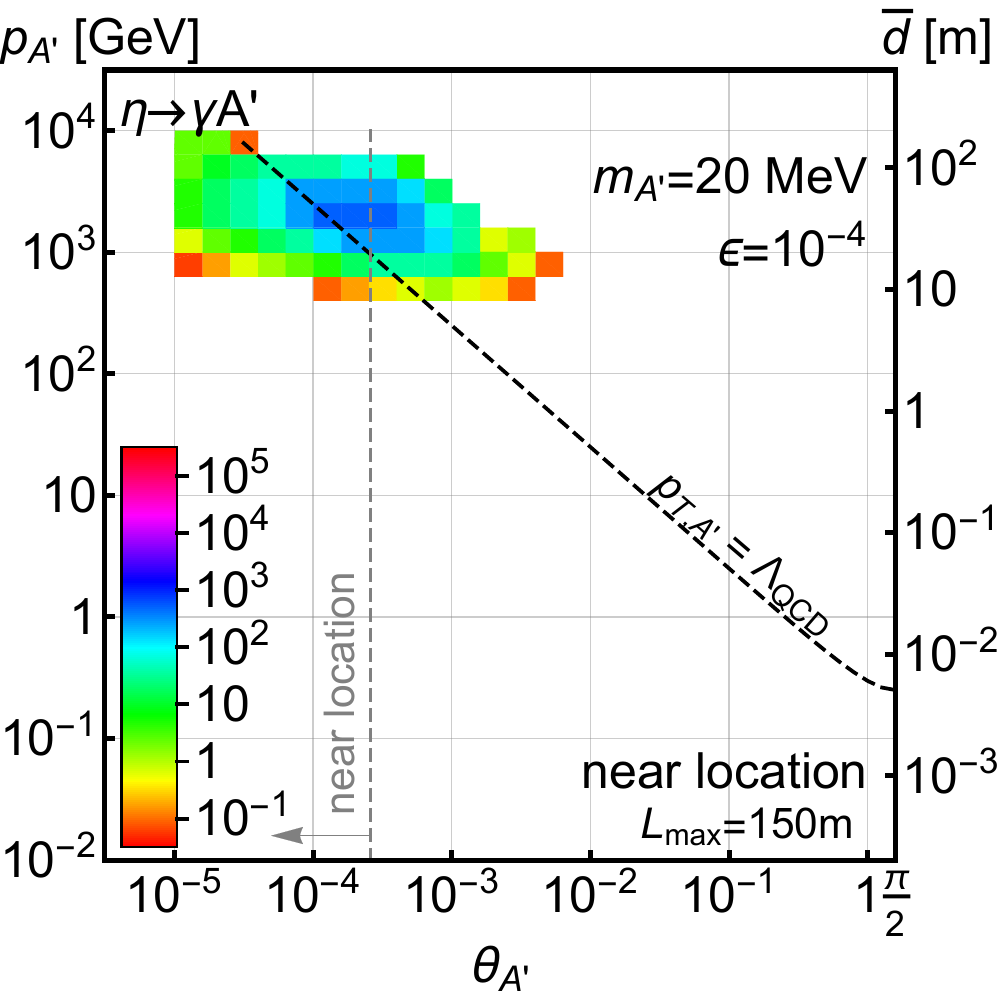} 
\includegraphics[width=0.32\textwidth]{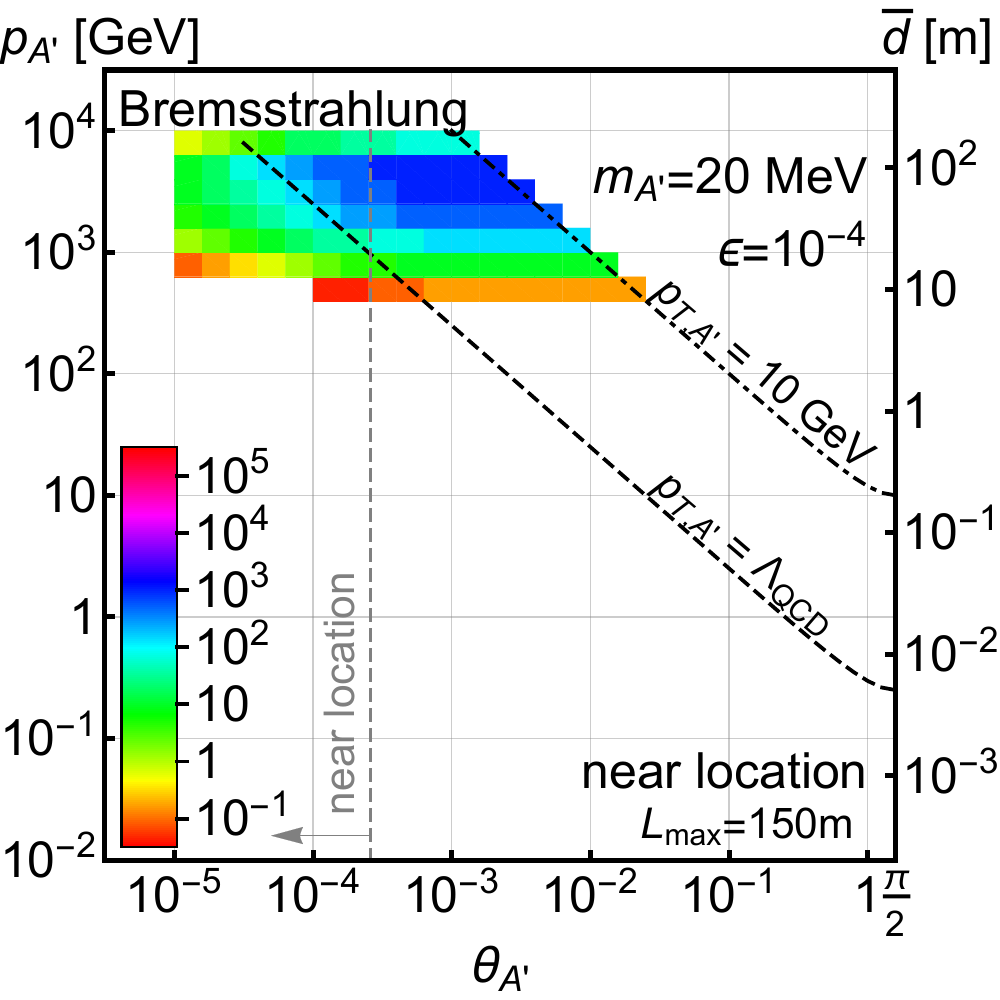}
\includegraphics[width=0.32\textwidth]{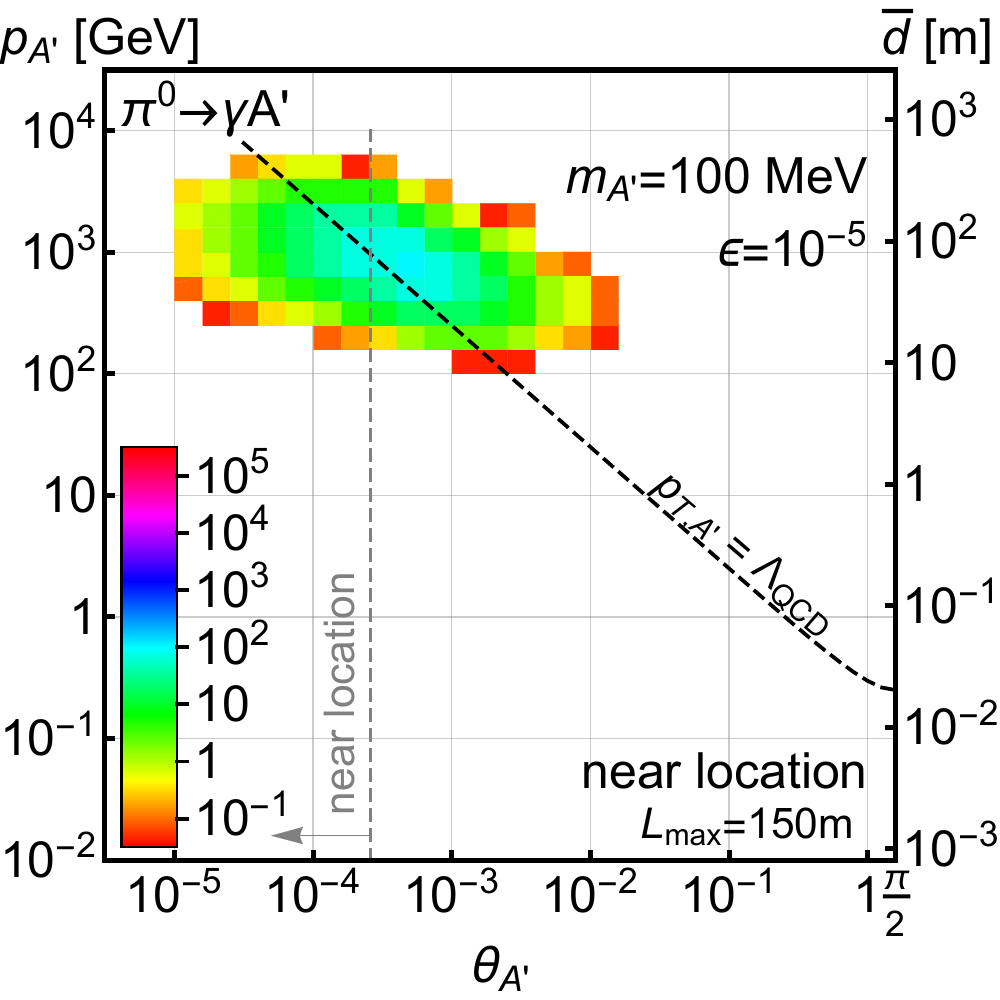}
\includegraphics[width=0.32\textwidth]{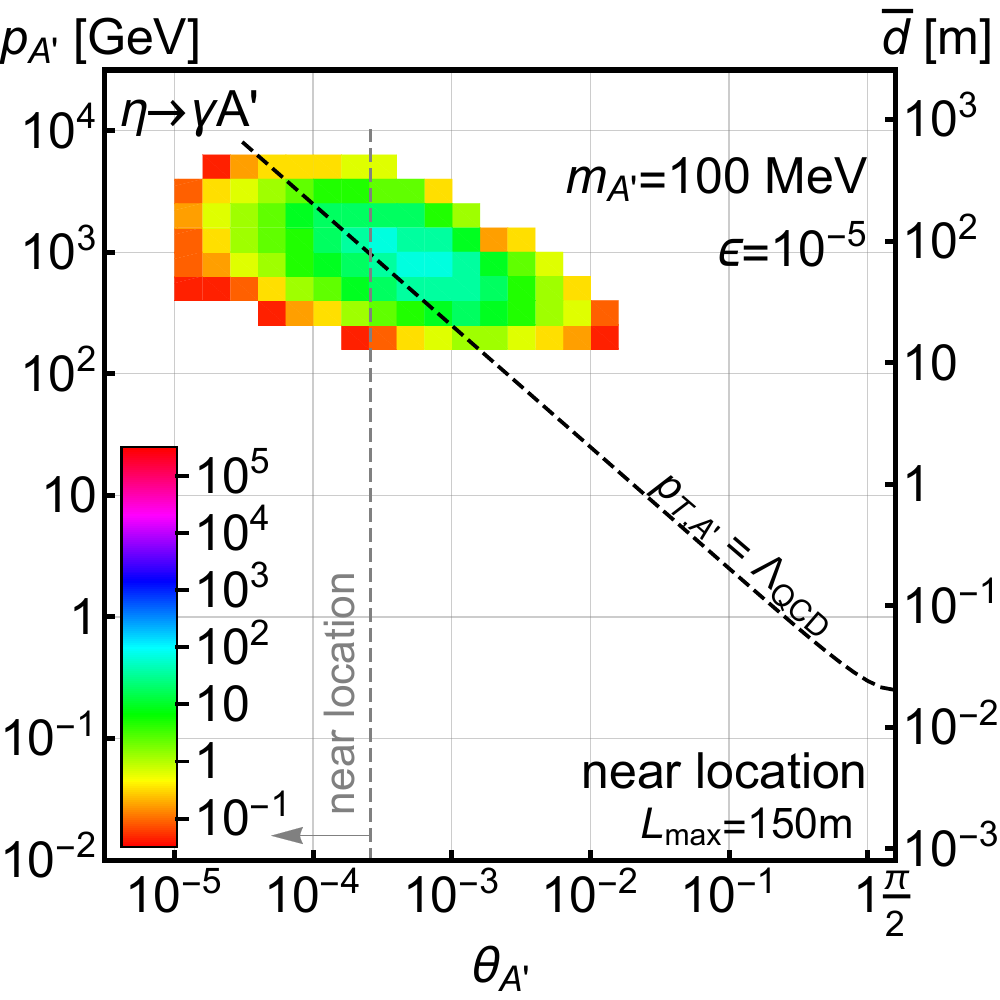}
\includegraphics[width=0.32\textwidth]{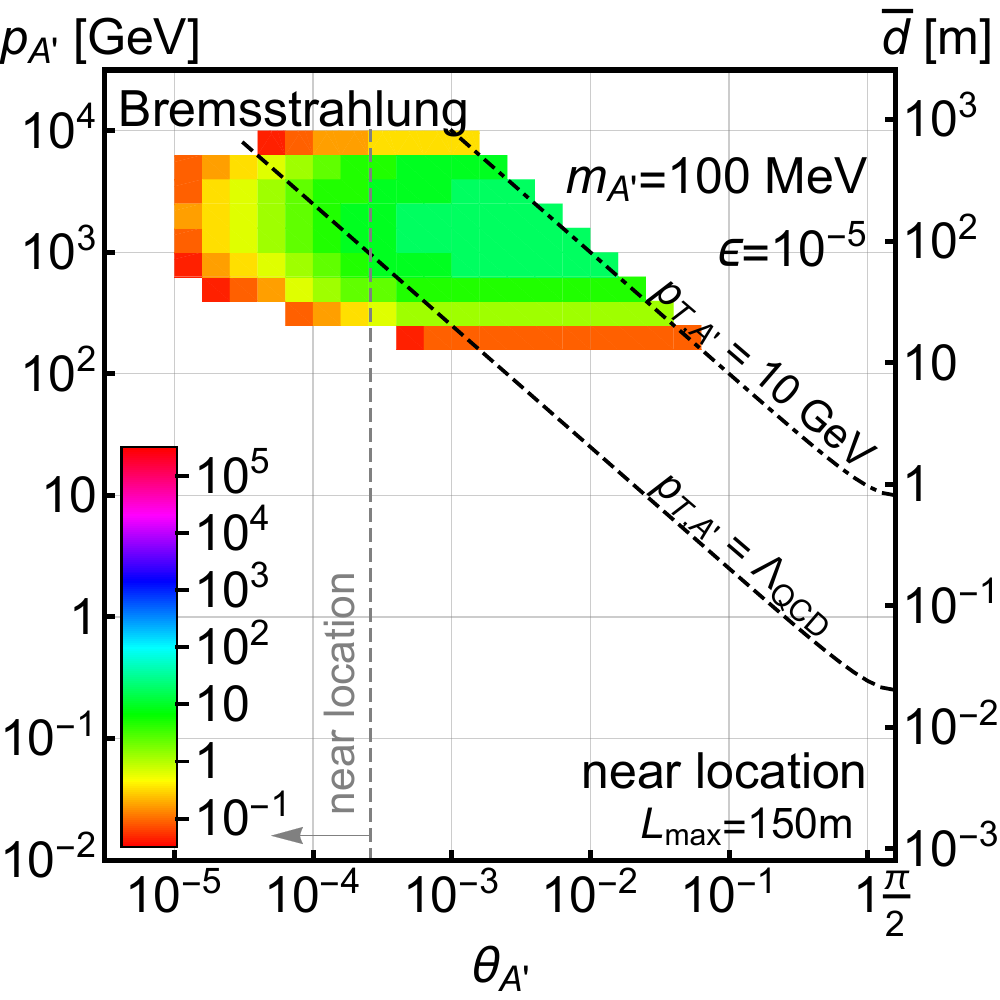}
\caption{Same as in \figref{aprimeptheta_weighted-onaxis}, but for the near detector location with $(\lmin, \lmax) = (145~\m, 150~\m)$. }
\label{fig:aprimeptheta_weighted_TAN}
\end{figure}

\subsection{Signal Characteristics and Track Separation}
\label{sec:signalcharacteristics}

The dark photon signature seen in \name\ consists of two highly energetic (often with energies above 500 GeV), opposite-charge tracks emerging from a vertex inside the detector volume. The combined momentum of the two tracks should point towards the interaction point.  We further expect that the dark photon decay kinematics produces two tracks with comparable energies (see the right panel of \figref{BGvsSIGmomentum}). A measurement of individual tracks with sufficient resolution and an identification of their charges is therefore imperative if the apparatus is to make use of kinematic features to distinguish signal from background. A \textit{tracking-based technology}, like a silicon strip pixel detector, would be optimal for such a task~\cite{Garcia-Sciveres:2017ymt}.

In signal events, the characteristic opening angle between the two tracks is typically $\theta_{ee} \sim m_{A'}/E_{A'}$.  For example, in the $m_{A'}=20~\mev$ case, the typical energy of dark photons that decay in the detector volume is $E_{A'}\sim2~\tev$, implying an opening angle $\theta_{ee}\sim10~\murad$. Over a length $\ell=1~\m$, the two tracks separate by $h_\ell \sim \theta_{ee} \ell \sim 10~\mu\m$. The pixel detectors currently in use at the LHC experiments have a typical size of $h_P \sim 100~\mu \m$ \cite{Aad:2008zzm,Chatrchyan:2008aa}, which would be insufficient to resolve the two tracks. Of course, in more optimistic cases, for example, with $m_{A'} = 100~\mev$, a similarly-energetic track that travels the length of the far detector would be separated by $500~\mu\m$, an observable separation.

Optimally, to achieve observable track separations for nearly all events that occur in the detector, a magnetic field may be used. Two initially collinear charged particles with energy $E$ and charges $\pm e$ that travel a distance $\ell$ along the $z$-axis through a magnetic field $B$ oriented along the $x$-axis separate by a distance 
\be
h_{B} \approx \frac{ec \ell^2  }{E} B   = 3~\mm \  
\left[\frac{1~\tev}{E} \right] \left[\frac{\ell}{10~\m} \right]^2 
\left[\frac{B}{0.1~\text{T}} \right] 
\ee
along the $y$-direction. A relatively small magnetic field with $B=0.1~\text{T}$ would therefore be sufficient to split most tracks and may be readily obtained by conventional magnets. 

\subsection{Extended Signal Sensitivity}
\label{sec:extended}

So far we have only discussed dark photons decaying inside the detector volume. However, \name\ may also be sensitive to dark photons that decay in the material in front of the detector. For example, for the far location, dark photons with mass $m_{A'} > 2m_\mu$ can decay to a muon pair in the region between the last LHC magnet in the intersection and \name\ which increases the effective detector volume. As we show in the next section, cosmic and beam-induced muon backgrounds do not tend to produce simultaneous tracks that can mimic the signal directionality characteristics. Such muon signal events may, then, increase the reach of \name\ beyond the estimates presented in this work in the parameter-space region where $m_{A'}> 2m_\mu$. 

Interestingly, for the near detector, a similar enhancement to the signal comes from secondary production of dark photons by SM particles hitting the TAN. Such processes are similar to those probed by beam dump experiments.  In particular, $A'$s can be produced by the scattering of incident SM photons off the electrons in the TAN, $\gamma\,e^- \to A' e^-$, or from decays of mesons produced in showers of particles induced by high-energy neutrons hitting the TAN. In this case, the distance between the production point of $A'$s and the detector is much shorter, since it is dictated by the length of the TAN ($3.5~\m$).  In principle, this production mechanism enables \name\ to probe shorter lifetimes and could therefore extend the reach to larger $\epsilon$ and $m_A'$. We leave a detailed discussion of this production mechanism for future work.

\section{Backgrounds}
\label{sec:background}

As discussed above, the signature of $m_{A'} \sim \mev - \gev$ dark photons in \name\ is highly-collimated, $e^+e^-$ or (for $m_{A'} > 2 m_{\mu}$) $\mu^+ \mu^-$ pairs with $\sim \tev$ energies that are produced in vacuum at a common vertex in the \name\ detector volume with no other particles, and whose summed momentum points back to the IP.  This is an extraordinary signature that has no SM analogue.  Of course, given realistic detectors and, particularly, the large particle fluxes at the near location, there are many SM processes that could, in principle, constitute backgrounds.  

In this section, we consider a variety of potential SM backgrounds that produce two high-energy, opposite-charge tracks that point back to the IP within the angular resolution of the detector and arrive simultaneously within the time resolution of the detector.  Such backgrounds are significantly more general than the signal, as they include charged tracks from charged hadrons, and also charged tracks that begin either inside or outside the detector.  As we see, a detector that can differentiate electrons, muons, and charged hadrons has greatly reduced background, especially if one is willing to consider only the electron signal.  Similarly, the ability to veto tracks that begin outside the detector and reconstruct vertices greatly suppresses the background.  We consider the more general class of backgrounds, however, because our aim is to determine to what extent these additional detector capabilities are required, and how well they must perform, to extract a signal.

Backgrounds at \name\ are greatly reduced by the natural and infrastructure shielding that exists at both the far and near locations.  For example, \name\ is protected by the $\sim 35~\km$ of rock that shields it from cosmic rays in the direction of the IP. The rate for coincident, opposite-charge, $\sim \tev$ cosmic muons that point back to the IP is therefore negligible.  Similarly, the background from charged particles produced at the IP is highly suppressed: such charged particles are typically bent away from \name\ by the D1 magnet, and electrons and charged hadrons are also absorbed before reaching \name, either by rock in the case of the far location or by the existing LHC infrastructure in the case of the near location.

We therefore expect the dominant backgrounds to be of two types:
\begin{itemize}
\item {\em Neutrino-Induced Backgrounds}.  Neutrinos that are produced through processes initiated at the IP can produce highly energetic, charged particles in \name\ that point back to the IP.  An adequate estimate of these physics backgrounds can be obtained, given well-known neutrino interaction rates.
\item {\em Beam-Induced Backgrounds}.  Beam-gas collisions and interactions of the beams and particles produced at the IP with LHC infrastructure can produce high-energy charged tracks that propagate into \name.  These backgrounds are more difficult to determine, and are best estimated with simulations, or better yet, from the experimental data themselves.  Here we extrapolate from published simulation results to obtain preliminary estimates.
\end{itemize}
We now consider these in turn.

\subsection{Neutrino-Induced Backgrounds}

Neutrinos produced through processes initiated at the IP are potentially serious backgrounds for our signal because they point back to the IP and are not absorbed.  Forward-going neutrinos are dominantly produced by the in-flight decays of mesons produced at the IP, particularly charged pions.  (The decays of heavier mesons also contribute to the neutrino flux, but do not significantly change our estimates~\cite{Park:2011gh}.) To create a charged track in an on-axis detector, such as \name, these charged pions must decay before they reach the D1 magnet or they will be deflected and the produced neutrinos will typically miss the detector.  The produced neutrino must then interact in \name.  (For reviews of neutrino-nucleus interactions see, e.g., Refs.~\cite{Formaggio:2013kya,Olive:2016xmw,Alvarez-Ruso:2017oui}.) Neutrino charged-current (CC) events $\nu_{\ell} N \to \ell X$ produce only a single charged lepton.  However, neutrinos can also produce two charged tracks when two CC events are coincident in time or through processes like $\nu N \to \mu^{\pm} \pi^{\mp} X$.  

Before presenting numerical results, we first obtain a rough analytic estimate of the neutrino event rate.  The distribution of charged pions produced at the IP is similar to the distribution of neutral pions shown in \figref{pionptheta}.  Requiring energies above 1 TeV and angles $\theta \alt 0.5$ mrad so that the produced neutrinos travel toward \name, we find roughly $N_{\pi^\pm} \sim 10^{15}$ in an integrated luminosity of $300~\text{fb}^{-1}$.  The probability that a given pion decays before the D1 magnet is
\be
P_\pi = 1-\exp\left(-\frac{L_{\text{D1}}\, m_{\pi^\pm}}{p_{\pi^{\pm}} \tau_{\pi^\pm}}\right) \approx 10^{-3} \left[\frac{\tev}{p_{\pi^\pm}}\right] ,  
\label{eq:ppion}
\ee
where $L_{\text{D1}} \approx 59-83~\m$ is the distance between the IP and the D1 magnet, $\tau_{\pi^\pm} \simeq 2.6 \times 10^{-8}~\s$, and $m_{\pi^{\pm}} \simeq 140~\mev$. The probability that the resulting neutrino interacts within the detector volume is 
\be
P_{\nu}\simeq
\Delta\,\,\sigma(E_\nu)\,\rho_{\text{det}}\,N_A \simeq 6 \times 10^{-12}\,\left[\frac{\sigma(E_\nu)}{10^{-35}~\cm^2}\right]\,\left[\frac{0.1~\m^2}{A_{\text{det}}}\right]\,\left[\frac{M_{\text{det}}}{1~\kg}\right] \ ,
\label{eq:neutrinoprod}
\ee
where $\rho_{\text{det}} = M_{\text{det}}/(A_{\text{det}}\,\Delta)$ is the average density of the target material within the detector, $M_{\text{det}}$ and $A_{\text{det}}$ are the mass and transverse area of the detector, respectively, $N_A = 6.02 \times 10^{23}~\text{g}^{-1}$, and $\sigma(E_\nu)$ is the neutrino-nucleus cross section.  We have normalized $M_{\text{det}}$ and $A_{\text{det}}$ to possible values for the \name\ target volume and $\sigma(E_\nu)$ to the CC cross section for neutrinos with $E_\nu\sim 200~\gev$~\cite{Formaggio:2013kya}, which is the average energy of neutrinos produced in the decay of TeV charged pions.  The number of charged leptons produced by $\sim 200~\gev$ neutrinos in \name\ is then $N_{\pi}P_\pi P_\nu \sim 10$ per kg of detector mass in $300~\text{fb}^{-1}$ integrated luminosity. 

A more precise numerical estimate can be obtained using our Monte Carlo sample of very forward pion events.  We assume charged pions travel in a straight line before the D1 magnet (neglecting possible defection by the quadrupole magnets), and also require that they do not hit the beam pipe before they decay.   The results are presented in the left panel of \figref{BGvsSIGmomentum}, where the red curve corresponds to the number of total CC events per $\kg$ of detector material that reach the detector and that are induced by neutrinos with energies larger than $E_{\nu,\text{min}}$. Comparing the numerical results for $E_{\nu, \text{min}} \sim 200~\gev$ with the analytic result derived above, we find excellent agreement.   Notably, although only high-energy neutrinos with $E_\nu\gtrsim 100~\gev$ could possibly mimic the signal, the background event yield decreases rapidly with $E_{\nu,\text{min}}$ and drops to $\sim 0.1$ for $E_{\nu,\text{min}} \sim \tev$. 

\begin{figure}[t]
\centering
\includegraphics[width=0.47\textwidth]{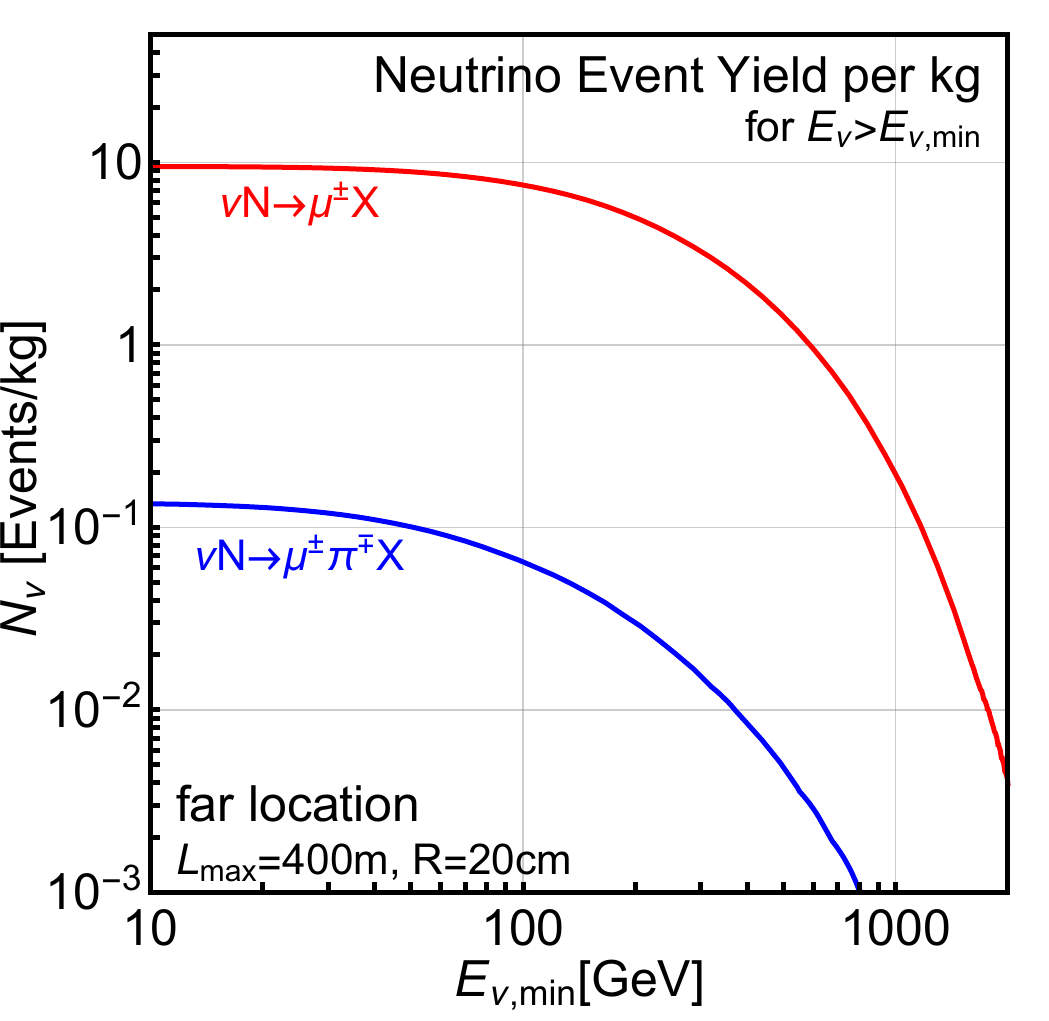} \quad
\includegraphics[width=0.47\textwidth]{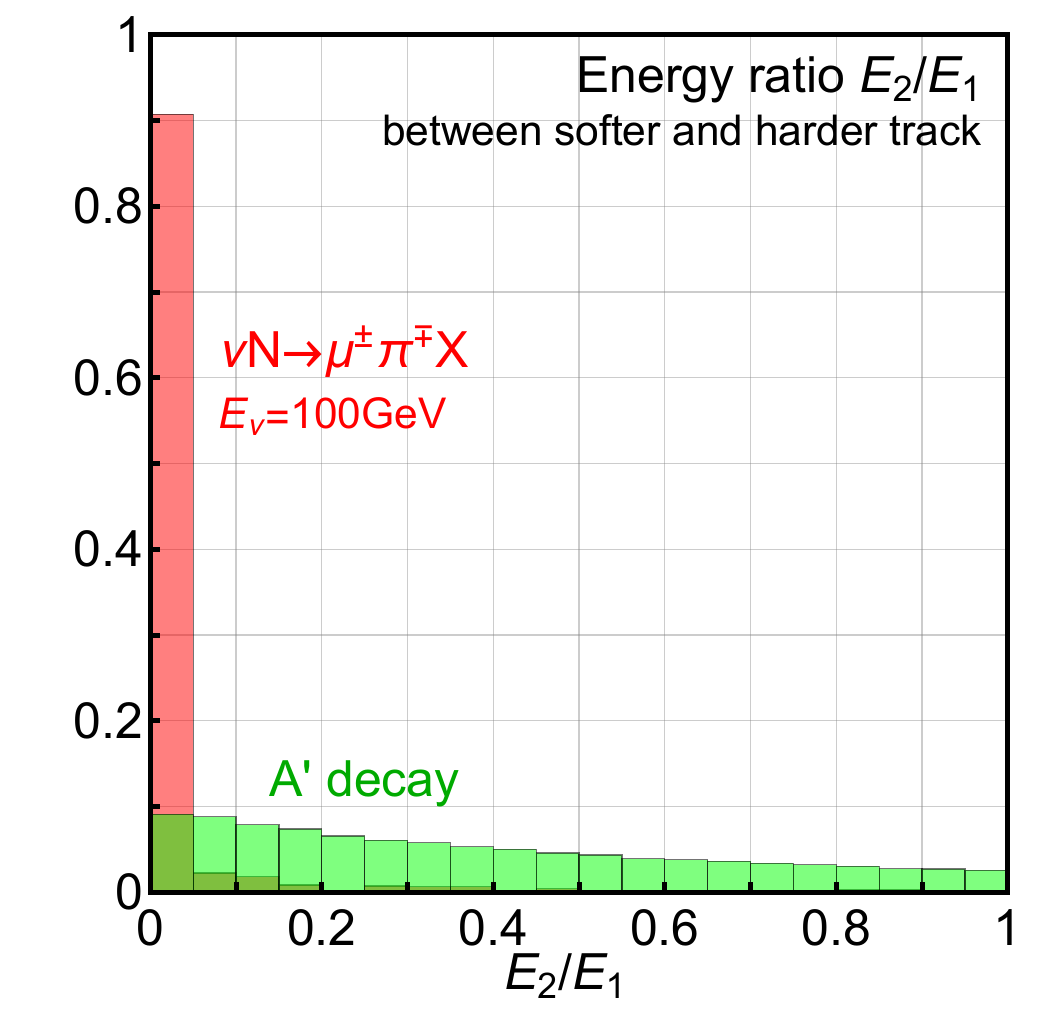}
\caption{Left: Number of expected events per kilogram of detector mass for the detector at the far location (see text) as a function of the minimal incident neutrino energy $E_{\nu,\text{min}}$. The red (blue) line corresponds to the total number of CC (single pion production) events induced by neutrinos with energies $E_\nu \geq E_{\nu,\text{min}}$. The plot assumes an integrated luminosity of $300~\ifb$.  Right: Ratio of the energies of the softer ($E_2$) and harder ($E_1$) tracks from $\nu_\mu N\to \mu^{\pm}\pi^{\mp}X$ with $E_\nu = 100~\gev$ (red histogram) and from $A' \to e^+e^-$ pair, assuming $E_{A'}\gg m_{A'} \gg m_e$, and unpolarized $A'$'s (green histogram).
}
\label{fig:BGvsSIGmomentum}
\end{figure}

We see that the number of coincident CC events mimicking charged lepton pairs per year is completely negligible.  We now discuss other neutrino-induced processes that can lead to a signal-like signature in \name. 

\subsubsection{Single Pion Production}

The process $\nu_\mu N\to \mu^{-}\pi^+X$ may produce a signature of two charged tracks. To estimate the rate, we use the GENIE Monte Carlo simulator~\cite{Andreopoulos:2009rq}.  The number of these events per $\kg$ of detector material (or rock) induced by neutrinos with energies larger than $E_{\nu,\text{min}}$, as a function of $E_{\nu,\text{min}}$, is shown in the blue curve of the left panel of \figref{BGvsSIGmomentum}.  For all $E_{\nu,\text{min}}$, the $\mu^{-}\pi^+X$ event rate is suppressed by $\sim 100$ relative to the total CC event rate.

We expect that this background can be reduced to negligible levels.  The event rate for $\nu_\mu N\to \mu^{-}\pi^+X$ in the detector, assuming $M_{\text{det}} \sim \kg$, is very small for $E_{\nu,\text{min}} \sim \tev$, even at the HL-LHC, and such events include much more activity than simply two charged tracks.  Potentially more troublesome are events that occur in the material just before the detector and propagate into the detector.  Such events can, however, be removed by veto-ing tracks that start outside the detector volume.

It is also interesting to note that the kinematic features of the signal and background allow for a clean separation, irrespective of veto-ing incoming tracks.  Typically, in the $E_\nu$ range of our interest, the neutrino-nucleon momentum transfer is much smaller than the neutrino energy. As a result, the pion in this process is typically much softer than the muon. In contrast, the distribution of energies in the $e^+e^-$ and $\mu^+ \mu^-$ pairs from dark photon decays is typically more symmetric. We show this in the right panel of \figref{BGvsSIGmomentum}, where we compare the energy ratio of the two charged tracks for both signal (green) and background (red).  In estimating the energy ratio distribution of the $e^+e^-$ pair, we have neglected the small effects of the $A'$ polarization, and assumed that dark photons are produced unpolarized. We see that requiring the ratio of the track energies to be $E_2 / E_1 > 0.1$ removes almost all of the background, while sacrificing little of the signal. 

\subsubsection{Neutral Kaon Backgrounds}

Another possible source of background is neutrino interactions that produce kaons. In particular, neutrinos that interact in the rock or detector can produce $K_S^0$ and $K_L^0$ mesons that decay in the detector, leaving a signature of a vertex with two tracks. In particular, $B(K_S^0\to\pi^+\pi^-) \simeq 0.7$, while the $K_L^0$ has dominant three-body decays into $\pi^\pm e^\mp\nu_e$ or $\pi^\pm \mu^\mp\nu_\mu$. In addition, these tracks would be more energetically symmetric than in the case of neutrino-induced single pion production.
Nevertheless, the energy distribution of $K^0$'s produced in neutrino-nucleus interaction in the $E_\nu$ range of interest resembles that of the single pion in the above discussion. 
Using GENIE, we find that the production rate of high-energy kaons is tiny inside the detector volume, but could be large in the rock or TAN in front of \name. In the latter case, however, these kaons often lose their energy in this material before reaching the detector. We have estimated this effect by employing the FLUKA code~\cite{Ferrari:2005zk,FLUKA2014} and find that only a tiny fraction of all produced kaons can mimic our signal, and the expected number of background events is smaller than 1.

\subsection{Beam-Induced Backgrounds}

\subsubsection{Beam-Induced Backgrounds at the Far Location}
\label{sec:beam_far}

We now discuss beam-induced backgrounds, beginning with the far location, positioned 400 m from the IP along the beam collision axis, outside the main LHC tunnel.  The distance from the beam at this location, $D\approx 2.6~\m$, exceeds the size of the gap of $\sim 1~\m$ between the beam pipe and the outer wall of the LHC tunnel.  In addition, particles traveling along the beam collision axis must travel through $\sim 50$ m of matter to reach \name\ at this location.  \name\ is therefore very well shielded from hadrons and electrons.  Provided electrons and muons can be distinguished, we are unable to find any significant backgrounds to the $e^+e^-$ dark photon signal, and we consider the $e^+e^-$ signal to be essentially background-free at the far location.

Muons, on the other hand, may pass through large amounts of matter without significant attenuation.  If two opposite-charge muons are produced through beam-gas collisions within the time resolution of the \name\ detector $\delta t$, they may be reconstructed as simultaneous tracks.  We expect that such muons are deflected by the magnets that curve the proton beams, but let us conservatively neglect this effect and determine the rate of coincident muons.  From Fig.~6 of \cite{Aad:2013zwa}, the flux of beam-induced, on-axis muons with $E_\mu\gtrsim 100~\gev$ is $\Phi \sim10^{-3}~\text{Hz}~\cm^{-2}$. These muons do not arrive uniformly in time, but are concentrated in time intervals corresponding to bunch crossings, where the rate is increased by a factor $t_{\text{spacing}}/t_{\text{bunch}}$, where $t_{\text{spacing}} \simeq 25~\text{ns}$ is the bunch spacing and $t_{\text{bunch}} \simeq 30~\cm/c = 1~\text{ns}$ is the bunch crossing time. In the following, we ignore minor corrections to our estimates arising from the fact that the actual average beam crossing frequency is slightly smaller than $1$ per $25~\ns$ due to a more complicated bunch train structure.  The probability for a muon to be in a given $\delta t$ interval is, then, 
\begin{equation}
P_{\delta t} = \Phi A_{\text{det}} \frac{t_{\text{spacing}}}{t_{\text{bunch}}} \, \delta t \sim 3 \times 10^{-9} \ ,
\end{equation}
where we assume a time resolution of $\delta t = 100~\text{ps}$~\cite{Apollinari:2015bam} and have used the cross sectional area $A_{\text{det}} \simeq 1300~\cm^2$ of the far detector. We therefore expect
\begin{equation}
N_{\mu^+ \mu^-} = P_{\delta t}^2 \, \frac{T}{\delta t}\,\frac{t_{\text{bunch}}}{t_{\text{spacing}}} \sim 0.1
\end{equation}
coincident two-muon events per year, where we have taken $T \sim 10^7~\s$ for an LHC year, and the factor $t_{\text{bunch}}/t_{\text{spacing}}$ accounts for the fact that, on average, we expect that background particles will arrive in \name\ only during times corresponding to bunch crossings. 

$N_{\mu^+ \mu^-}$ is linearly proportional to $\delta t$; if time resolutions of $\delta t = 10~\text{ps}$ can be realized~\cite{CTPPS}, $N_{\mu^+ \mu^-}$ would be reduced further by an order of magnitude.  The number of $\mu^+ \mu^-$ background events can also be greatly suppressed by requiring energies significantly above 100 GeV, by vetoing muons that arrive from outside the detector, and by requiring that the two tracks reconstruct a vertex.  We conclude that, for \name\ placed in the far location, the $e^+ e^-$ and $\mu^+ \mu^-$ signatures of dark photons can be completely distinguished from the beam-induced backgrounds we have considered, and the far location therefore provides a potentially background-free environment for such new physics searches.

\subsubsection{Beam-Induced Backgrounds at the Near Location}
\label{sec:beam_near}

The near location is, of course, a far more challenging environment for new physics searches.  At this location, there are large backgrounds from neutral particle-TAN interactions, and the resulting particles are not bent away from the \name\ detector.  An accurate picture of the size of these backgrounds requires a dedicated simulation, using tools such as the FLUKA~\cite{Ferrari:2005zk,FLUKA2014} and MARS~\cite{Mokhov:1995wa,Mokhov:2012ke} packages, or the experimental data themselves, but this is beyond the scope of this study.  

Some important observations are possible, however, given results in the literature.  For example, MARS simulation results have been presented in Ref.~\cite{Mokhov:2003ha}.  In Fig.~32 differential fluxes $d\Phi/dE$ are presented for protons, neutrons, mesons, photons, electrons, and muons at positions just before the TAN and inside the TAN, and in a $4~\cm \times 4~\cm$ square centered on the beam axis.  Similar results, but after the TAN and at radii of $13-46~\cm$ from the beam line, are presented in Fig. 41.  From these figures, we see that the beam-induced flux drops very rapidly as one moves away from the beam line and, for hadrons and electrons, also as one moves through the TAN.  

In contrast to the electron and hadron fluxes, the muon flux presented in Fig.~32 of Ref.~\cite{Mokhov:2003ha} may be safely assumed to be the flux seen in the after-TAN near location for \name.  We see that for $E_{\mu} \agt 100~\gev$, the flux is $\Phi \sim 10^3~\text{Hz}~\cm^{-2}$.  
Given the near location cross sectional area $A_{\text{det}} \simeq 50~\cm^2$, and following the analysis of \secref{beam_far}, in which the number of background events scales as $(\Phi A_{\text{det}})^2$, we expect $N_{\mu^+ \mu^-} \sim 10^{8}$ beam-induced, high-energy muon pair events per year at the near location.  There may also be correlated muon pair backgrounds, for example, from $J/\psi \to \mu^+ \mu^-$. Clearly for the $\mu^+ \mu^-$ signal, it is important to reduce these backgrounds by veto-ing tracks that start outside the detector and requiring that the tracks reconstruct a vertex, point back to the IP, and have symmetric energies.  For the $e^+ e^-$ signal, it is important to distinguish muons from electrons. 

Assuming sufficient muon discrimination, the leading backgrounds are from charged hadrons and electrons.  The charged hadrons dominate the electron flux, and at the after TAN location, at least for radii of $13-46~\cm$, are far below muons and roughly of the order of $\Phi \sim 10^{-1}~\text{Hz}~\cm^{-2}$, implying a coincident background of $\sim 1$ event per year. Additionally there might be a sizable rate of neutral kaon decays $K_S \to \pi^+ \pi^-$ and $K_L \to \pi^\pm \ell^\mp \nu$ before and inside the detector. Such a background level is tolerable, provided it is well-estimated, given the possibility of very large signal rates that we see below.  It may also be reduced by requiring a veto on tracks that start outside the detector, that the tracks point back to the IP and reconstruct a vertex, and by the ability to differentiate charged hadrons from electrons.  The electron background is negligible relative to the charged hadrons.

\section{Expected Reach and Results}
\label{sec:results}

We now estimate the reach in dark photon parameter space of the detectors we have discussed above.  For dark photons masses $m_{A'}<1~\gev$, the branching ratio into two, opposite-charge particles ($ee$, $\mu\mu$ and $\pi^+\pi^-$) is almost $100\%$~\cite{Buschmann:2015awa}.\footnote{An exception is the region around the narrow $\omega$ resonance at $m_{A'}=782.6 \pm 8.5~\mev$, in which the dark photon can mix with the $\omega$ and the decay mode to $\pi^+ \pi^- \pi^0$ becomes sizable.} We therefore focus on the signature of two opposite-charge high-energy tracks that was discussed in detail in the previous sections. Following the discussion in \secref{signalanddetector}, we assume a detector that includes a high-resolution tracking system and a magnetic field. We further assume $100\%$ efficiency in detecting and reconstructing the  dark photon signature. The number of signal events $N_{\text{sig}}$ is then equal to the number of dark photons that decay within the detector volume. 

\Figref{eventrate} shows contours of $N_{\text{sig}}$ in the $(\epsilon, m_{A'} )$ plane. The three contour types correspond to the three dark photon production sources discussed above: $\pi^0$ decay, $\eta$ decay, and dark photon bremsstrahlung. The gray-shaded regions represent parameter space that has already been excluded by previous experiments. The left and right panels assume the far and near detector benchmark designs of \eqsref{eq:far_benchmark}{eq:near_benchmark}, respectively. In estimating $N_{\text{sig}}$, we have employed a cut on the dark photon momentum, $p_{A'} > 100~\gev$, which is anyway effectively imposed by the requirement that the dark photons propagate to the detector locations considered.  As we see, 1 to $10^5$ dark photon events may be detected by \name\ in currently viable regions of dark photon parameter space with $m_{A'} \sim 10~\mev - 1~\gev$ and $\epsilon \sim 10^{-7} - 10^{-3}$.   

\begin{figure}[t]
\centering
\includegraphics[width=0.47\textwidth]{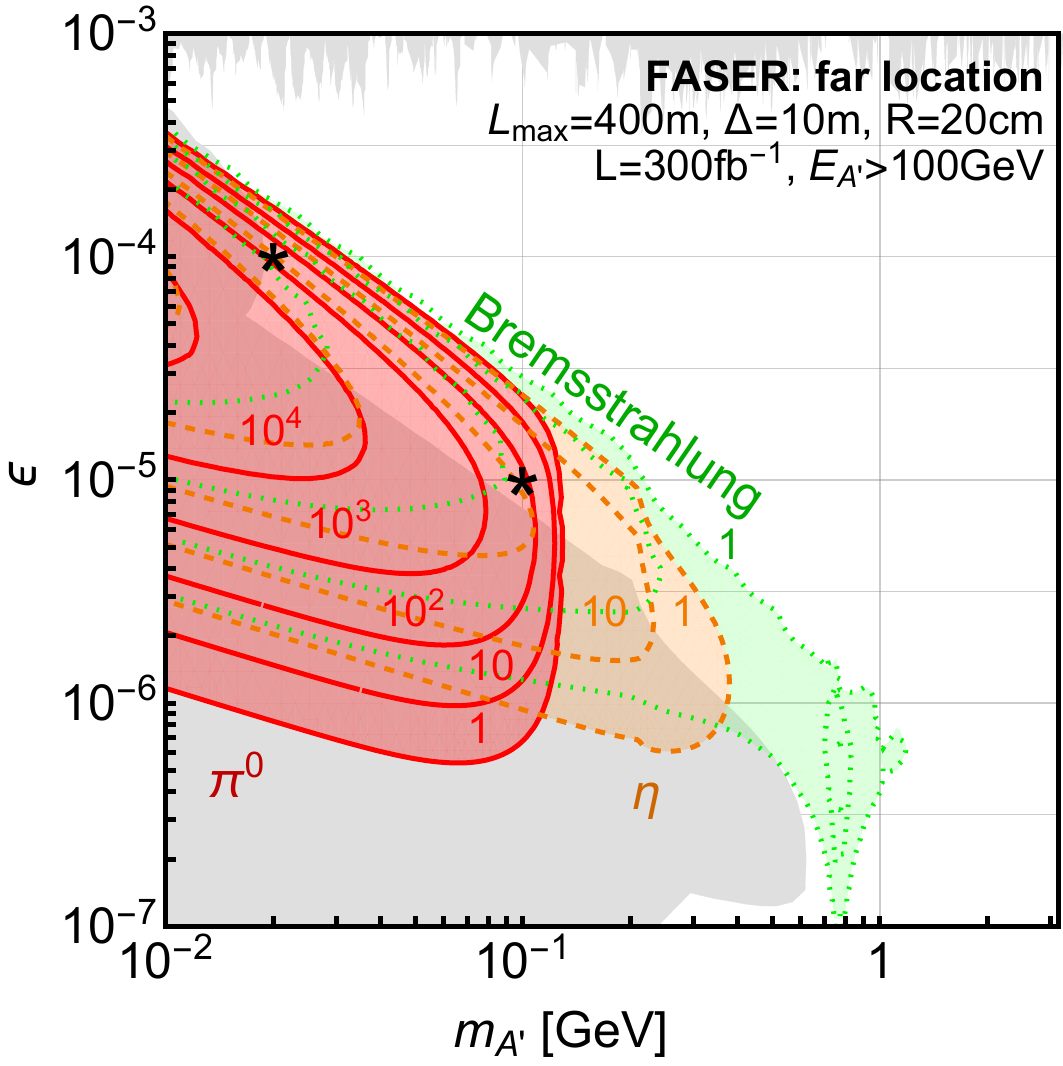} \quad
\includegraphics[width=0.47\textwidth]{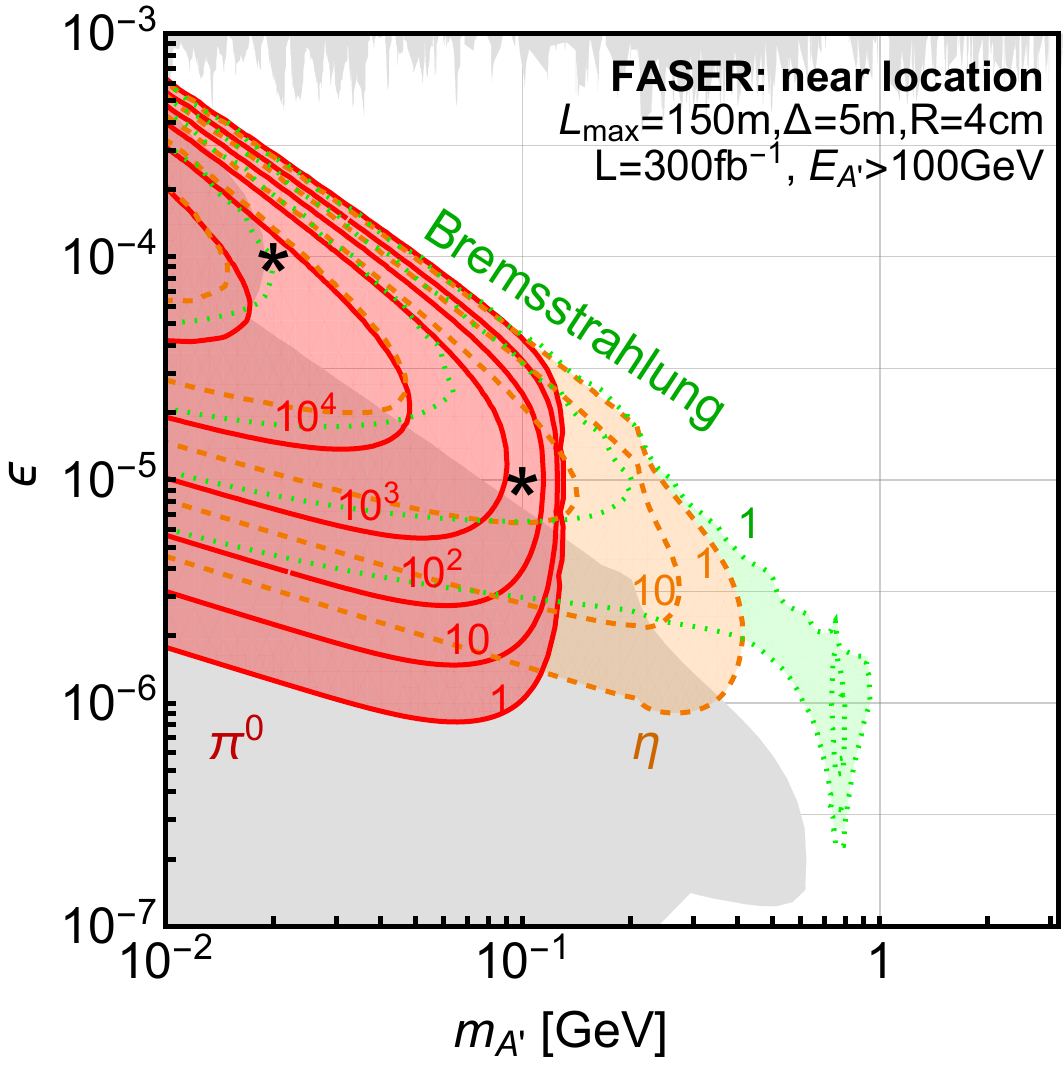}
\caption{Number of signal events in dark photon parameter space for the far (left) and near (right) detector locations, given an integrated luminosity of $300~\ifb$ at the 13 TeV LHC. The different colors correspond to the three production mechanisms: $\pi^0\to A' \gamma$ (red), $\eta\to A' \gamma$ (orange), and proton bremsstrahlung (green). Contours represent the number of signal events $N_{\text{sig}}$. The gray shaded regions are excluded by current experimental bounds. The black stars correspond to the representative parameter-space points of \eqref{eq:ps_points}.}
\label{fig:eventrate}
\end{figure}

The region of parameter space probed by \name\ has interesting implications for dark matter. If the dark photon couples to a hidden sector particle $X$ with $m_X \sim m_{A'}$, the $X$ annihilation cross section is $ \sigma (XX \to A' \to \text{SM}) \sim \epsilon^2 \alpha \alpha_D / m_{A'}^2$, where $\alpha_D$ is the hidden sector's fine structure constant.  $X$ can then be a WIMPless dark matter candidate with the correct thermal relic density if $\epsilon^2 \alpha \alpha_D / m_{A'}^2 \sim \alpha_{\text{weak}}^2/ m_{\text{weak}}^2$~\cite{Feng:2008ya}.  Assuming $\alpha_D \sim 1$, this implies $\epsilon \sim m_{A'} / m_{\text{weak}}$, that is, for $m_{A'} \sim 10-100~\mev$ one obtains $\epsilon \sim 10^{-5} - 10^{-4}$. Therefore, provided that the invisible decay channel, $A'\to XX$, is kinematically forbidden, \name\ probes regions of parameter space where, in simple scenarios, hidden matter has the correct thermal relic density to be dark matter.

\Figref{reach} shows the exclusion reach for the far (left) and near (right) detector design benchmarks for an integrated luminosity of $\mathcal{L}=300~\ifb$ (solid) and $\mathcal{L}=3~\iab$ (dashed). It is based on the assumption that background can be distinguished from signal by employing a combination of the cuts discussed in \secref{background} and that the systematic uncertainty of the signal rate is small. Given this assumption, $95\%$ C.L. exclusion contours correspond to $N_{\text{sig}}=3$ contours. It is important to note, though, that even a relatively large number of background events above the simple estimates from \secref{background} would not drastically reduce the reach in parameter space, provided the background is well-understood, especially in the upper part of the exclusion regions with the kinetic mixing parameter $\epsilon \sim 10^{-3}-10^{-5}$ that is of most interest to us. This is because, in this region of the dark photon parameter space, the number of expected events grows exponentially with decreasing $(\epsilon\,m_{A'})^2$ as discussed below. For comparison, in \figref{reach} we also show the expected reach of other proposed searches for dark photons with small $\epsilon$, namely LHCb~\cite{Ilten:2015hya,Ilten:2016tkc}, HPS~\cite{Moreno:2013mja}, SeaQuest~\cite{Gardner:2015wea}, and SHiP~\cite{Alekhin:2015byh}.

\begin{figure}[t]
\centering
\includegraphics[width=0.47\textwidth]{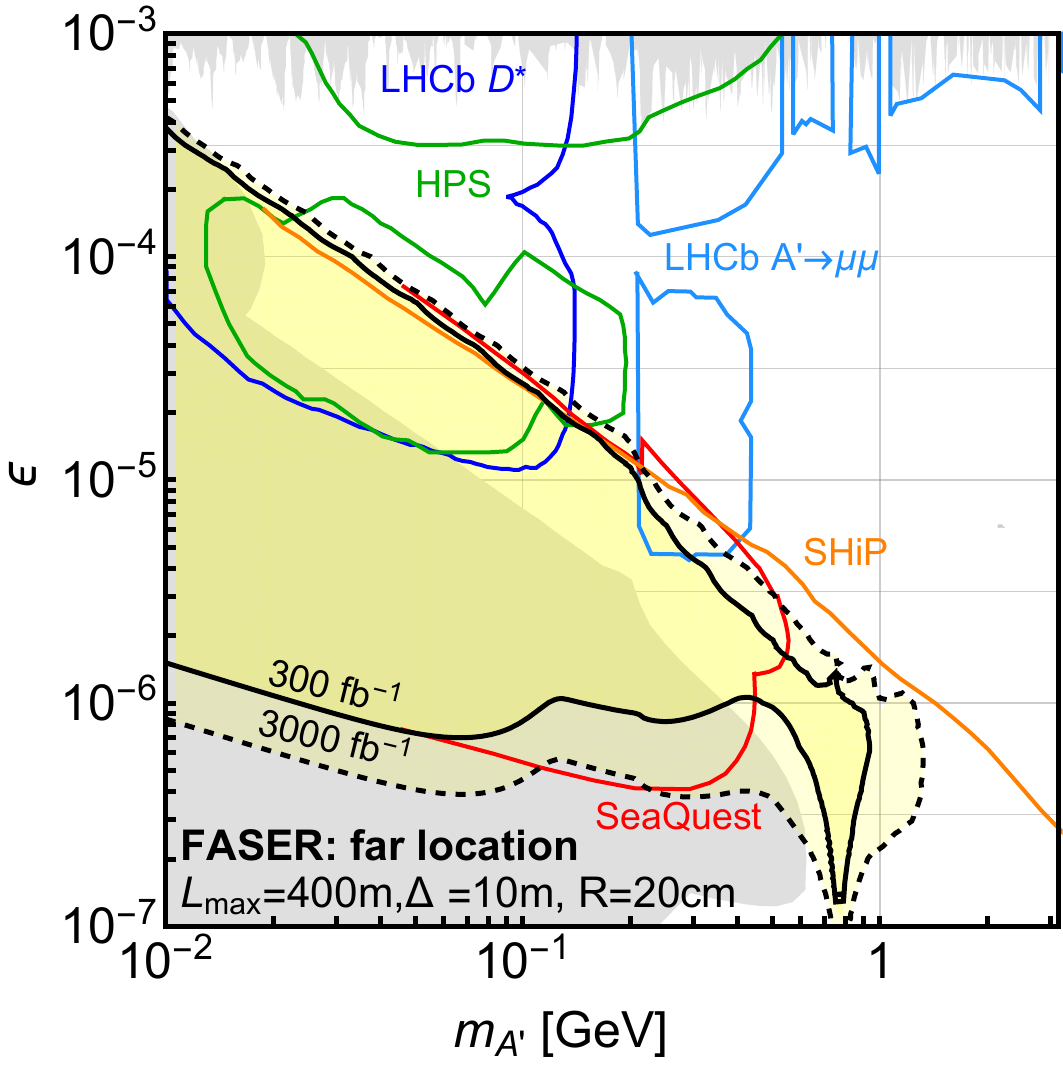} \quad
\includegraphics[width=0.47\textwidth]{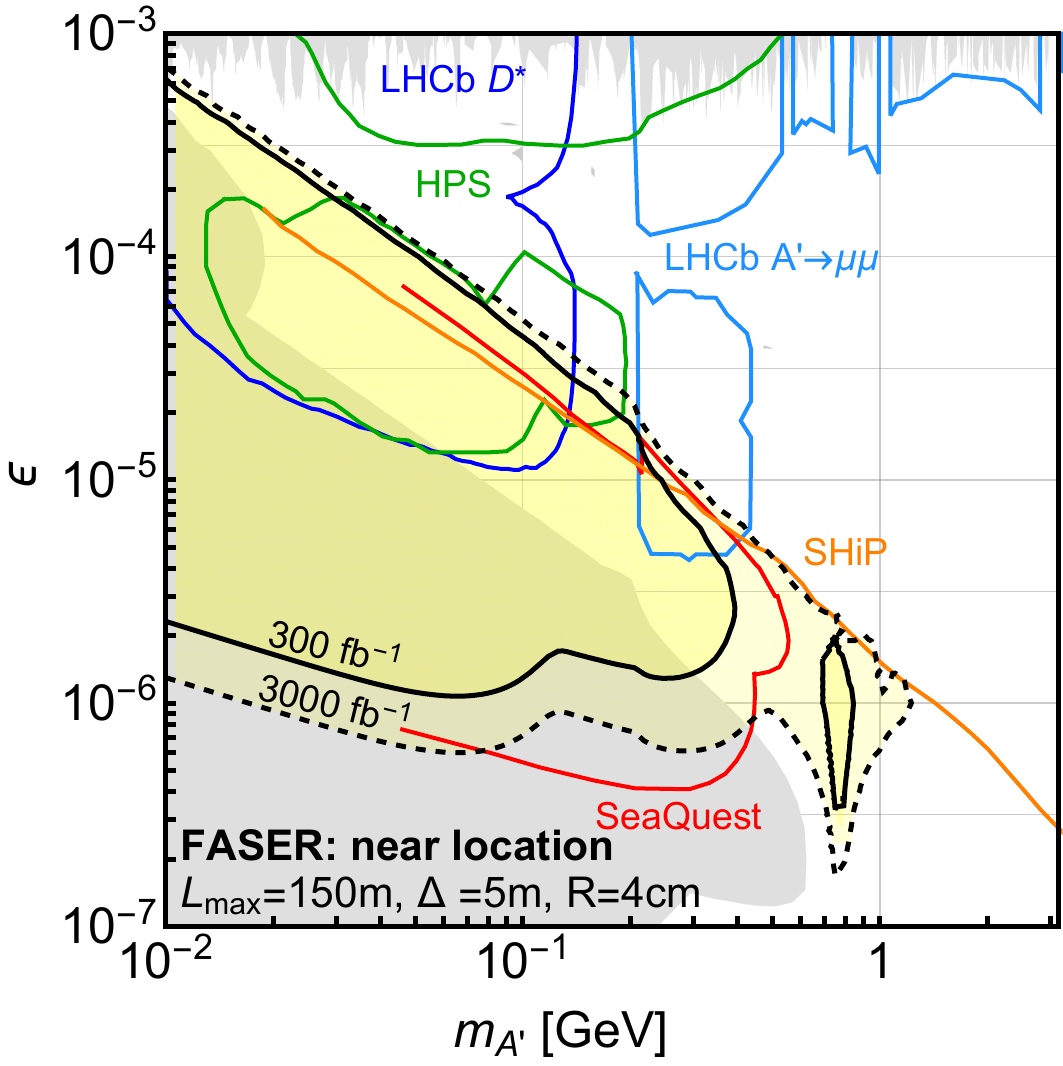}
\caption{Combined 95\% C.L. exclusion reach on dark photon parameter space for the far (left) and near (right) detector design benchmarks for an integrated luminosity of $300~\ifb$ (solid line) and $3~\iab$ (dashed line). The gray shaded regions are excluded by current experimental bounds, and the colored contours represent projected future sensitivities of LHCb~\cite{Ilten:2015hya,Ilten:2016tkc}, HPS~\cite{Moreno:2013mja}, SeaQuest~\cite{Gardner:2015wea}, and SHiP~\cite{Alekhin:2015byh}. 
}
\label{fig:reach}
\end{figure}

To better understand the results shown in \figref{reach}, it is helpful to analyze the dependence of $N_{\text{sig}}$ on the dark photon parameters $m_{A'}$ and $\epsilon$ in various limits. To this end, let us recall that
\be
N_{\text{sig}} = {\cal L}^{\text{int}}\sigma_{pp\to A' X} \mathcal{P}_{A'}^{det} (p_{A'},\theta_{A'})\propto {\cal L}^{\text{int}}\,\epsilon^2\,e^{-\lmin/\bar{d}}\,\left[ 1-e^{-(\lmax-\lmin)/\bar{d}}\right] \ ,
\ee
where $\bar d\sim \alpha^{-1}_{EM} p_{A'} \epsilon^{-2} m_{A'}^{-2}$ and, therefore, we obtain
\be
N_{\text{sig}} \propto\left\{\begin{array}{cc}
{\cal L}^{\text{int}}\,\epsilon^2\,e^{-\lmin/\bar{d}} 
& \text{for\ \ }\bar{d}\ll \lmin\\
\vspace{-0.2cm}
& \\
{\cal L}^{\text{int}}\,\epsilon^2\,\frac{\lmax-\lmin}{\bar{d}} 
& \text{for\ \ }\bar{d}\gg \lmin \ .
\end{array}
\right.
\ee
In the upper part of the exclusion regions in \figref{reach}, the characteristic dark photon decay length drops below the distance to the detector, i.e., we are in the regime where $\bar{d} \ll L_{\text{near}}$. In this case,  $N_{\text{sig}}$ scales linearly with ${\cal L}^{\text{int}}$, but is exponentially suppressed for increasing $(\epsilon\, m_{A'})^2$.  The contours of $N_{\text{sig}}$ in \figref{eventrate} are therefore very tightly spaced in $(m_{A'}, \epsilon)$ parameter space, and the reach shown in \figref{reach} is not improved much by going from ${\cal L}^{\text{int}} = 300~\ifb$ to $3~\iab$.  On the other hand, a change in the detector location, $\lmin$, or maximum dark photon momentum, $p_{A'}^{\text{max}}$, can have a significant effect on the reach. This is because, requiring the characteristic decay length to be similar to the distance to the detector, $\bar{d}\sim \lmin$, implies $\epsilon\, m_{A'} \propto \sqrt{p_{A'}^{\text{max}}/\lmin }$, and so the reach in the parameters $m_{A'}$ and $\epsilon$ is quite sensitive to changes in $p_{A'}^{\text{max}}$ and $\lmin$. We use this feature to compare \name\ to the aforementioned beam dump experiments below.

In the opposite limit, at small $\epsilon$, i.e., for $\bar{d}\gg \lmin$, we obtain $N_{\text{sig}}\propto \epsilon^4\,m_{A'}^2$. The number of events is now only suppressed as a power of $\epsilon$, not exponentially, as $\epsilon$ decreases.  Contours of $N_{\text{sig}}$ in \figref{eventrate} are therefore less tightly spaced in $(m_{A'}, \epsilon)$ parameter space, and the reach shown in \figref{reach} is significantly improved by going from ${\cal L}^{\text{int}} = 300~\ifb$ to $3~\iab$.   

\Figref{reach} shows that the sensitivity contours of \name, SeaQuest, and SHiP have fairly similar boundaries at high $\epsilon$. This is as expected, given the discussion above: both SeaQuest and SHiP have luminosities (protons-on-target) that are several orders of magnitude larger than \name, but the reach at high $\epsilon$ is mainly determined by the $\lmin/p_{A'}^{\text{max}}$ ratio, and this is similar for the far detector location, SHiP, and SeaQuest:\footnote{In the case of the SeaQuest experiment the distance $\lmin$ is not strictly defined due to the details of the detector design. We follow Ref.~\cite{Gardner:2015wea} in our estimate.}
\begin{eqnarray}
\textbf{\name, far detector:} \quad
\frac{\lmin}{p_{A'}^{\text{max}}} &= \frac{390~\m}{6500~\gev}  
&= \ 0.060 \ \frac{\m}{\gev}  \\
\textbf{SHiP:} \quad
\frac{\lmin}{p_{A'}^{\text{max}}} &= \; \; \frac{63.8~\m}{400~\gev} 
&= \ 0.160\  \frac{\m}{ \gev} \\
\textbf{SeaQuest:} \quad
\frac{\lmin}{p_{A'}^{\text{max}}} &= \; \; \frac{4~\m}{120~\gev}   
&= \ 0.033\ \frac{\m}{ \gev} \ .
\end{eqnarray}
The larger luminosities of SeaQuest and SHiP do improve their reach at low $\epsilon$ relative to \name, but the corresponding parameter space is already largely excluded by other experimental constraints.  Note that the extended reach of SHiP at large masses $m_{A'}\gtrsim 500~\mev$, is mainly due to the hard QCD contribution to dark photon production, which we have not included in our analysis for the reasons discussed in \secref{directproduction}. If  one focuses on the contributions from meson decays and bremsstrahlung only (see, e.g., Ref.~\cite{Graverini:2016}), the reaches are similar.

The comparison between SeaQuest, SHiP, and \name\ is, of course, dependent on many factors; our only goal in this simple discussion is to explain why their reaches are very roughly comparable.  One may, however, more precisely compare the near and far detector designs for \name\ by noting that $\lmin/p_{A'}^{\text{max}}$ is smaller for the near detector:
\begin{eqnarray}
\textbf{\name, near detector:} \quad
\frac{\lmin}{p_{A'}^{\text{max}}} &= \frac{145~\m}{6500~\gev}  
&= \ 0.022 \ \frac{\m}{\gev} \ .
\end{eqnarray}
This implies an improved reach at large $\epsilon$, which is indeed apparent in the right panel of \figref{reach}. 
At small $\epsilon$, the near detector exhibits a slight loss of sensitivity, because of its smaller angular acceptance. Dark photons with $\bar d$ comparable to the near detector location typically have lower momentum than in the far detector case. The discussion of the $(\theta, p)$ distributions in \secref{production} shows that such dark photons are produced at relatively large angles, which are not covered by the smaller near detector. Of course, the low $\epsilon$ boundary is already excluded by other experiments, and so on the whole, the near detector probes more virgin territory in dark photon parameter space.

\section{Conclusion and Outlook}
\label{sec:conclusions}

Although the ATLAS and CMS experiments have focused primarily on searches for heavy new particles at high-$p_T$, the LHC also provides an exceptional environment to search for light, weakly-coupled new physics. Such particles may well be found not at high $p_T$, but at low $p_T$.  For an integrated luminosity of 300 $\ifb$, the LHC is expected to produce about $2.3 \times 10^{16}$ inelastic $pp$ scattering events, allowing for the production of a sufficient number of light particles even if they are extremely weakly coupled, and most of these are in the very forward direction. 

Because of their small couplings, these particles may travel a macroscopic distance before decaying. We propose to place a new experiment, \fullname, or \name, in the very forward region, downstream of the ATLAS or CMS IP.  We consider two locations for \name, both on the beam axis: one 400 m from the IP after the LHC tunnel starts to curve, and another 150 m from the IP, right behind the TAN and before the D2 magnet inside the straight part of the LHC tunnel.  \name\ would operate concurrently within the LHC infrastructure. 

As a new physics example, we have considered dark photons with mass $m_{A'}\sim \mev - \gev$. In this mass range, dark photons can be produced in light meson decays or via proton bremsstrahlung, and they decay predominantly into two (meta-)stable charged particles. Only the most energetic dark photons with $E_{A'}\agt 1~\tev$ are expected to reach the \name\ detector. Equipping the detector with a tracking system and a magnetic field would allow \name\ to identify the signal and distinguish it from background. For the far location, we expect that the backgrounds are negligible.  A small cylindrical detector with an outer radius of just $20~\cm$ and a length of $10~\m$ (total volume $\sim~1~\m^3$) at the far location may be sufficient to discover dark photons in a large region of unprobed parameter space with $m_{A'}\sim10~\mev-1~\gev$ and kinetic mixing $\epsilon \sim 10^{-7}-10^{-3}$.  The near location is a much more challenging environment, but given the ability to veto tracks originating outside the detector and distinguish electrons from muons and charged hadrons, the backgrounds may also be highly suppressed, and the signal reach is potentially even better than for the far location.  

The reach of \name\ is potentially comparable to the projected reach of the proposed SHiP experiment at the high $\epsilon$ boundary.  As discussed in \secref{results}, this boundary is set largely by the ratio $\lmin/ p_{A'}^{\text{max}}$, which is similar for \name\ and SHiP.  The projected reach of SHiP at the low $\epsilon$ boundary, $\epsilon\sim10^{-9}$, is much greater, given its $10^4$ times larger number of collisions, but other previous experiments already exclude most of this region.  The projected reach of SHiP also extends to larger $m_{A'}$, based on estimates of direct dark photon production processes, which we have not included here for the reasons given in \secref{directproduction}. 

We believe this study significantly motivates future work on a detector for new physics searches in the very forward region of the ATLAS and CMS experiments.  There are many interesting future directions to explore.  On the experimental side, clearly the feasibility of the proposed detector locations and designs must be carefully examined.  In particular, the near detector behind the TAN should be integrated into the LHC infrastructure, and the beam-induced backgrounds should be estimated more carefully, or better yet, measured experimentally. It is also possible that other detector locations may be promising.  For example, an off-axis position downstream from the D1 magnet, considered in \appref{off_axis_detector}, could be useful to probe other new physics scenarios.  Alternatively, one could consider putting \name\ after the D2 magnet; charged particles produced in collisions with the TAN would then be deflected, reducing background.  We note also that it may be fruitful to consider other types of particle collisions such as proton-lead and lead-lead. Although these collisions are explored at lower luminosities by the LHC program~\cite{Goddard:1629486}, the typical cross sections are larger~\cite{Oyama:2013xra, Abelev:2014epa, Khachatryan:2015zaa} and could lead to significant signal rates.

On the theoretical side, we have focused primarily on dark photon production in meson decays and through proton bremsstrahlung, where the rates are well-understood. As discussed in \secref{directproduction}, however, direct production of dark photons and other light states is expected to be one of the dominant production mechanisms.  Many current parton distribution functions are unable to adequately describe the corresponding kinematic region. In this sense, the results presented in this paper should be seen as conservative estimates, and it would be good to include estimates for direct production in the reach projections.  Our sensitivity contours are also conservative because they do not include the possible signal of dark photons decaying to muon pairs in the material before the detector, as well as secondary production in the TAN, which may possibly extend the reach, as discussed in \secref{extended}.

Finally, we have considered dark photons as one popular example of light, weakly-coupled new physics. \name\ will also probe many other interesting physics topics.  As already pointed out in Ref.~\cite{Park:2011gh}, a very forward detector might be able to detect a sizable number of neutrino events and measure the corresponding cross sections at high energies.  In addition, there is a vast array of other new physics scenarios that can be probed with \name, and it would be interesting to study the potential of \name\ to discover new physics in these frameworks. Examples include other mediators that induce couplings between WIMPless dark matter and the SM~\cite{Feng:2008ya}, as discussed in \secref{results}; the parameter regions of SIMP/ELDER models that reduce to WIMPless models~\cite{Kuflik:2015isi}; axion-like particles that mix with pions and decay to $e^+ e^-$ pairs with $\sim 10-100$ m decay lengths~\cite{Dolan:2014ska}; heavy neutral leptons that mix with active neutrinos and decay via $N \to e^+ e^- \nu$ with long lifetimes~\cite{Gorbunov:2007ak,Atre:2009rg}; co-annihilating light dark matter scenarios, which are difficult to probe through direct and indirect detection, but where the heavier dark state decays to the lighter one with long lifetime through $\chi_2 \to \chi_1 e^+ e^-$~\cite{Izaguirre:2015zva}; and dynamical dark matter, where dark matter consists of an ensemble of particles with a variety of masses and lifetimes~\cite{Dienes:2011ja,Dienes:2011sa}.

In summary, we look forward to discovering new physics at the LHC!

\acknowledgments

We thank John Campbell, David Casper, Susan Gardner, Joey Huston, Ben Kreis, Andrew Lankford, Nikolai Mokhov, Tanguy Pierog, and Jordan Smolinsky for useful discussions, David Cohen for help with cluster computing, and, particularly, Mike Albrow for many useful insights. This work is supported in part by NSF Grant No.~PHY-1620638. J.L.F. is supported in part by Simons Investigator Award \#376204. I.G. and F.K. performed part of this work at the Aspen Center for Physics, which is supported by NSF Grant No.~PHY-1607611.  S.T. is supported in part by the Polish Ministry of Science and Higher Education under research grant 1309/MOB/IV/2015/0.  

\appendix

\section{Off-Axis Detector Location}
\label{sec:off_axis_detector}

The main goal of this paper was to explore the potential of downstream on-axis detectors to discover very forward, long-lived particles. In this appendix, we entertain the possibility of an off-axis detector, which could be placed even closer to the IP than the on-axis near detector considered above, while still being shielded at some level from SM particles created at the IP. In particular, a careful choice of the azimuthal position of such a detector could significantly reduce the flux of charged particles that are deflected by the D1 magnet into \name, as can be seen in \figref{infrastructure}.

As with the on-axis detectors, we consider cylindrical shapes for the off-axis detector, but with inner and outer radii $R_{\text{in}}$ and $R_{\text{out}}$, respectively. When presenting the sensitivity plots below we assume for simplicity that the off-axis detector is a full hollow cylinder that surrounds the LHC infrastructure. In more realistic setups that account for the aforementioned reduction of the SM background, this should be replaced by a slice of azimuthal angular size $\phi$, which would reduce the signal rate by a factor of $\phi/2\pi$. In the simplest case of the hollow cylinder, the decay in volume probability is
\be
\label{eq:P_decay_in_volume_off_axis}
\mathcal{P}_{A'}^{det} (p_{A'},\theta_{A'})
= ( e^{-\lmin/\bar{d}} - e^{-\lmax/\bar{d}} ) \
\Theta(R_{\text{out}} \! - \! \tan\theta_{A'}\lmax) \
\Theta(\tan\theta_{A'}\lmax \! -\! R_{\text{in}}) \, .
\ee

In choosing $R_{\text{in}}$ and $R_{\text{out}}$, we note that the TAS only absorbs particles at angles above $\theta=0.9~\mrad$.  To use the TAS as a shield, we therefore consider the following detector geometry as a benchmark design for the off-axis detector:
\be
\label{eq:off_axis_benchmark}
\textbf{off-axis detector: } \lmax=100~\m,\, \Delta=10~\m,\, R_{\text{out}} \! =20~\cm,\, R_{\text{in}} \! =10~\cm \, .
\ee

\Figref{aprimeptheta_weighted_off_axis} shows the $(\theta, p)$ distribution for dark photons that decay within the $(\lmin , \lmax)$ values given in \eqref{eq:off_axis_benchmark} for the off-axis design, neglecting the angular requirements.  By virtue of its closer location, this detector design benefits from capturing less-energetic dark photons with $E_{A'}\sim\text{few}~100~\gev$. Unfortunately, most of these dark photons travel at small angles $\theta_{A'}<1~\mrad$ relative to the beam axis, and therefore fall outside the angular coverage of the off-axis detector, which is indicated by the gray dashed lines. 

\begin{figure}[t]
\includegraphics[width=0.32\textwidth]{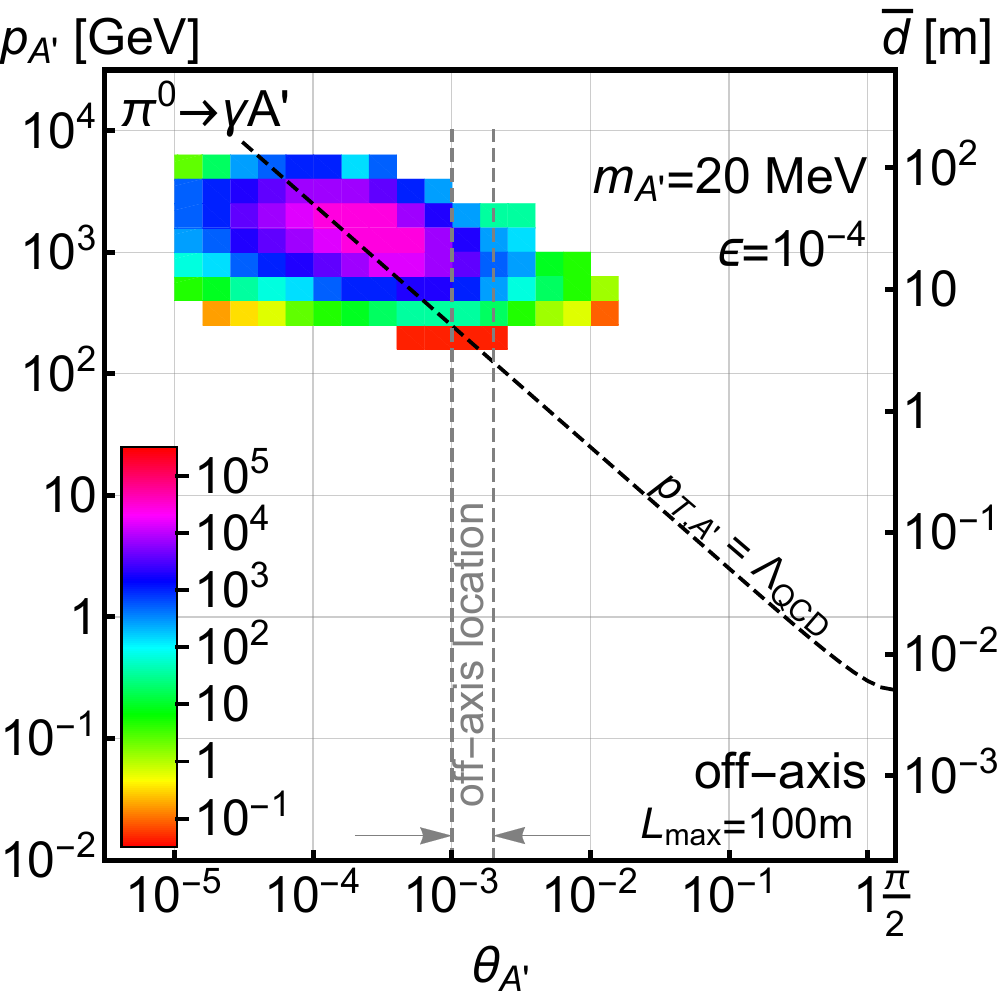}
\includegraphics[width=0.32\textwidth]{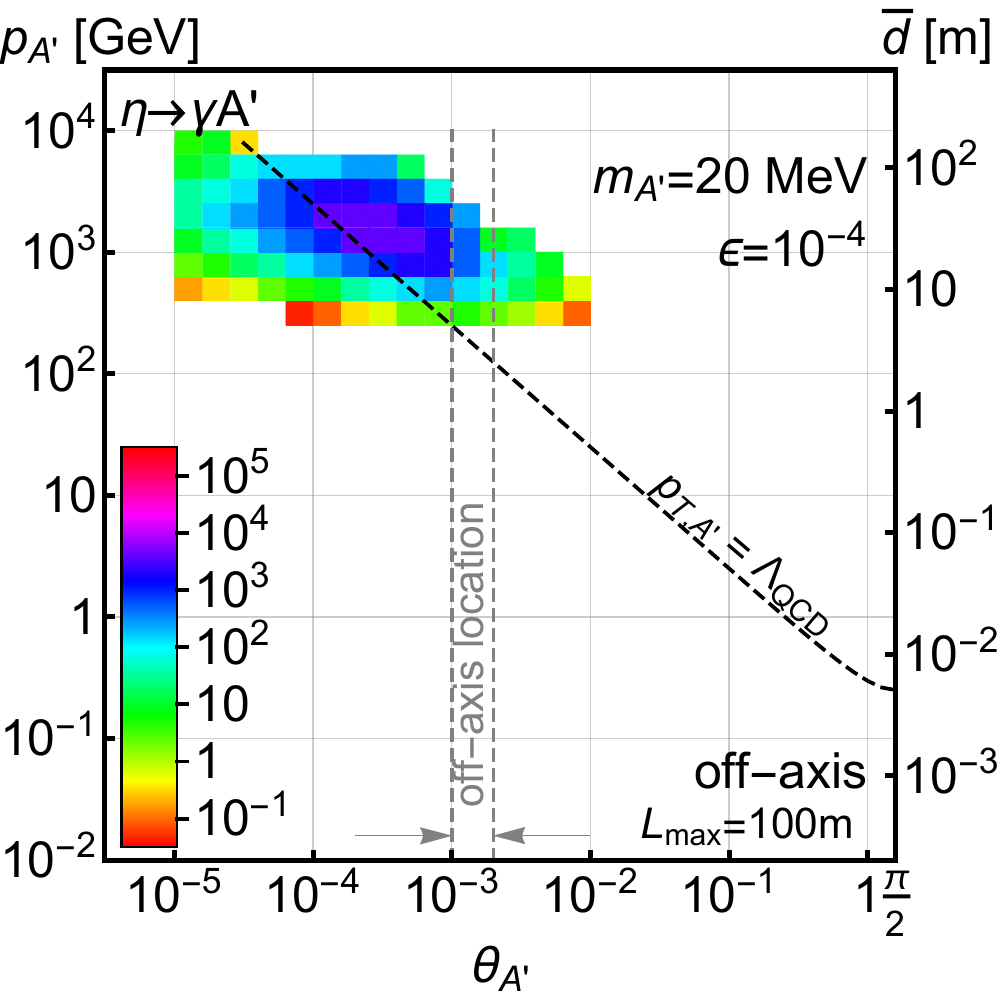} 
\includegraphics[width=0.32\textwidth]{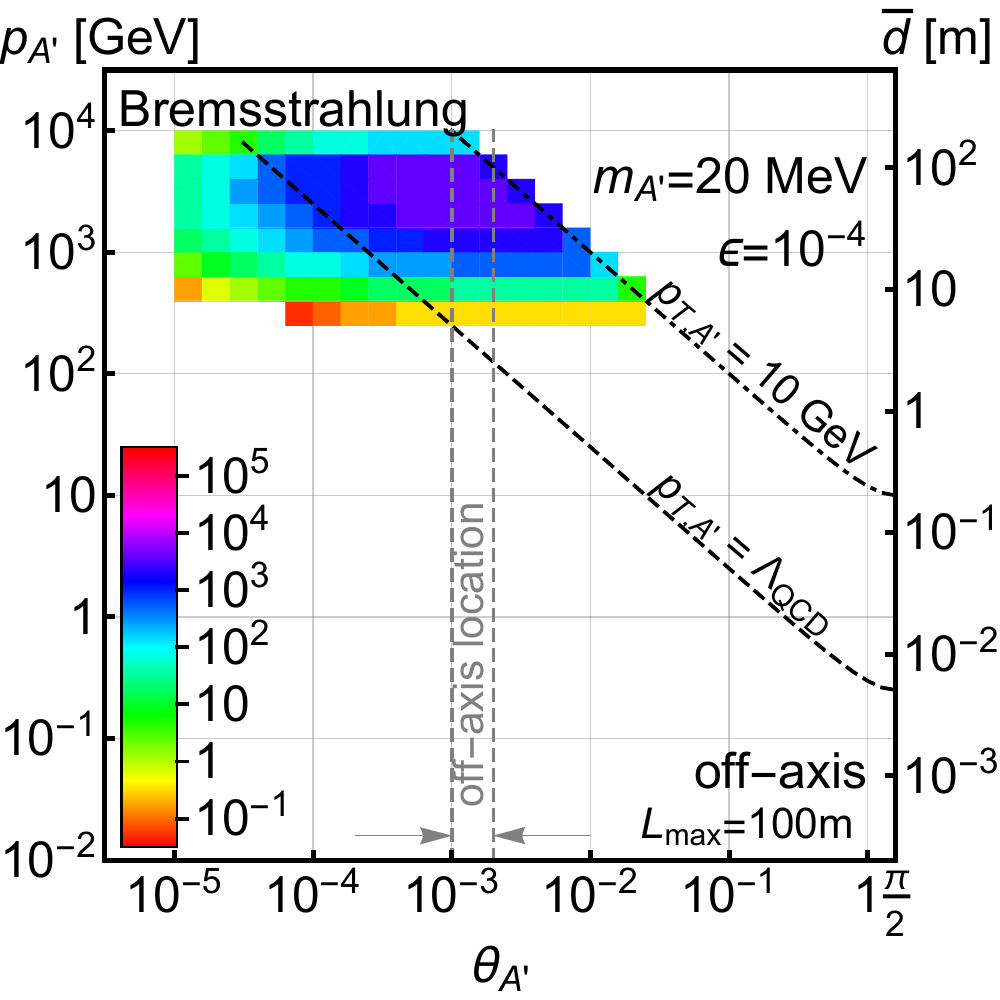}
\includegraphics[width=0.32\textwidth]{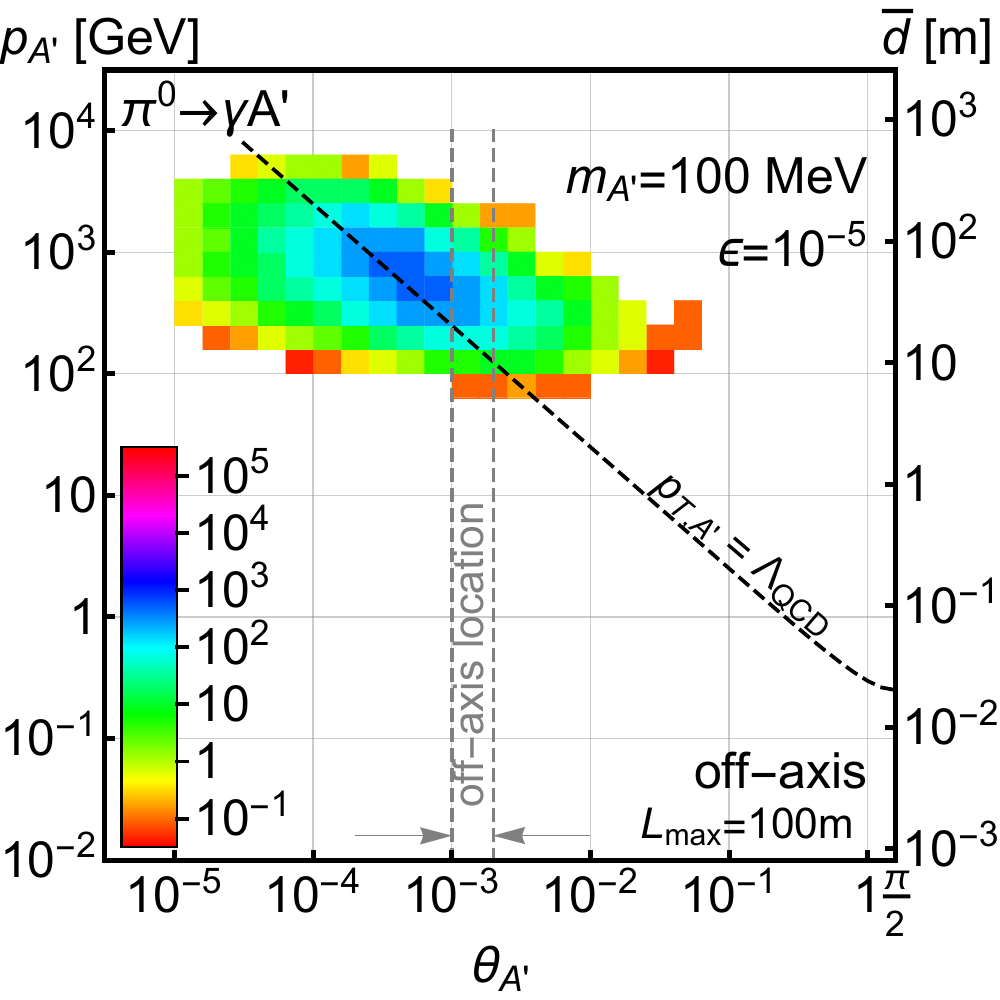}
\includegraphics[width=0.32\textwidth]{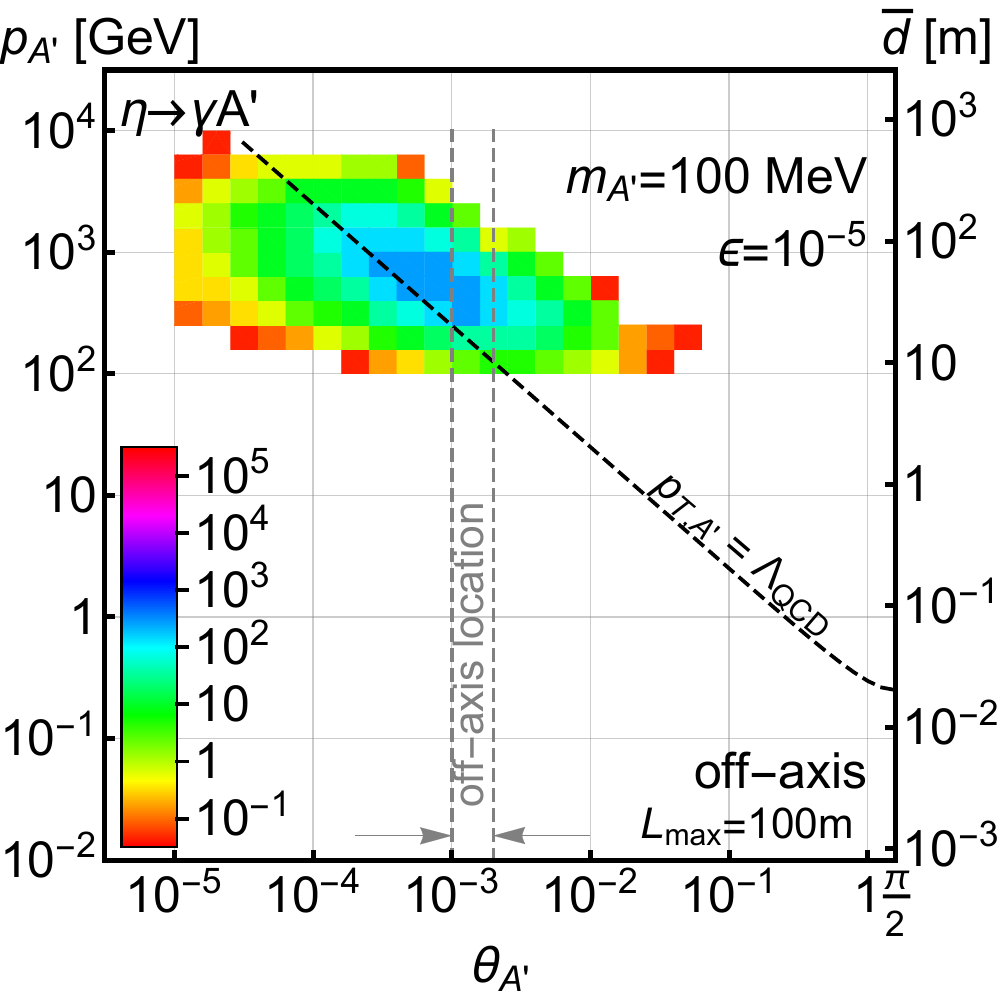}
\includegraphics[width=0.32\textwidth]{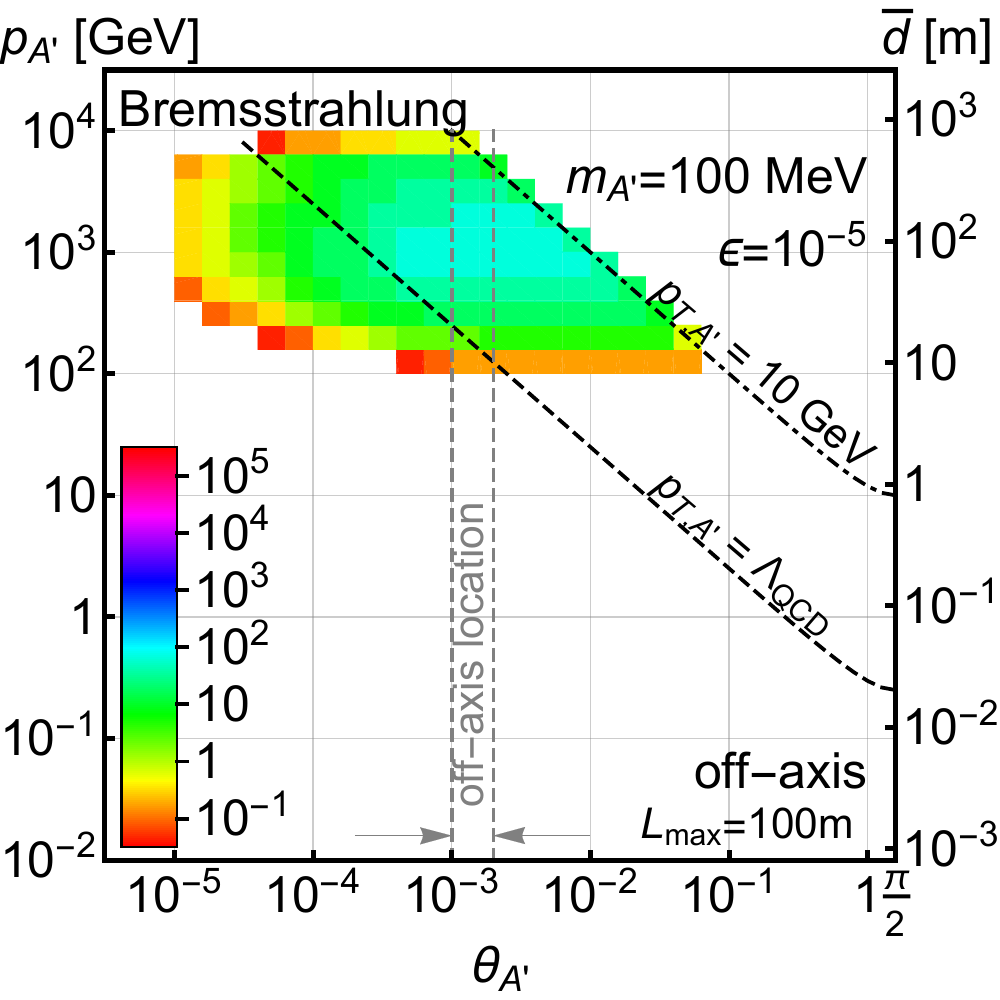}
\caption{Same as in \figref{aprimeptheta_weighted-onaxis}, but for the off-axis detector location with $(\lmin, \lmax) = (90~\m, 100~\m)$.}
\label{fig:aprimeptheta_weighted_off_axis}
\end{figure}

In \figref{reach_off_axis} we summarize the results for the off-axis design as in \secref{results}. For the reach, we have assumed negligible background. This is a strong assumption for the off-axis case, given the proximity of the off-axis detector to the IP.  A dedicated background estimation is required using tools like, for example, FLUKA~\cite{Ferrari:2005zk,FLUKA2014} and MARS~\cite{Mokhov:1995wa,Mokhov:2012ke}. 

\begin{figure}[h]
\centering
\includegraphics[width=0.47\textwidth]{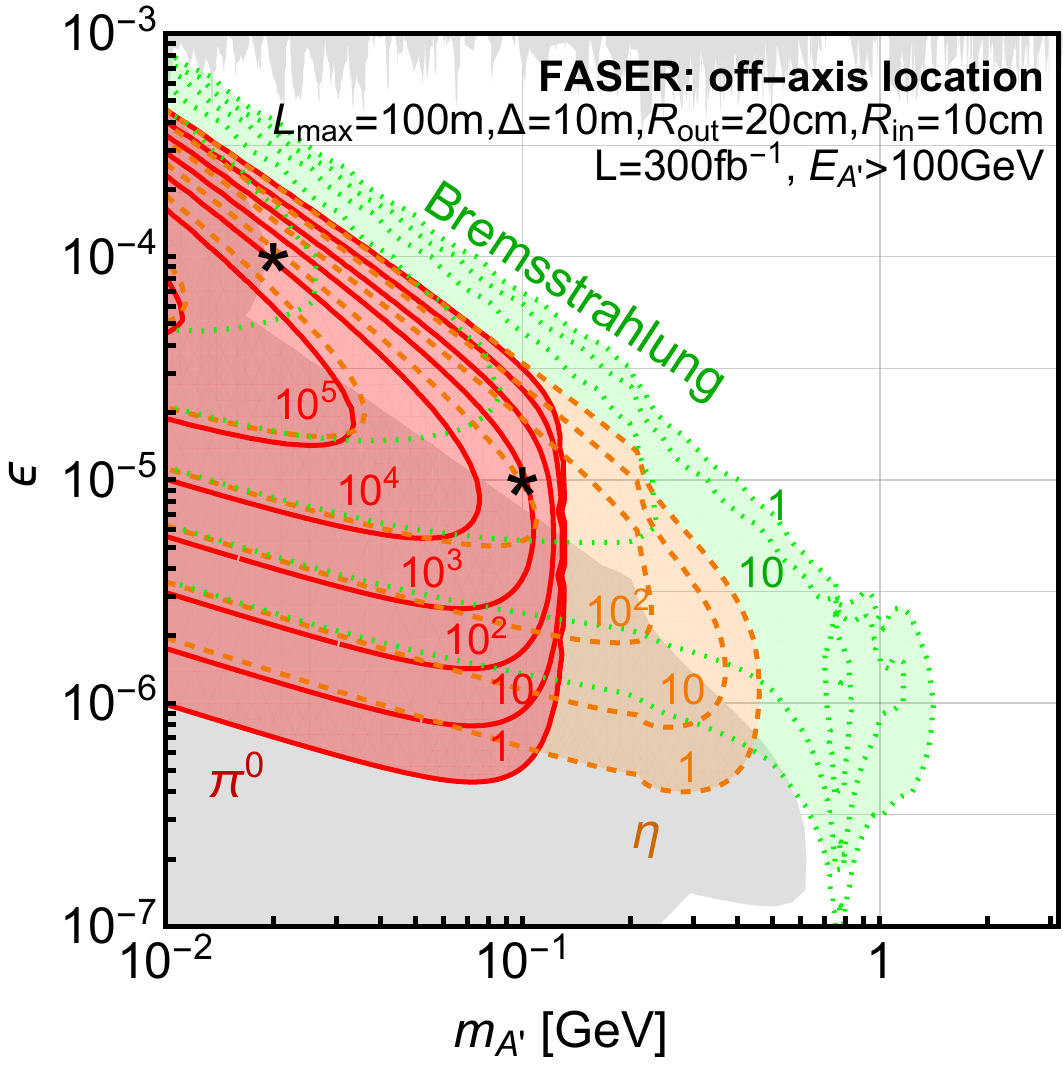} \quad
\includegraphics[width=0.47\textwidth]{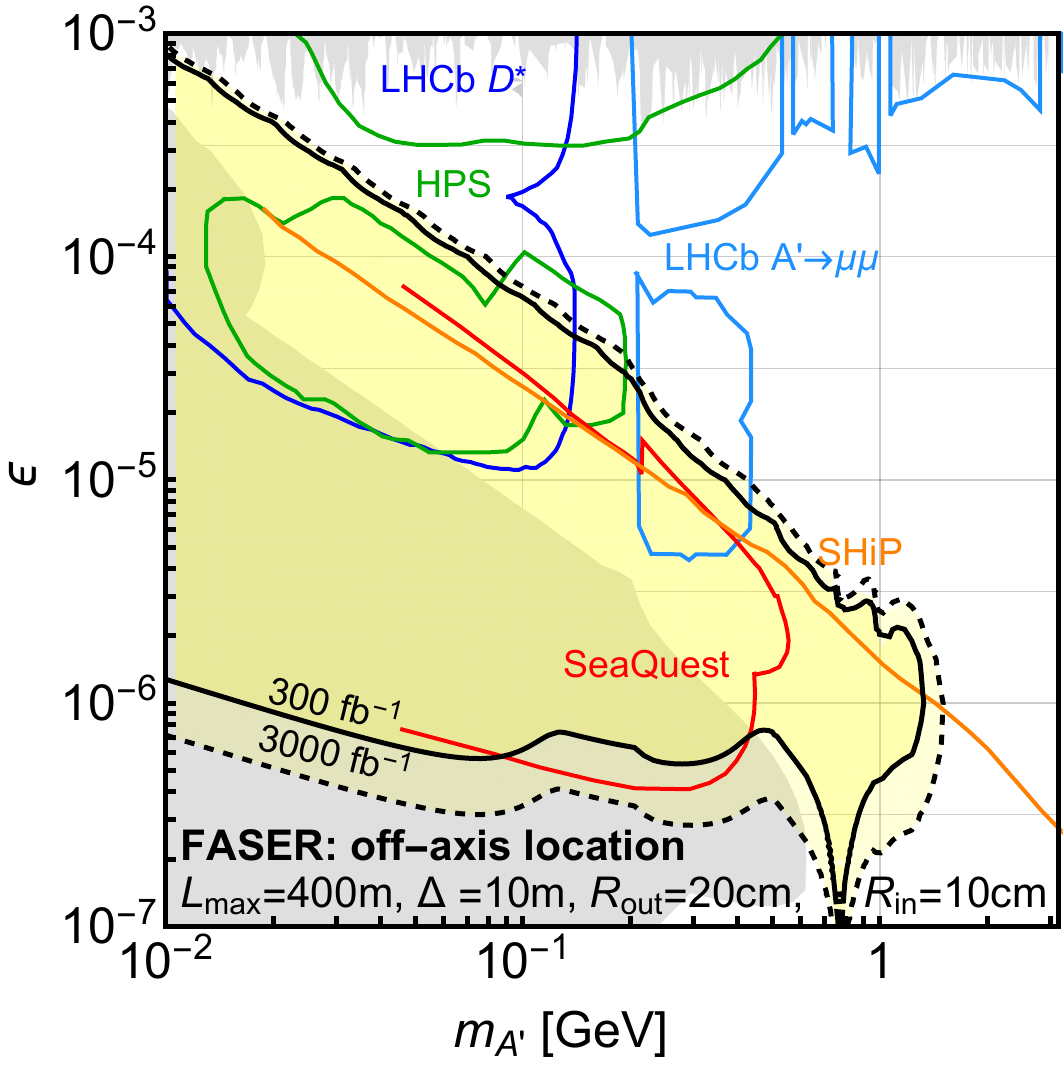}
\caption{Results for the off-axis detector design in dark photon parameter space.  The gray shaded regions are excluded by current experimental bounds. Left: Number of signal events given an integrated luminosity of $300~\ifb$ at the 13 TeV LHC. The different colors correspond to the three production mechanisms $\pi^0\to A' \gamma$ (red), $\eta\to A' \gamma$ (orange), and proton bremsstrahlung (green).  Contours represent the number of signal events $N_{\text{sig}}$.  Right: Combined 95\% C.L. exclusion reach for an integrated luminosity of $300~\ifb$ (solid line) and $3~\iab$ (dashed line), assuming negligible background.  The colored contours represent projected future sensitivities of LHCb~\cite{Ilten:2015hya,Ilten:2016tkc}, HPS~\cite{Moreno:2013mja}, SeaQuest~\cite{Gardner:2015wea}, and SHiP~\cite{Alekhin:2015byh}.
}
\label{fig:reach_off_axis}
\end{figure}

\section{Dark Photons from Proton Bremsstrahlung}
\label{sec:brem}

To estimate the cross section for the $2\to 3$ bremsstrahlung process $pp \to p A' X$, we use the Fermi-Weizsacker-Williams method of virtual quanta~\cite{Fermi:1924tc,Williams:1934ad,vonWeizsacker:1934nji} (see also, e.g., Refs.~\cite{Bjorken:2009mm,Liu:2017htz} for recent discussions for $ep$ collisions).  In this approach, one effectively divides this process into a splitting $p\to p'+A'$ and a hard $pp$ scattering. The splitting function, $w(z,p_T^2)$, is convoluted with the following $pp$ scattering cross section that corresponds to the reduced $s' = 2m_p(E_p-E_{A'})$ which takes into account the energy of the emitted dark photon. The total cross section is~\cite{Kim:1973he}
\begin{equation}
\sigma_{pp\to pA'X} = \int{dz}\int{dp_T^2}\,w(z,p_T^2)\,\sigma_{pp}(s') \ ,
\label{eq:sigmabrem}
\end{equation}
where we integrate over $p_T$, the transverse momentum of the dark photon, and $z = p_{A',z}/|\vec{p}|$, the fraction of the initial momentum carried away by the dark photon in the direction of the beam, $p_{A',z}$, with respect to the proton beam momentum $|\vec{p}|$. 

In the rest frame of one of the initial state protons ($p_1$), one treats the other (moving) proton ($p_2$) as an effective source of a cloud of virtual photons, $\gamma^\ast$, that interact with the proton $p_1$ at rest. The weighting function $w(z,p_T^2)$ is then determined based on the matrix element of the $2\to 2$ scattering
\begin{equation}
p_1(p) + \gamma^\ast(q)\to p(p') + A'(p_{A'}) \ ,
\label{eq:2to2}
\end{equation}
where we have denoted the particle momenta in brackets. One typically requires $|q^2|$ to be small, i.e., comparable to $\Lambda_{\text{QCD}}^2$, so as not to break the proton $p_1$ apart.  Dark photons can be emitted from both protons, but, given the experimental energy and angular cuts on the dark photons imposed after boosting to the lab frame, typically only bremsstrahlung from one of the colliding protons plays a non-negligible role. (See the corresponding discussion of the SM photon bremsstrahlung in heavy-ion fixed-target experiments in Ref.~\cite{Mikkelsen:2015dva}.)  From the matrix element of \eqref{eq:2to2}, one obtains the splitting function~\cite{Kim:1973he,Tsai:1973py,Blumlein:2013cua}
\begin{eqnarray}
w(z,p_T^2) & = & \frac{\epsilon^2\alpha}{2\pi\,H}\left\{\frac{1+(1-z)^2}{z} - 2z(1-z)\left(\frac{2m_p^2+m_{A'}^2}{H}-z^2\frac{2m_p^4}{H^2}\right)\right.\nonumber\\
& & \left. +2z(1-z)(z+(1-z)^2)\frac{m_p^2m_{A'}^2}{H^2} + 2z(1-z)^2\frac{m_{A'}^4}{H^2}\right\},
\label{eq:wsplitting}
\end{eqnarray}
where $H = p_T^2 + (1-z)\,m_{A'}^2 +z^2\,m_p^2$. 

When integrating $w(z,p_T^2)$ in \eqref{eq:sigmabrem} to obtain the total cross section, one needs to impose cuts on both $z$ and $p_T^2$ that guarantee that the FWW approach is valid.  The validity conditions can be summarized as $E_{A'}, E_{p}, E_{p'} \gg m_{A'}, m_p, p_T$ in the rest frame of one of the protons~\cite{Blumlein:2013cua}. As discussed in detail in Refs.~\cite{Kim:1973he,Tsai:1973py}, the dominant contribution to the integral comes from regions of phase space where the $\gamma^*$ has minimal virtuality, that is, where
\begin{equation}
|q^2_{\text{min}}| \approx \frac{1}{4\,E_p^2\,z^2\,(1-z)^2}\,\left[p_T^2 + (1-z)m_{A'}^2 + z^2\,m_p^2\right]^2 \ ,
\label{eq:qmin}
\end{equation}
is minimal, where $E_p$ is the incident proton energy in the rest frame of the other proton. We therefore require $|q^2_{\text{min}}| < \Lambda_{\text{QCD}}^2$ as a hard-cut requirement (implemented with the Heaviside function). The requirement on $|q^2_{\text{min}}|$ implies that $z$ cannot be too close to $0$ or to $1$. 

On the other hand, $p_T$ itself is not constrained much by requiring $|q^2_{\text{min}}|<\Lambda_{\text{QCD}}^2$, since it appears in \eqref{eq:qmin} in the ratio $p_T^4/E_p^2$, and $E_p$ is large in our case.  However, the additional condition that one needs to take into account when calculating the number of dark photons going toward \name, is to require the $p_T$ to satisfy
\begin{equation}
\frac{p_T}{p_{A', \text{lab}}^z} < \frac{r}{\lmax} \quad (= \tan\theta_{\text{max}}\approx \theta_{\text{max}}) \ ,
\label{eq:condgeom}
\end{equation}
where $r$ is the radius of the detector, and $\lmax$ is the distance between the IP and the far end of the detector. For our default detector locations, the geometrical acceptance requirement of \eqref{eq:condgeom} introduces an upper limit $\theta_{\text{max}} = 0.5~\text{mrad}$ ($0.27\,~\text{mrad}$) for the on-axis detector at the far (near) location, and $\theta_{\text{max}} = 2~\text{mrad}$ for the off-axis detector design discussed in~\appref{off_axis_detector}. This constrains $p_T$ to values that are small enough for the FWW approximation to be valid.\footnote{The conditions for detector geometrical acceptance imply the following upper limit on the transverse momentum: $p_T\lesssim p_{\text{beam}}\,z_{\text{max}}\,\theta_{\text{max}}$. As examples, in Ref.~\cite{Blumlein:2013cua}, this detector requirement implies $p_T<1~\gev$, while for the SHiP detector~\cite{Graverini:2016}, where the beam energy is $p_{\text{beam}}=400~\gev$, $z_{\text{max}}=0.86$, and $\theta_{\text{max}} \simeq 22\,\text{mrad}$, the requirement implies $p_T \lesssim 8~\gev$. In our case, e.g., for the on-axis detector at far location, it implies $p_T\lesssim 3~\gev$.} 

Other than the geometrical acceptance condition of \eqref{eq:condgeom}, one needs to remember that the FWW approximation depends crucially on dark photons being emitted in the forward direction with small $\theta_{A'} \simeq p_T/E_A' \ll 1$. In our case, since the signal region typically has $E_{A'}\gtrsim 100~\gev$ (see Figs.~\ref{fig:aprimeptheta_weighted-onaxis} and \ref{fig:aprimeptheta_weighted_TAN}), we require $p_T\lesssim 10~\gev$ when presenting results in \secsref{production}{signalanddetector}. This has, however, a negligible impact on our final sensitivity reach plots shown in \secref{results}, because for the detector designs that we study, the geometrical acceptance condition already imposes stronger constraints on $p_T$.

In the above discussion, we considered a coherent emission of a dark photon from a proton. However, for large dark photon masses, and hence large momentum transfers $p_{A'}^2=m_{A'}^2$, the dark photon will be able to feel the proton's internal structure. Following the extended vector meson dominance model~\cite{Faessler:2009tn,deNiverville:2016rqh}, we include a timelike form factor $F_1(p_{A'}^2)$, which also incorporates mixing with the $\rho$ and $\omega$ vector mesons:
\begin{equation}
F_1(p_{A'}^2) = \sum_{V=\rho\, \rho' \rho'' \omega\, \omega'  \omega''} \ \frac{f_V \ m_V^2}{m_V^2-p_{A'}^2-i m_V \Gamma_V} \quad
\end{equation}
where $f_{\rho}= 0.616$, $f_{\rho'}= 0.223$, $f_{\rho''}= -0.339$, $f_{\omega}= 1.011$, $f_{\omega'}= -0.881$, $f_{\omega''}= 0.369$. This effectively cuts off the contribution to the dark photon signal from large $m_{A'} \agt 1~\gev$ where coherent scattering might not be a valid approximation. 

From \eqref{eq:sigmabrem}, it follows that the $A'$ production event rate per one $pp$ scattering is
\begin{equation}
\frac{dN}{dz\, dp_T^2} = \frac{\sigma_{pp}(s')}{\sigma_{pp}(s)}\, w(z,p_T^2) \ ,
\end{equation}
where $s = 2m_pE_p$, $s'=2m_p(E_p-E_{A'})$, and the inelastic cross section is taken from Ref.~\cite{Olive:2016xmw}. The total expected number of events is, then,
\begin{eqnarray}
N_{\text{ev}} &= &N_{\text{tot}}\times |F_1(m_{A'}^2)|^2\times \label{eq:Nevbrem} \\
& &\int{dz}\int{dp_T^2}\,\frac{\sigma_{pp}(s')}{\sigma_{pp}(s)}\,w(z,p_T^2)\,\Theta\left(\Lambda^2_{\text{QCD}}-q^2\right)\,\Theta_{\text{geom}}\left(\frac{r}{L}-\frac{p_T}{p_{A',lab}^z}\right)\,\mathcal{P}_{A'}(\epsilon,m_{A'},p_{A'}) \ ,\nonumber
\end{eqnarray}
where $N_{\text{tot}}\simeq 2.3\times 10^{16}$ is the total number of $pp$ collisions, $\Theta$ denotes the Heaviside function, and $\mathcal{P}_{A'} =  e^{-\lmin/\bar{d}} - e^{-\lmax/\bar{d}} $ is the probability that a dark photon decays within the required distance. In \eqref{eq:Nevbrem}, the geometrical acceptance conditions encoded in $\Theta_{\text{geom}}$ and $\mathcal{P}_{A'}$ are evaluated in the lab (collider) frame.

\bibliography{faser}

\end{document}
\grid